\documentclass[fleqn,usenatbib]{mnras}

\usepackage{newtxtext,newtxmath}
\usepackage{float}

\usepackage[T1]{fontenc}

\usepackage{multicol}
\usepackage{subcaption}
\usepackage{booktabs}
\usepackage{caption}
\usepackage{longtable}
\usepackage{enumitem}
\usepackage{bm}
\usepackage{xurl}
\DeclareRobustCommand{\VAN}[3]{#2}
\let\VANthebibliography\thebibliography
\def\thebibliography{\DeclareRobustCommand{\VAN}[3]{##3}\VANthebibliography}

\usepackage{graphicx}	
\usepackage{amsmath}	

\title[Exploring NGC 7419 membership and variability]{Exploring membership and variability in NGC 7419: An open
cluster rich in super giants and Be type stars}

\author[A. Chakraborty, J. Jose and A. C. Carciofi]{
Arghya  Chakraborty,$^{1}$\thanks{E-mail: 	arghyachakraborty@students.iisertirupati.ac.in}
Jessy Jose, $^{1}$\thanks{E-mail: jessyvjose1@gmail.com}
Alex C. Carciofi$^{2}$\thanks{E-mail: : carciofi@usp.br}
\\
$^{1}$Department of Physics, Indian Institute of Science Education and Research Tirupati, Yerpedu, Tirupati - 517619, Andhra Pradesh, India\\
$^{2}$Instituto de Astronomia, Geofísica e Ciências Atmosféricas, Universidade de São Paulo, São Paulo, SP 05508-090, Brazil\\
\\
}

\date{Accepted XXX. Received YYY; in original form ZZZ}

\pubyear{2025}

\begin{document}
\label{firstpage}
\pagerange{\pageref{firstpage}--\pageref{lastpage}}
\maketitle

\begin{abstract}
NGC 7419 is a young open cluster notable for hosting five Red Supergiants and a high abundance of Classical Be (CBe) stars. CBe stars are main sequence non-supergiant B-type stars, which 
exhibit or have exhibited Balmer line emissions in their spectra. We perform a membership analysis using Gaia DR3 data and machine learning techniques like Gaussian Mixture Models (GMM) and Random Forest (RF) and determine the cluster's mean distance to be ${3.6^{+1.0}_{-0.6}}$ kpc. We identify 499 Gaia-based members with a mass above $\sim$ 1.2 M$_\odot$, and estimate the cluster's age to be $21.1 ^{+1.6}_{-0.6}$ Myr. Using our revised  $H\alpha$ excess-based analysis, we find 42 CBe stars containing many known CBe stars, bringing the total number of CBe stars in NGC 7419 to 49 and the fraction of CBe to (B+CBe) members to 12.7\%. We investigate the variability of the candidate members from ZTF and NEOWISE data using Standard Deviation, Median Absolute Deviation, and Stetson Index (J), and their periodicity using the Generalized Lomb Scargle Periodogram variability. We find that 66\% of CBe stars are variable: 23\% show periodic signals, typical of pulsation/rotation, 41\% display variability characteristic of disk dynamics or binarity, and 14\% exhibit long-term changes, consistent with disk dissipation/formation. We also find that all pulsating CBe stars are early-type, while 50\% of stars with long-term variations are early-type, and the other 50\% are mid-type. Our results agree with previous findings in the literature and confirm that CBe stars display variability through multiple mechanisms across different timescales. 
\end{abstract}

\begin{keywords}
membership -- CBe stars -- variability -- periodicity 
\end{keywords}


\raggedbottom
\section{Introduction}

 Classical Be  (CBe) stars are a special class of B type stars on the main sequence. \citet{collins_1987} gave the definition of a Be star as "a non‐supergiant B star whose spectrum has, or had at some time, one or more Balmer lines in emission". The first Classical  Be star was $\gamma$ Cassiopeiae, observed in 1866 by Angelo Secchi, the first star ever observed with emission lines.
In general,  CBe stars are found to have high rotational velocity resulting in mass loss. This loss of mass paired with rapid rotations results in a \textit{decretion disk} (opposite to accretion to signify the direction of mass transport) in these stars. This is thought to be the mechanism for the presence of circumstellar material in these stars, which manifests as emission lines, mostly Balmer lines and some Fe lines,  in their optical spectra. The viscous decretion disk model  (\citealt{1991Lee,Carciofi:2011fc}) is considered to be the best description for the circumstellar disk of a Be star, where the disk around Be stars are sustained by the discrete events of mass ejection from the stellar surface, known as 'outburst' (\citealt{1997kroll,kee2015pulsationalmassejectionstar,1998rivinius, Grundstrom2011,2018Labadie}).

CBe stars often show variability in different timescales (\citealt{2013Rivinius,2003PorterCBe}) and references therein). These periodic signals may arrive from different mechanisms: a) Variations in brightness on timescales of weeks to decades due to disk formation or loss episodes \citep{Rimulo2018}, b) quasi-cyclic variations due to wavy motions in the disc on timescales of months to years \citep{okazaki2006,carciofi2009}, c) Variations arising from binarity in the timescales of weeks to months \citep{2016Panoglou,2018Panoglou} and d) short term periodic or stochastic variations due to stellar pulsation and/or rotation on timescales of 0.2 to 3 days \citep{2003Rivinius}. \cite{CBemultiperiod} presented a comprehensive analysis of the various types of variability observed in CBe stars, also confirming the presence of grouped multi-periodicity that was previously believed to be present in these stars. More recently, \cite{2025Labadie} reported detailed photometric and spectroscopic observations of discrete mass ejections, confirming that they are universally asymmetric in nature--i.e., they likely originate in a small region along the stellar equator.

Although it is known that disk evolution in CBe stars is driven primarily by viscosity, the mechanism of the formation of such disks still remains a mystery. Recent results show that the formation of such disks due to mass ejection is linked to viscosity \citep{Carciofi:2011fc}. For a long time, near-IR excess compared to normal B-type stars has been detected in CBe stars, the first being in \cite{1967Johnson}. This excess was attributed to free–free and bound-free processes in the circumstellar material (gaseous disks) around these Be stars \citep{1994dougherty}. In this context, \cite{2015Viera,Viera2017} comprehensively studied the dynamical evolution of disks by developing the concept of \textit{pseudo-photosphere}. \cite{2018Granada} shows that the trends for a Be star's disk activity, spectral type, and variability can be obtained from its color excess in a color-color diagram. \cite{2024A&A...682A..59J} found their results to be in accordance with this work. This can help bridge the gap in understanding the evolution of disks in CBe stars. It is well documented in previous literature, like \cite{2013Rivinius} and references within, that CBe stars are mostly early B-type stars.

\begin{figure}
\includegraphics[scale=0.6,trim=30 0 40 20,clip]{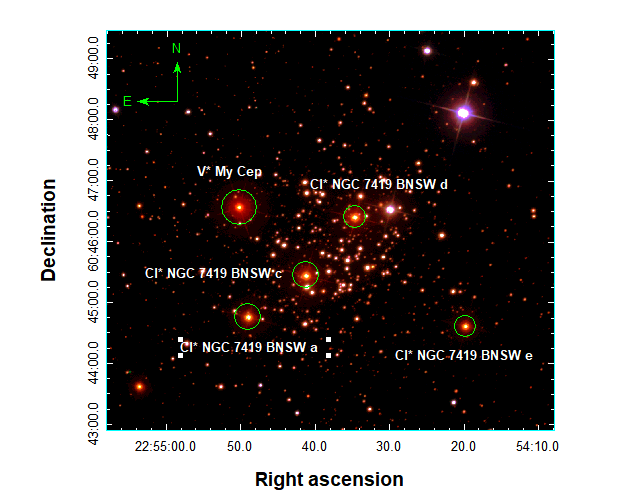}
\caption{RGB color image of NGC 7419 using  observations by filters $g$, $r$ and $i$ of the SDSS telescope for an area of 7 arcmin x 6 arcmin. The 5 Red Supergiants of this cluster are marked with green circles.}
\label{NGC 7419}
\end{figure}

In this paper, we focus on the young cluster NGC 7419 (Fig. \ref{NGC 7419}), located at a distance of $\sim$ 2.9 -- 4 kpc (\citealt{Marco_2013,subramaniam,Beauchamp}) with previously reported ages ranging from 14 -- 15 Myr (\citealt{Marco_2013,Beauchamp}) to 22 -- 25 Myr (\citealt{subramaniam,Joshi}). These studies also show that NGC 7419 is rich in CBe stars. \citet{Marco_2013} shows that NGC7419 has a very high fraction ($\sim$ 40 $\%$) of Be stars among the early B type stars, whereas very few Be type stars at later spectral class. They also report strong variability in the emission characteristics of Be-type stars in the cluster. The abundance of CBe stars in NGC 7419 suggests that there should be some mechanism at work by which a large fraction of the early B-type stars possess high rotational velocity \citep{subramaniam}. NGC 7419 also has 5 notable Red Supergiants, with no blue supergiants, the highest in any cluster till the end of the 20th century. The Red Supergiants in the cluster are i)  Cl* NGC 7419 BNSW d,
ii) V* MY Cep,
iii)Cl* NGC 7419 BNSW c,
iv) Cl* NGC 7419 BNSW a,
v)  Cl* NGC 7419 BNSW e,
with V*  My Cep being the most luminous one. 

Although NGC 7419 has a large number of CBe stars and 5 Red Supergiants, a detailed membership analysis of the cluster and the characterization of its CBe stars are still missing. 
In this paper, we perform the membership analysis of cluster NGC 7419  using Gaia DR3 data and also perform a variability analysis of the bright members of the cluster in optical and IR wavelengths. Added goals of this work \textbf{are} to analyze the properties of the CBe-type stars, search for variability, and possibly classify them into periodic and non-periodic variables. 
Also, Red Supergiants often show slow, irregular variability with no detectable periodicity \citep{2006}, and in this paper, we analyze the variability of Red Supergiants of the cluster given their lightcurves satisfy certain quality criteria.  

The paper is organized as follows. Section \ref{Section2} deals with various data sets used for the analysis, Section \ref{membershipsection} describes the membership analysis, and Section \ref{IPHASsection} is on finding possible CBe candidates based on sources that show $H\alpha$ excess. Sections \ref{ztfvarsec} and \ref{neovarsec} the variability analysis of the candidate members in the cluster using optical and infrared time series data. Section \ref{discussion} discusses the correlation between WISE CMDs and global trends in Be star's disk activity, spectral type, and variability, and the observation of different time scales of variability due to different mechanisms seen in CBe stars. Finally, Section \ref{Summary} concludes the results.

\section{Data Sets Used}\label{Section2}
Various archival data sets from the following surveys are used for membership analysis as well as  to search for variable stars within the cluster NGC 7419. Details of individual data set are  given below.  

\subsection{Gaia DR3}\label{gaia data}
Gaia, a European Space Agency observatory launched in 2013, was originally set to operate from 2014 to 2019 and was extended to operate until around January 10th, 2025. It measures star positions, distances, and motions with high precision. The third data release (Gaia DR3) (\citealt{gaia1,gaiadr3paper}) offers five-parameter astrometry and three-band photometry for about 1.5 billion stars, significantly improving upon Gaia DR2 \citep{gaiadr2paper} with a 30\% increase in parallax precision, double the accuracy for proper motions, and reduced systematic errors.
The photometry also features increased precision but, above all, much better homogeneity across color, magnitude, and celestial position. A single passband for G, BP, and RP is valid over the entire magnitude and color range, with no systematics above the 1 percent level. Vizier \citep{vizier2000} was used to retrieve the necessary Gaia DR3 data.
 We use Gaia DR3 data to perform the membership analysis and determine the fundamental parameters of NGC 7419. 
 Given the tidal radius of the cluster reported to be around 5 arcmin \citep{NGC7419Radius}, we extracted the Gaia DR3 data, which is centered around the coordinates resolved by SIMBAD, i.e., RA = 22:54:18.96; Dec = +60:48:50.4 for a conservative area within 12 arcmin radius along with the following constraints: $0.1$\,mas < $\pi$ < $0.8$\,mas and non-null values for PMRA (proper motion in right ascension), PMDEC (proper motion in declination), ra (right ascension), dec (declination), Gmag (broadband NUV to NIR), BPmag (blue) and RPmag (red). We consider sources with renormalized unit weight error (RUWE < 1.4, \citealt{lindegren2021}) to avoid blended objects. The above parallax range was chosen,  which encompasses the majority of the known members of the cluster (see details below), including the supergiants, as well as to exclude the field contaminants with extreme parallax values. A total of 2913 sources are retrieved within these criteria.
 
\subsection{ZTF (Zwicky Transient Facility) DR18}

The Zwicky Transient Facility is a wide-field sky astronomical survey using a new camera attached to the Samuel Oschin Telescope at the Palomar Observatory in California, United States \citep{2018}. It was commissioned in 2018 and is named after the astronomer Fritz Zwicky. With time series observations in $r$, $g$ and $i$ bands, the Zwicky Transient Facility is designed to detect transient objects that rapidly change in brightness, for example supernovae, gamma ray bursts, and collision between two neutron stars, and moving objects like comets and asteroids \citep{Bellm_2018}. 

For candidate members of NGC 7419, we use the ZTF DR18, made available on July, 2023. This release adds two months of observations to the DR17 data, up to May 7th, 2023, for the public portion of the survey and private survey time before Jan 4th, 2022. We use the ZTF $r$ band data in the fields with ID 1837 and 831 within a 12' radius of NGC 7419 as it had the highest number of crossmatches with the candidate members identified from Gaia DR3 (see Section \ref{membershipsection}). The other two bands,  $g$ and $i$,  had very low average number of observations per lightcurve, and hence we do exclude them. We used the $r$ band data for only those sources which had more  than 50 good quality observations (catflags < 32768). 
Data within 100 days in each lightcurve was considered as one epoch, and only magnitudes in each epoch that fall within the range of median value $\pm$ 2 Standard Deviations of that epoch were considered. The median cadence for all crossmatched lightcurves was $\approx$ 1 day.

\subsection{Wide-field Infrared Survey Explorer (WISE) and NEOWISE Reactivation Database (2023)}

We used the mid-IR photometric data obtained by the Wide-field Infrared Survey Explorer (WISE; \citealt{WISE}). WISE was launched in
2009 and performed its cryogenic all-sky survey for about a year in four bands: W1 (3.4 $\mu$m), W2 (4.6 $\mu$m), W3 (12 $\mu$m), and W4 (22 $\mu$m). We use data in the W1 and W2 filters, which have saturation limits of W1 = 8 mag and W2 = 7 mag \citep{2012wise.rept....1C}.

NEOWISE utilizes the space telescope Wide-Field Infrared Survey Explorer (WISE). The WISE spacecraft was brought out of hibernation in September 2013 and renamed as NEOWISE with a mission to detect and characterize asteroids and comets, and to learn more about the population of near-Earth objects that could pose an impact which could be hazardous to the Earth. The NEOWISE 2023 Data Release includes the single-exposure images and photometry in 3.4 and 4.6 $\mu m$ (W1 and W2) bands that were acquired between 13th December, 2020 and 13th December, 2021 UTC, along with data gathered from the previous 8 years  (\citealt{Mainzer2011,2014ApJ...792...30M}) in multiple epochs. An epoch in NEOWISE data typically spans 180 days, with each epoch containing anywhere from 4 -- 30 observations with a near uniform cadence of 0.066 days.

We considered unsaturated sources with SNR (Signal to Noise Ratio) greater than 3 and having at least 30 good-quality observations. The low value of SNR was opted to obtain multi-epoch photometry for a maximum number of targets as possible. 

\subsection{The INT Photometric $H\alpha$ Survey of the Northern Galactic Plane (IPHAS)}

The INT Photometric $H\alpha$ Survey of the Northern Galactic Plane (IPHAS) is a 1800 square degrees CCD survey of the northern Milky Way spanning the latitude range -5$^\circ$ < b < +5$^\circ$ \citep{Drew}. The data in this survey consists of observations in the $H\alpha$ narrow-band along with Sloan r$^\prime$ and i$^\prime$ broad-band filters. Any source with SNR>5 is included here. This survey has the saturation limits at 13, 12, and 12.5 magnitudes in  r$^\prime$ ,  i$^\prime$ , and  $H\alpha$ bands, respectively \citep{IPHASDR2}.

\section{Membership analysis of NGC 7419 using Gaia DR3 and machine learning techniques}\label{membershipsection}

The groundbreaking studies by \cite{sanders1971} and \cite{vasilevskis} utilized proper motion measurements of stars to verify their membership. They employed a bivariate Gaussian mixture model (GMM) to represent the distribution of stars in the vector point diagram (VPD). Subsequently, \cite{kozhurina} enhanced the membership probability assessments by incorporating the celestial coordinates of the stars along with their proper motions.
Then, in recent history, photometric and astrometric measurements, celestial coordinates were used together for unsupervised and supervised algorithms in the works of \citealt{Sarro2014,galli2020,Statistics01randomforests,pedregosa2018scikitlearnmachinelearningpython,Das_2023,gupta2024}, to name a few.
We use a combination of an unsupervised learning algorithm, Gaussian mixture model (GMM), and a supervised learning algorithm, Random Forest (RF), in order to identify the membership of the sources within 12' radius of the cluster using Gaia DR3 data (see \citealt{Das_2023,2025Das,gupta2024} for details). The steps for the analysis are detailed below. 

\subsection{Membership using Gaussian Mixture Model}\label{Membership}

A Gaussian mixture model (GMM) is a probabilistic model that assumes all the data points are generated from a mixture of a finite number of Gaussian distributions with unknown parameters. 
This algorithm is based on unsupervised learning and is a soft probabilistic clustering, giving probabilities of cluster membership instead of just membership. 
The main difficulty in learning Gaussian mixture models from unlabeled data is that they usually do not know which points came from which cluster (component) \citep{Pattern}.  To solve this problem, an iterative statistical model, known as expectation maximization, is used, which assumes random Gaussians for the data points to get a preliminary clustering and then uses the obtained clustering to form a new Gaussian. This process is repeated iteratively until convergence criteria like "no further change in cluster assignments" are reached \citep{Probabilistic}.

\cite{NGC7419Radius} reports a tidal radius of 5 arcmin for NGC 7419. Thus,  we used a reasonable set of 1031 sources within a 5 arcmin radius for GMM. Choosing a small radius like this will ensure that the astrometric and photometric properties of the stars in this region should represent the whole complex, and also, at the same time, the field-star contamination would be as minimal as possible. After running a GMM analysis of the 1031 sources, we find 525 high-probability members with a probability greater than 0.9. This is reasonable as previous works along these lines use probability of 0.8 as their membership cutoff (\citealt{Das_2023,2025Das,gupta2024}). The various features used for this training set were PMRA, PMDEC, parallax, ra, and dec.  

The Gaussian Mixture Model classified 4 of the 5 supergiants as members. The remaining supergiant (Cl* NGC 7419 BNSW a) was left out probably because it had an anomalously higher PMRA value compared to the other supergiants. We now have a training set consisting of high probability members and non members that can be used in the Random Forest classifier.

\subsection{Membership using Random Forest (RF) analysis}

Compared to GMM, Random Forest (RF) \citep{Statistics01randomforests} is a form of Supervised Learning that requires a prior training set (see \citealt{Das_2023, 2025Das,gupta2024} for details). It uses an ensemble of uncorrelated decision trees for prediction. Since the tree models are not correlated, an ensemble of such uncorrelated tree models will be less biased than a single such model. Each individual tree in the random forest gives a class prediction, and the class with the most votes becomes our model’s prediction. These decision trees are trained using the training set, and to decrease correlation between trees, random sampling with replacement on the training set is performed. This sampling is done on data points themselves as well as features for classification.

We use the output from the GMM analysis to create a training set for the RF algorithm. We consider the 2913 sources within a 12 arcmin radius of the cluster as a test set for the RF algorithm and the 5 arcmin data obtained from GMM analysis consisting of high probability members and non-members as the training set.

After running the RF algorithm, 562 sources are classified as members, which satisfies the probability criteria greater than 90\%. The algorithm classifies 3 out of the five supergiants in the cluster as members, whereas the remaining two supergiants are manually included in the member list as they are already well-known cluster members from previous studies (\citealt{Marco_2013,Beauchamp}) and had a probability of being cluster members of 0.86 and 0.87 respectively. This takes the number of probable members within a 12 arcmin radius to 564.

\begin{figure}
\includegraphics[width=\columnwidth]{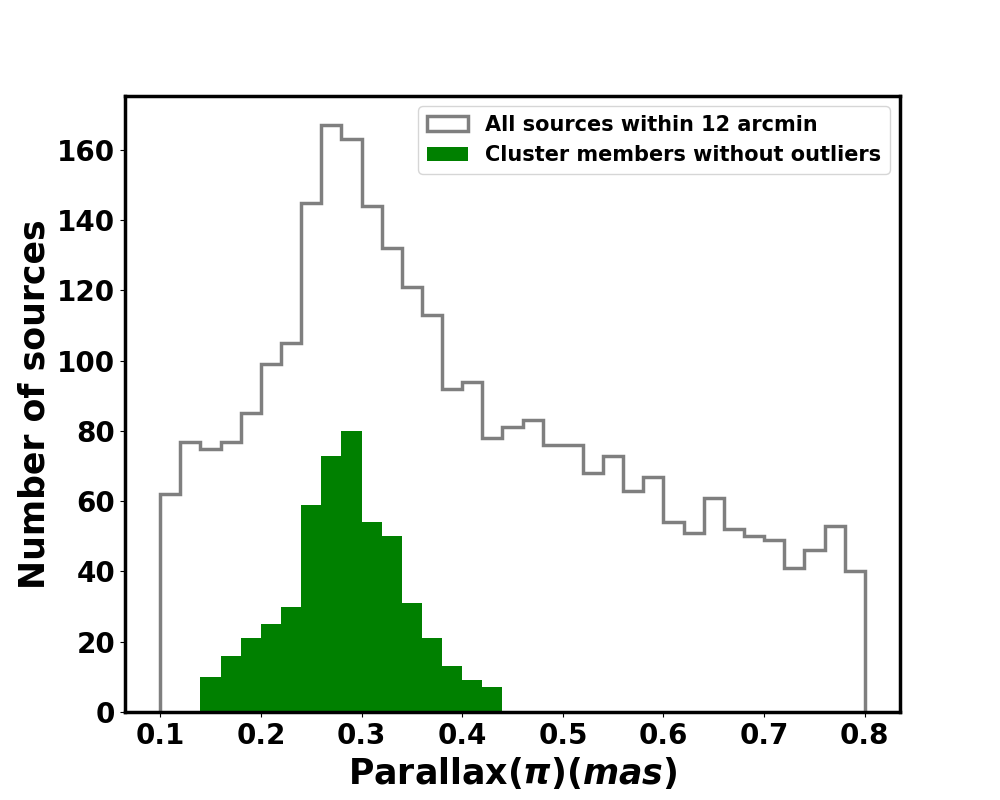}
\caption{Histogram distribution of parallax of all the sources  (grey) within 12 arcmin radius of the cluster NGC 7419 along with the candidate members (green) selected based on the  Random Forest classification. See text for details.}
\label{2}
\end{figure}

\begin{figure}
\includegraphics[width=\columnwidth]{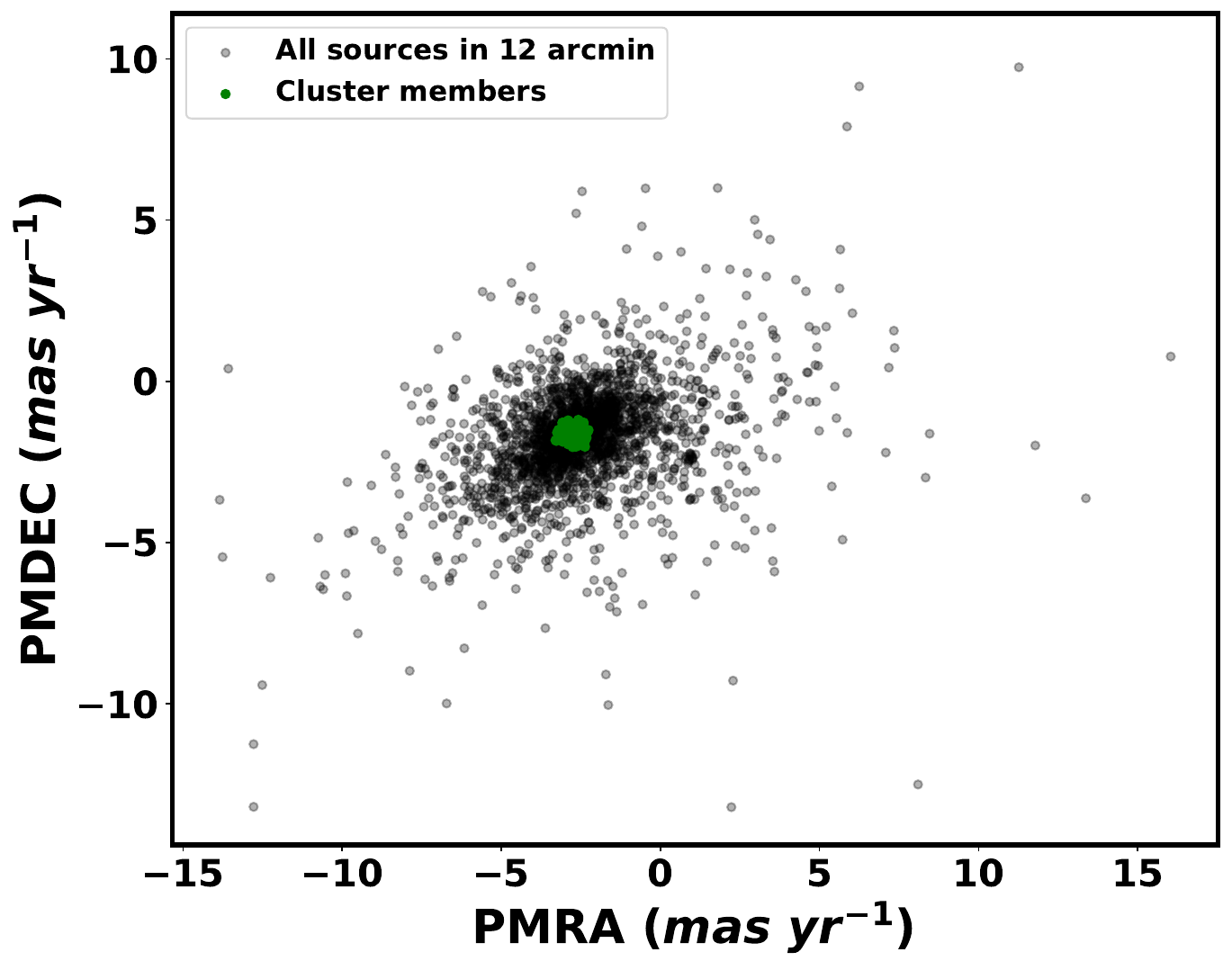}
\caption{PMRA vs. PMDEC distribution of all the sources within 12 arcmin radius of the cluster (grey dots) along with the probable candidate members selected based on the Random forest classification (green dots).}
\label{3}
\end{figure}


From RF analysis, the candidate members we identify have a parallax distribution of 0.11 -- 0.52 mas, which corresponds to a huge range in the distance, from 2000 to 9000 pc, with the majority of them peaking around 3400 pc. In order to further filter out the possible field contaminants, we perform a box-plot analysis, and the outliers are removed by considering only sources within 1.5 times the inter-quartile range (IQR) of the median parallax value. The resultant sources have parallaxes within 0.14 -- 0.43 mas with a mean value of 0.283 mas. After removing the outliers, 499 candidate members remained. Among these members, 70\%  were present within 5 arcmin of the center (coordinates in Section \ref{gaia data}), and 90\% were present within 10 arcmin from the center. The candidate members with their parallax, proper motion, and magnitude information are given in Table \ref{members}. They have mean proper motions of -2.73 $\pm$ 0.15 ${\mathrm{mas \ yr^{-1}}}$ and -1.60 $\pm$ 0.15 ${\mathrm{mas \ yr^{-1}}}$  in RA and DEC, respectively. 

Figure \ref{2} represents the parallax ($\pi$) distribution of all the sources within 12 arcmin radius of the cluster, along with parallax for the candidate members highlighted in green.
  
Figure \ref{3} represents the PMRA vs. PMDEC plot for the RF-selected candidate members against all the sources within 12 arcmins of the cluster. 
Both Figures \ref{2} and \ref{3} show that the above algorithms efficiently identify the candidate members of the cluster from a large pool of field stars, which otherwise would have been difficult. 

We compare the candidate members identified in the above analysis with the spectroscopically identified membership of the cluster by  \citet{Marco_2013}. They had  141 spectroscopically confirmed B-type members between  5 and 14 M$_\odot$. Cross-matching our candidate members with their sample within a match radius of 1 arcsec, we found 127 common sources in both catalogs. Thus, our membership includes 90\% of cluster members found in \cite{Marco_2013}. Figure \ref{8} is a BP-RP vs. G color-magnitude diagram (CMD) of all the Gaia DR3 sources within a 12 arcmin radius of the cluster, along with the 499 candidate members highlighted in green and the 127 common sources found in \citet{Marco_2013} in red.   
Lastly, the supergiants are not present in the 127 common sources since the spectroscopy study had identified only B-type stars. 
From the CMD in Fig. \ref{8}, the newly identified cluster members overlap well with previously known sources, and the current analysis is detecting candidate members 2 mag deeper than the previous spectroscopic analysis.

\subsection{Distance and age of the cluster}\label{distance age section}

The mean and Standard Deviation of the parallax of the 499 candidate members identified in Section \ref{membershipsection} is 0.28 $\pm$ 0.06 mas, which corresponds to the mean distance to the cluster as ${3.6^{+1.0}_{-0.6}}$ kpc.

Although Gaia provides high-quality photometric data, including parallax, simply inverting the parallax to get the distance leads to biases that tend to get larger for larger parallax uncertainties as well as distant objects \citep{2021bailerjones}. Hence, a proper statistical treatment is needed to avoid this biasing, the results of which are compiled in the Bailer-Jones Catalogue \citep{2021yCat.1352....0B}. Geometric distances are computed using direction-dependent priors, while photogeometric distances are computed using colors and apparent magnitude. We also estimate the distance to the cluster using this catalog to compare it with the above-estimated distance. In the Bailer-Jones Catalogue, we prefer photogeometric distance over geometric distance as NGC 7419 is quite a distant cluster, which implies higher uncertainties associated with parallax data as described in \citet{2021yCat.1352....0B}. 
After removing the outliers, we estimate the distance to the cluster using this catalog to be 3.3 $\pm$ 0.6 kpc, which agrees with the above measurement, within uncertainty. 

\citet{subramaniam} and \citet{Marco_2013} estimated average distances of 2.9 kpc and 4 kpc, respectively, for the cluster. Both these studies used zero-age main sequence fit to the CMD to get the distance after accounting for the reddening towards the cluster. Since the distance estimated from the mean parallax method is in agreement with the \citet{2021yCat.1352....0B}  catalog, 
hereafter we consider our calculated distance to the cluster as the cluster distance, which is ${3.6^{+1.0}_{-0.6}}$ kpc.
In order to obtain the average age of the cluster, we obtain isochrones of several ages from 8 to 30 Myr (based on previous literature) from PARSEC models \citep{2012MNRAS.427..127B} and correct them for the cluster distance of 3.6 kpc and extinction, $A_V$. We take the extinction, ${A_V}$, towards the cluster to be 5.2 mag from previous literature \citep{bhatt1993,subramaniam,Marco_2013} and correct each isochrone using equation 1 and the extinction coefficients provided in \cite{gaiacoefficients}. We interpolate the data points in the isochrone.  We then did a chi-square calculation (see Fig. \ref{chi-square isochrone} in Appendix) using the interpolated magnitudes and observed magnitudes in the following way:

\begin{equation}
    \chi^2 = \sum_{i=1}^{N} \frac{(O_i - E_i)^2}{\sigma_i} \,,
\end{equation}

\noindent where $O_i$'s are the observed magnitudes and $E_i$'s are the interpolated isochrone magnitudes, and $\sigma_i$'s are the corresponding uncertainties in the observed magnitudes.

The best-fitting average age of the cluster corresponding to the minimum chi-square value comes out to be around $21.1 ^{+1.6}_{-0.6}$ Myr. We determine the uncertainty by taking the upper and lower bounds within 5\% of the best-fit age. This age estimate does not include the effects of parallax uncertainty. 
The uncertainty in the parallax measurement leads to an average spread in the distance modulus of around 0.45 mag. 
However, this spread in the distance modulus does not seem to have a significant effect on the age estimation from our chi-square minimization method. This can be because the chi-square method is not sensitive enough, and a Monte-Carlo method might be better suited for this \citep{2019GaiaageDR2}. The isochrone of 21.1 Myr fits well with the five supergiants lying towards the upper right of the CMD as well as the main sequence distribution of the members, which is evident in Fig \ref{8}. Since Gaia G band magnitudes have a lower limit of around 20 mag (\citealt{gaiadr3paper,2023Marton}), we can find members with masses above $\sim$1.2 M$_\odot$ (for an age of 21.1 Myr, $A_V$ of 5.2 mag, a distance of 3.6 kpc, PARSEC models). Our estimated age is comparable to those reported in previous literature (\citealt{Joshi,subramaniam}).  We estimate the masses of the cluster members by interpolating over the best-fit stellar isochrone (21.1 Myr) using the interp1d function of \textit{scipy} \citep{2020SciPy-NMeth} with the apparent G band magnitudes of the cluster members as input. We compare their estimated masses with the modified table\footnote{\label{mamajek}\url{https://www.pas.rochester.edu/~emamajek/EEM_dwarf_UBVIJHK_colors_Teff.txt}} from \cite{2013PecautMamajek} to derive the spectral type for the members that do not have the previous spectral classification, and stellar radii for all members. The reader is cautioned that these values are only approximate for stars significantly displaced from the isochrone. Combining our estimates with previously known spectral types of members (\citealt{2003Caron, subramaniam,2011subramaniam,Marco_2013}), we find that nearly 75\% of these stars are B type with masses ranging from around {$2 \ \mathrm{to} \ 13\  \mathrm{M}_\odot$. 
Since the cluster is relatively evolved and distant, and because of the limited sensitivity of the Gaia data,  we do not trace most of the pre-main sequence branches in the CMD. For these B-type stars, we calculate their individual critical or break-up velocities using their traditional definition in the framework of Roche Approximation (\citealt{1963ApJ...138.1134C,2013Rivinius, Townsend}),

\begin{equation}
    v_{\mathrm{crit}} = \sqrt{\frac{2}{3}\frac{GM}{R_\mathrm{p}}}\,,
\end{equation}

\noindent where $G$ is the gravitational constant, $M$ is the stellar mass, and $R_\mathrm{p}$ is the polar radius, which is related to the equatorial radius ($R_\mathrm{e}$) by the relation $R_\mathrm{p}$=$\frac{2}{3}R_\mathrm{e}$. One thing to remember is that the radii we use is an average stellar radius instead of the polar radius; hence, our estimates of the break-up velocities are not exact and can be considered lower limits. The results are shown in Table \ref{B-type star table}.

\begin{figure}
\includegraphics[width=\columnwidth]{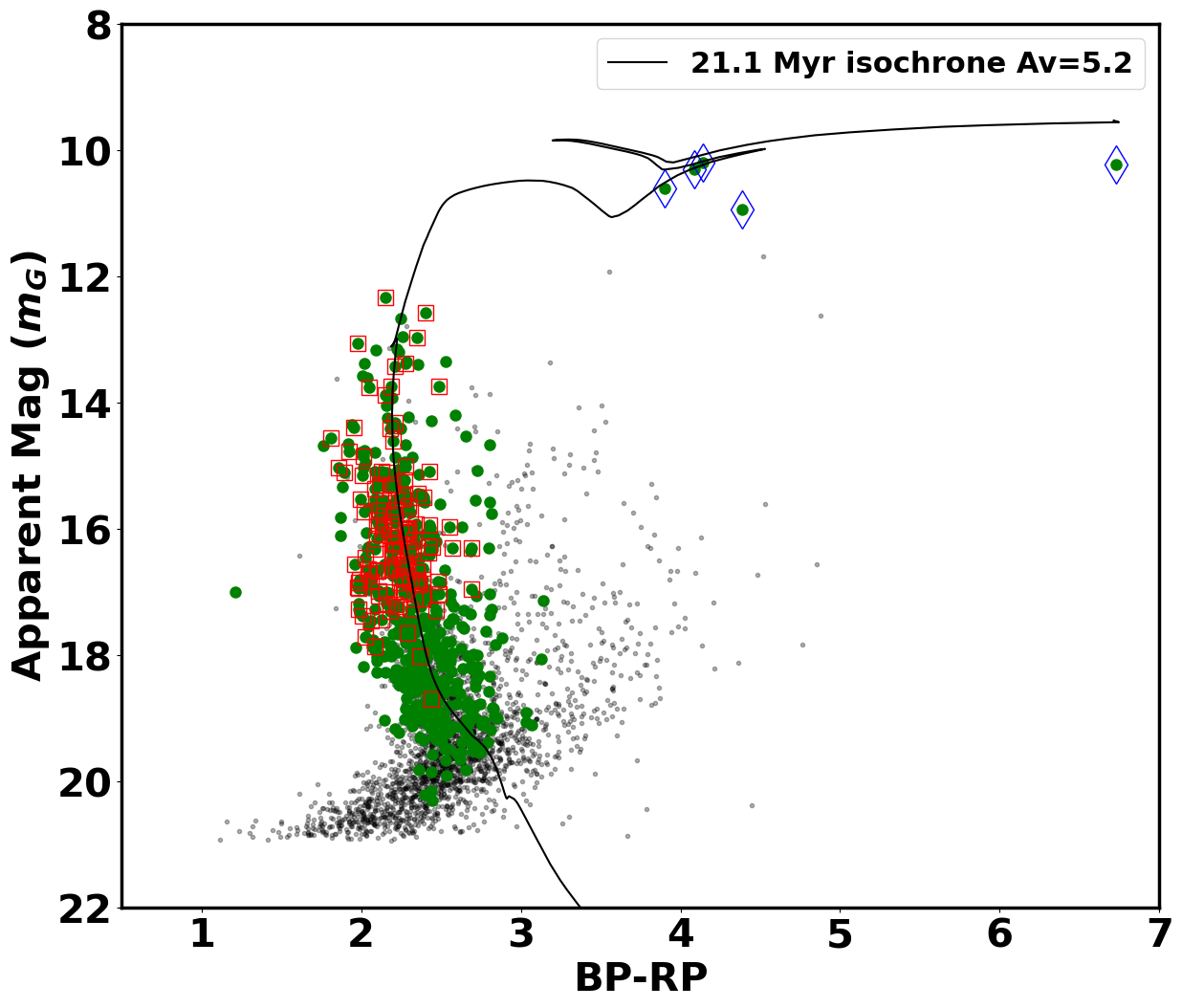}
\caption{Gaia BP-RP versus G band Color-magnitude diagram for all the sources within 12$^\prime$ radius of the cluster NGC 7419 (grey dots). Green dots represent the candidate members selected based on the GMM and RF analysis after removing the outliers. Open red squares are the common sources between this analysis and the spectroscopic member list from \citet{Marco_2013}. Open blue diamonds indicate Red Supergiants within the cluster. The solid curve represents the 21.1 Myr isochrone from PARSEC models \citep{2012MNRAS.427..127B} after correcting for the distance and extinction  $A_{V}$ = 5.2 mag.}
\label{8}
\end{figure}

\section{Search for $H\alpha$ emission line sources}\label{IPHASsection}

We use the IPHAS photometry in r$^\prime$, i$^\prime$, and $H\alpha$ bands to identify the candidate emission line sources within NGC 7419 (\citealt{Barensten,2015Dutta,2024Damian}). Of the 499 candidate cluster members, 490 common sources are obtained in the  IPHAS catalog. We follow the method described in  \citet{Barensten} to identify the $H\alpha$ sources. Figure \ref{iphas} shows the r$^\prime$-i$^\prime$ vs. r$^\prime$-$H\alpha$ color-color diagram for candidate members of NGC 7419. The synthetic main sequence tracks given in \citet{Drew} for ${E(B-V)=0}$, corrected for the cluster reddening $A_V$= 5.2 mag using the extinction relations given in \cite{Barensten}, are shown as a dotted curve in Fig. \ref{iphas}. Those sources lying above the 5 sigma threshold from the main sequence are considered to be $H\alpha$ excess or emission line sources, where $\sigma$ was determined from the photometric uncertainty in the following way (\cite{Barensten}):
\begin{equation}
    (r'-H\alpha)_{\mathrm{star}}- (r'-H\alpha)_{\mathrm{threshold}}>5\sigma\,,
\end{equation}
where, 
\begin{equation}
    \sigma= \sqrt{\sigma^2_{(r'-H\alpha)}+ m^2\sigma^2_{(r'-i')}} \,,
\end{equation}

\noindent and $m$ is the local slope of the threshold.
We find a total of 42 sources to satisfy this criterion, of which 30 are already known CBe stars from previous spectroscopic studies (\citealt{2003Caron,subramaniam,2011subramaniam,Marco_2013}). Hereafter, we update the list of candidate CBe members of NGC 7419 with these 42 sources and the non-duplicate previously known CBe members from the literature. We also see that using a method similar to \cite{2024Navarete} for identifying sources showing $H\alpha$ excess results in nearly identical numbers. This brings the total number of unique candidate CBe members in NGC 7419 up to 49, of which we classify 47 as members in Section \ref{membershipsection}. The 2 CBe stars, which we do not classify as members, did not have Gaia data available at or near their position. Nevertheless, we use the 49 CBe stars in the later sections as the list of CBe stars in the cluster. Given the cluster's age ($\sim$ 21 Myr), these stars cannot be Herbig AeBe types, which typically are of age $\sim$ 1 -- 2 Myr \citep{Herbig}. From previous studies (\citealt{2003Caron, subramaniam, 2011subramaniam,Marco_2013}) and our earlier spectral type estimation in Section \ref{distance age section}, we find 85\% of CBe stars to be early-type (B0.5V-B3V), 9\% (B4V-B6V) to be mid-type, and 6\% (B7V-B9V) to be late-type. The CBe to (B+CBe) fraction for members is 47/371, or 12.7\%. Specifically, the CBe/(B+CBe) fraction for members is 24.8\% for early-type stars, 10.5\% for mid-type stars, and 1.7\% for late-type stars. These fractions align closely with the findings in \cite{2024Navarete}, where it is also observed that higher mass stars exhibit a larger fraction compared to lower mass stars.

\begin{figure}
\includegraphics[width=\columnwidth]{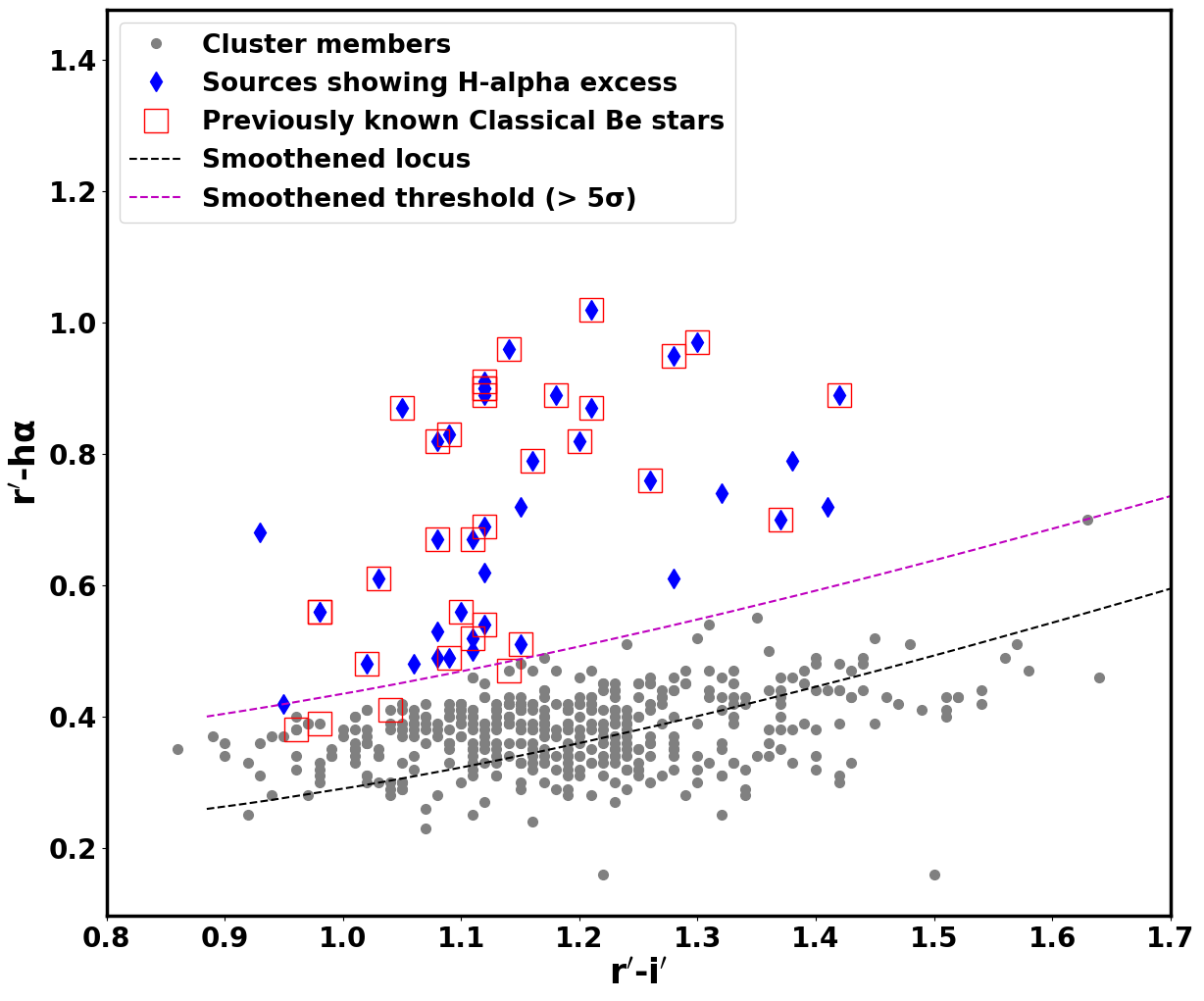}
\caption{IPHAS color-color diagram with main sequence locus (black curve) and 5 sigma threshold (purple curve) showing sources (blue diamonds) with $H\alpha$ excess. The previously known CBe stars are marked in red boxes. }
\label{iphas}
\end{figure}

\section{Identification of variable sources using ZTF data} \label{ztfvarsec}

We obtained 476 crossmatches out of 499 from the ZTF survey in the $r$ band in the fields with ID 1837 and 831. We did not use any other bands as sufficient data points are not available. For the $r$ band, only sources with 50 or more good-quality observations are considered. Among these 476 crossmatches, duplicate lightcurves for the same source position are present, and hence we choose the field with the greatest number of observations for the same. Figure \ref{13} shows the distribution of mean uncertainty in ZTF lightcurves as a function of mean magnitude, calculated as the average for all available epochs, for all the above sources. It is clear that photometric uncertainty becomes exponentially high ($>$ 0.04 mag) for sources fainter than 18 mag in the $r$ band. Hence, only those sources with less than 18 mag are considered further for the variability analysis. All of this combined reduced the number of sources for variability analysis using ZTF data to 288. We filter the individual lightcurves to include only data within 2 Standard Deviations from the median in each epoch group (100 days for ZTF). This choice of 100 days is motivated by the fact that we primarily want to investigate the periodicity of CBe star variability in the short term (from stellar pulsation on timescales of a few hours to days or disk formation or dissipation and wave motions in the \textit{decretion} disk on timescales of a few months). This evidently excludes very long-term periodicity, which is also observed in CBe stars due to disk growth or dissipation \citep{2013Rivinius} and will be investigated in a future paper.  We find 44 CBe counterparts after cross-matching these 288 sources with the updated CBe star list.

Following this, we use three different methods for our variability analysis: Standard Deviation, Median Absolute Deviation, and Stetson Index \citep{1996stetson}.

\begin{figure}
\centerline{\includegraphics[width=\columnwidth]{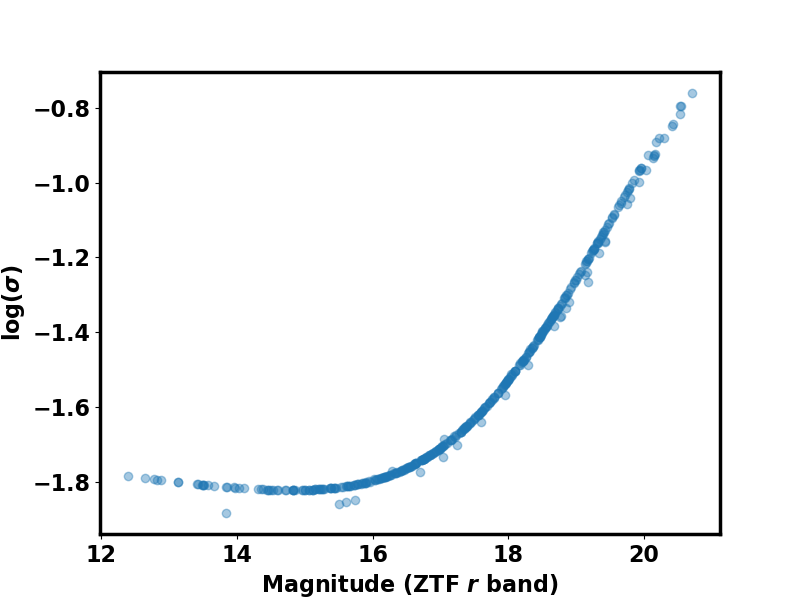}}
\caption{Mean magnitude vs. mean error($\sigma$) in magnitude for sources retrieved from ZTF $r$ band observations. We plot $\mathrm{log_{10}(\sigma)}$ in the $y$ axis for better visibility. }
\label{13}
\end{figure}

\subsection{\textbf{Variability from Standard Deviation and Median Absolute Deviation}}

Standard Deviation (SD) and Median Absolute Deviation (MAD) are fairly common methods for identifying variability \citep{Sokolovsky_2016}. Standard Deviation and Median absolute deviation are defined below, respectively:
\begin{equation}
    \mathrm{Standard \ Deviation}= \sqrt{\frac{\sum_{i=1}^{n}(m_i-\overline{m})^2}{n-1}}\,,
\end{equation}
\begin{equation}
     \mathrm{MAD} = \mathrm{median}(abs(m_{i} - \mathrm{median}(m))\,,
\end{equation}

\noindent where $\overline{m}$ is the mean magnitude and $m_{i}$ is the $i^{\mathrm{th}}$ magnitude of a given lightcurve.

The Standard Deviations (SD) and Median absolute deviations (MAD) are plotted against the respective Mean Magnitudes of the 288 ZTF counterparts in Fig. \ref{MAD and SD ztf}. A main locus has been traced using a least squares fit, which uses the median ($\sigma$) SD and MAD values of the sources instead of the mean so as not to be affected by outliers. The 3$\sigma$ threshold is shown in Fig. \ref{MAD and SD ztf}. Any source lying above the 3$\sigma$ line has been considered as a variable. 
Thus, 113 probable variables are obtained from the SD method, and 105 probable variables are obtained from the MAD method, with 99 crossmatches between the two methods.

\begin{figure*}
\begin{multicols}{2}
     \subcaptionbox{Standard Deviation}{\includegraphics[width=\linewidth]{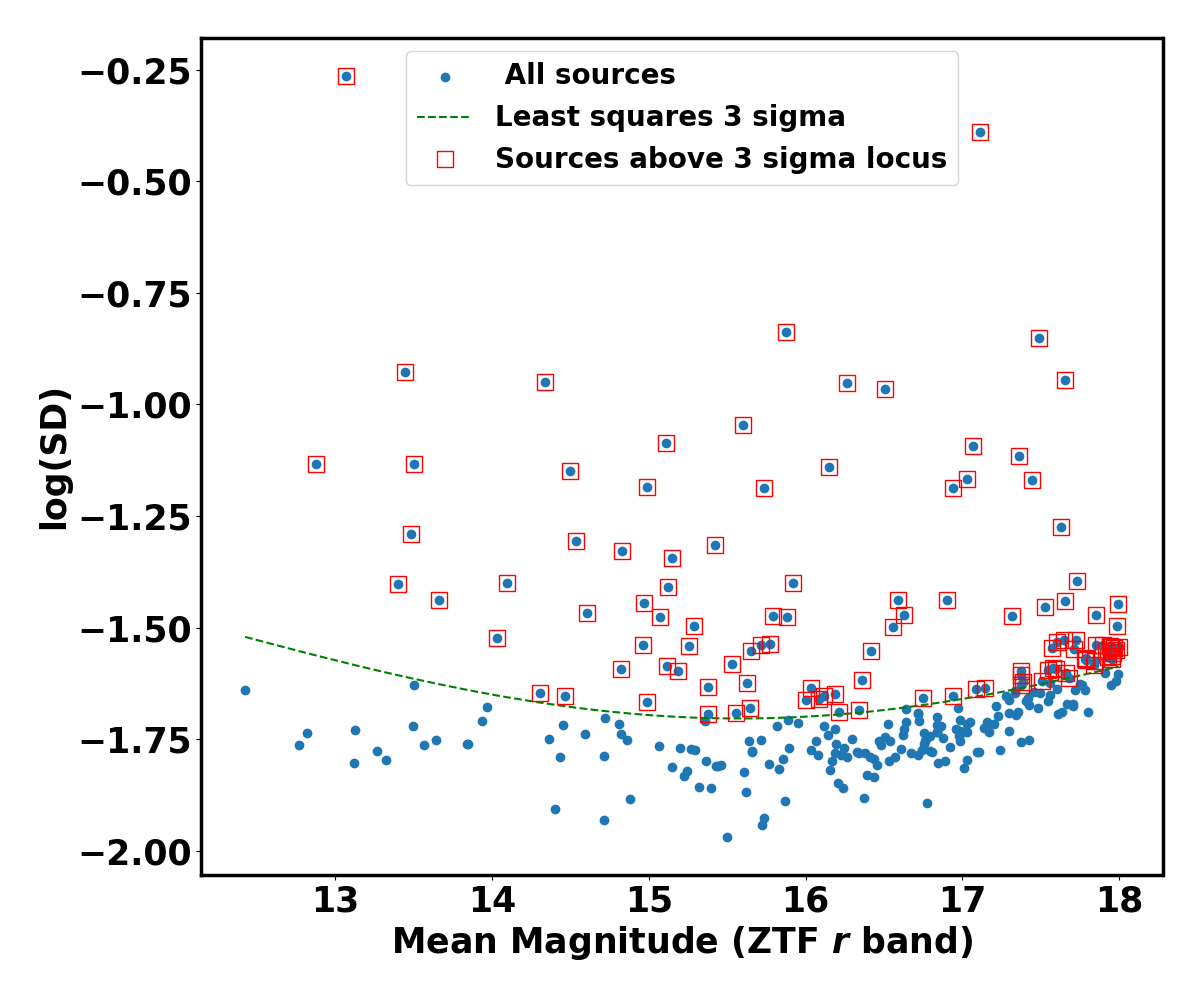}}\par 
     \subcaptionbox{MAD}{\includegraphics[width=\linewidth]{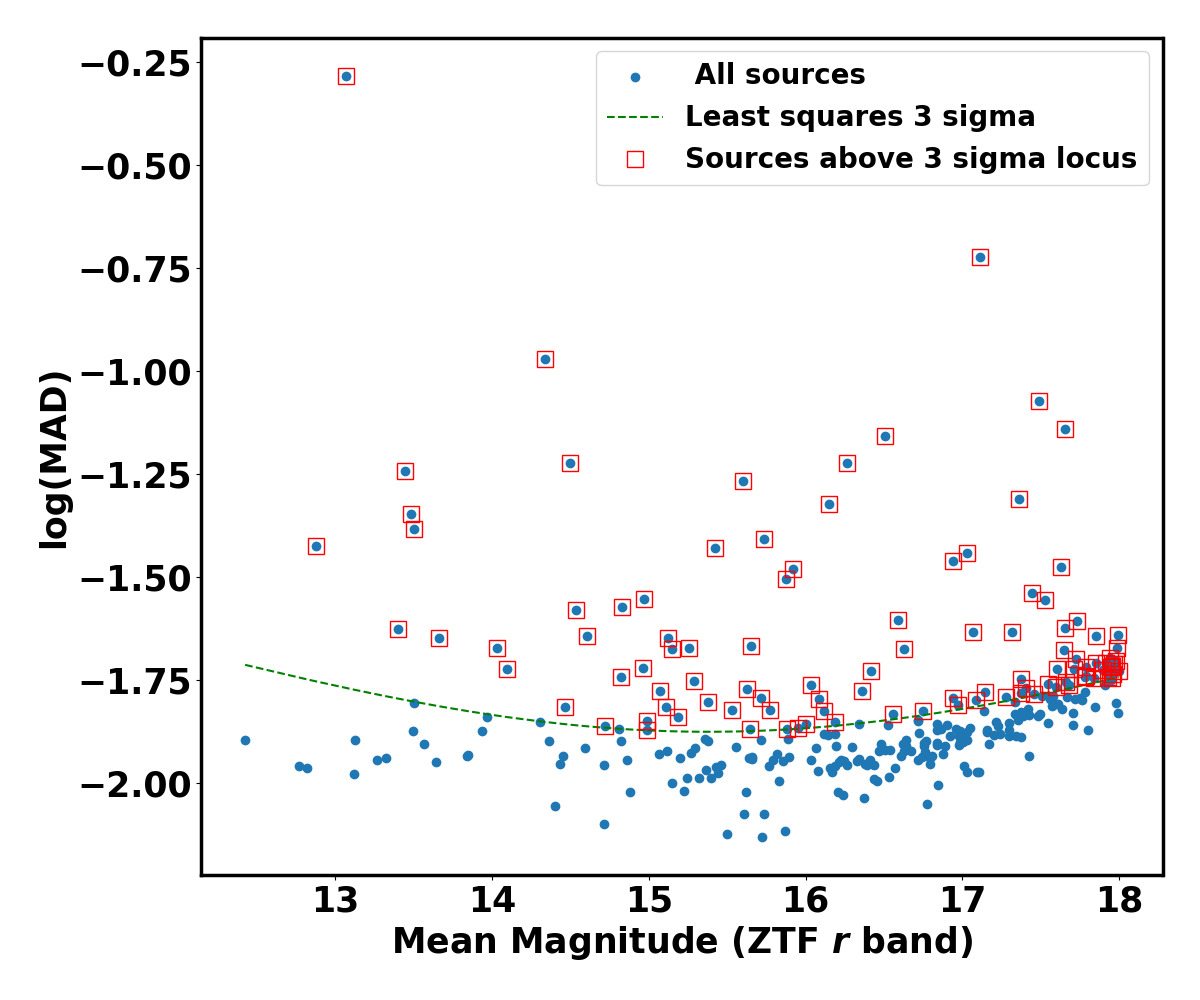}}\par 
    \end{multicols}

\caption{Standard Deviation and MAD of sources in ZTF data as a function of $r$ band magnitude, along with 3$\sigma$ locus, where $\sigma$ is the median SD/MAD value for the ZTF counterparts. We plot the $\mathrm{log_{10}}$ of SD and MAD for better visibility.}
\label{MAD and SD ztf}
\end{figure*}

\subsection{\textbf{Variability from Stetson Index}}\label{ZTFstetsonsection}

The robust version of the Stetson Index (\citealt{1996stetson,cody2014}) has been computed for the sources with ZTF $r$ band data. It uses quasi-simultaneous two-band data to determine the correlation between them. The expression for Stetson Index (J)  is given by 

\begin{equation}
   \textrm{J}= \sum_{i=1}^{n}\mathrm{sgn}(P_i)\sqrt{|P_i|}\,,
\end{equation}

where, 
\begin{equation}
P_i= { \frac{n}{n-1}}(\delta b_i \cdot \delta v_i)\,,
\end{equation}

\noindent and $n$ is the number of observations in the lightcurve, sgn is the signum function that gives the sign for the corresponding values of $P_i$. $b_i$ and, $v_i$ are the two band magnitudes at a given epoch and $\sigma_{b_{i}}$ and $\sigma_{v_{i}}$ are the defined as the respective uncertainties.

\begin{equation}
     \delta b_i = \frac{b_i- \overline{b}}{\sigma_{b_{i}}} \,,  \delta v_i = \frac{v_i- \overline{v}}{\sigma_{v_{i}}}\,,
\end{equation}

\noindent Here, $\delta _b $ and $\delta _v $  are called the residuals. For a nonvariable star, the photometric errors, $\sigma_b$  and $\sigma_v$  are random, and hence the residuals, $\delta_ b $ and $\delta_ v $  are uncorrelated, and for large $n$ their Stetson Index tends to zero. 

For variables, the residuals $\delta _b $  and $\delta _v $ are correlated since the phenomenon that governs the variation will change the brightness in the same direction for both the bands, and so the Stetson Index for variables will be some positive constant. This method is useful only for cases where the time between successive observations is small compared to the variation period \citep{1993AJ....105.1813W}. The ZTF data we consider here is only for a single band, whereas the computation of the Stetson Index requires two-band data. To overcome this, a method similar to that described in \cite{10.1093/mnras/stw2262} is adopted. We chose a $\Delta T_{\mathrm{max}}$=1 day between corresponding observations of the \textit{odd} and \textit{even} lightcurves (see \citealt{10.1093/mnras/stw2262} for details) for deciding which data points will be considered for Stetson Index computation. This cutoff is chosen after analyzing a histogram of $\mathrm{log_{10}}$($\Delta T$) (see Fig. \ref{ZTF delt Hist} in Appendix) for all ZTF lightcurves for our cluster. The single-band data is split into two by taking the odd indexed data points as one band and the even indexed data points as the second. A heuristic cutoff is chosen based on the median value ($\sigma$) of the index and accepting any source with J value greater than 3$\sigma$ to be a candidate variable.

From the computation of the Stetson Index for 288 sources, those with the Stetson Index greater than 3 times the median value (J > 146) for all sources are considered as variables. We choose the median value instead of the mean so as not to let our analysis be affected by outliers. Using this method, 34 sources are classified as variables. Based on the Stetson Index value, we subdivide the 34 variables into 3 categories: Strong, Moderate and Low with the corresponding cutoffs being J > 5$\sigma$, J $\in$ (4$\sigma$, 5$\sigma$] and J $\in$ (3$\sigma$, 4$\sigma$] respectively, where $\sigma$ refers to the median Stetson Index value. 
In Fig. \ref{18}, the distribution of the Stetson Index as a function of \textit{r}-band magnitude is shown, where most sources have a J value around 0 corresponding to non-variables. We identify 34 variable sources that have Stetson Index values in the range of 149.7 to 3261 and are marked in different colors.  
\begin{figure}
\includegraphics[width=\columnwidth]{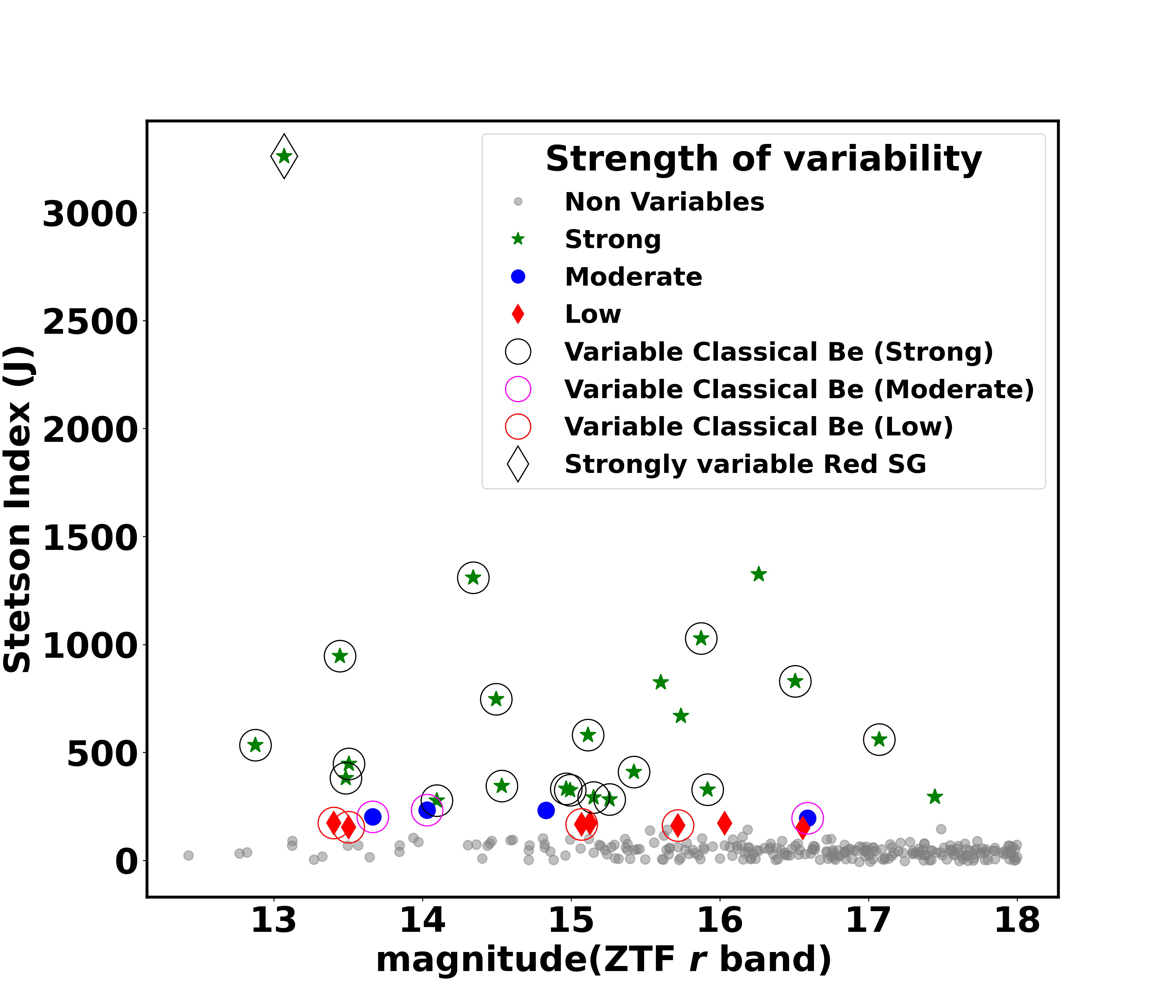}
\caption{Distribution of Stetson Index (J) as a function of Magnitude (ZTF $r$ band). Stars with J greater than 3 times the median Stetson Index value are classified as variables. Different colors represent the strength of their variability, which are listed as legends.}
\label{18}
\end{figure}
A total of 33 sources are variable across all three methods. Hence, from here onwards, we have these 33 sources as the final list of candidate variables in optical, which forms the most reliable list of variables based on 3 different methods.

Cross-matching the 33 variables we identify from ZTF data with the 49 CBe stars in the region, we find 24 CBe stars to be common. Thus, almost 50$\%$ of the CBe type stars in NGC 7419 are showing variability in optical wavelengths. Furthermore, most of the variable CBe are classified as Strong variables. One out of the five Red Supergiants (V* My Cep) in the cluster is also present in the list of these variables with a Stetson Index (J) value of 3261 and is highlighted in Fig. \ref{18}.

\subsection{Periodicity Analysis of ZTF light curves using Lomb Scargle Periodograms }\label{ztfls}

We further perform periodic and non-periodic classification using the Generalised Lomb Scargle Periodogram (GLSP), which is a robust version of the original (\citealt{1982ApJ...263..835S,1976Ap&SS..39..447L,1982ApJ...263..835S,vanderplas,GLS2009A&A...496..577Z}) on the 288 ZTF sources. The periodogram obtained for each source often contains aliased signals whose frequencies depend on the cadence of the window function \citep{vanderplas}. The window function describes the Fourier transform of the sampling times (with unit weights), and acts as a kind of convolution kernel in the frequency domain. Using a masking method similar to that described in \cite{Aliasremovalalgo} and \cite{aliascleaning}, we compute periodograms, phase lightcurves and best period of the sources after running the Lomb Scargle Periodogram for a period range of 0.025 to 150 days. For the frequency grid step width, we choose a reasonable value of $n_o=10$ \citep{vanderplas}. The lower bound for this period range is taken from \cite{2020ZTFPercatalog} since ZTF lightcurves are unevenly sampled (Fig. \ref{ZTF delt Hist}), and thus, using the Nyquist frequency limit (which may or may not exist for unevenly sampled data) is not feasible \citep{vanderplas}. The upper bound is chosen due to computational constraints. However, our method still incorporates the typical timescales of the variability of CBe stars, which are objects of primary focus for our work due to pulsation/rotation, alternating disk growth and decay, binarity, and wave motions in the disk, leaving out very long-term variations caused by the formation and dissipation of disks \citep{2013Rivinius}, which we discuss in Section \ref{CBemultperiod}. This type of variation is often visible in the raw lightcurve as \textit{bursts} or \textit{drops}  
\citep{Rimulo2018}.  The overall algorithm is summarized below:
 \renewcommand{\theenumi}{\arabic{enumi}.}%
 \begin{enumerate}[left=0pt]
 
  \item Compute window and observed power spectrum from cleaned 
   lightcurves.
  \item To get window peaks, those peaks with Lomb Scargle power above 5$\sigma$ ($\sigma$ = Standard Deviation) of the baseline are considered. The baseline power considered here is zero.
  \item In order to get observed power spectrum (OPS) peaks, we use the argrelmax function from \textit{scipy} \citep{2020SciPy-NMeth} to get local maxima (significant peaks) with the \textit{order} parameter set to 1000. We choose the minimum height of each such peak to be greater than or equal to 15\% of the maximum peak in the OPS.
  \item Using the window function, we remove window peaks with periods within 5\% of the periods of \textit{significant} peaks in the OPS.
  \item We then ensure that the remaining peaks are above the False alarm probability \citep{baluev2008} threshold of 0.005. False alarm probability denotes peaks that are caused purely by the coincidental alignment of random errors (\cite{vanderplas}).
  \item Finally, we classify a source as periodic if the least squares sinusoid fit has an amplitude higher than the Standard Deviation of the magnitude of that source. Any source that fails to satisfy these criteria is classified as a Non-Periodic or Uncertain.
  \end{enumerate}
  
It should be noted that masking the window aliases may also cause true peaks to get masked if they happen to lie within 5\% of the aliases. However, it is still better than naively taking the highest peak as the correct one, as emphasized in \cite{vanderplas} and \cite{Aliasremovalalgo}. The latter also confirms that only 1 -- 2\% of objects are affected by this issue. An example of the result for the star UCAC4 755-075651 using the above algorithm is shown in Fig. \ref{LSexample}. It clearly shows that the highest peak is not always the correct one. The highest peak is at around 1 day, but it and its higher frequency aliases are clearly an imprint of the window function on the observed power spectrum and are hence masked out.

\begin{figure*}
\includegraphics[width=2\columnwidth]{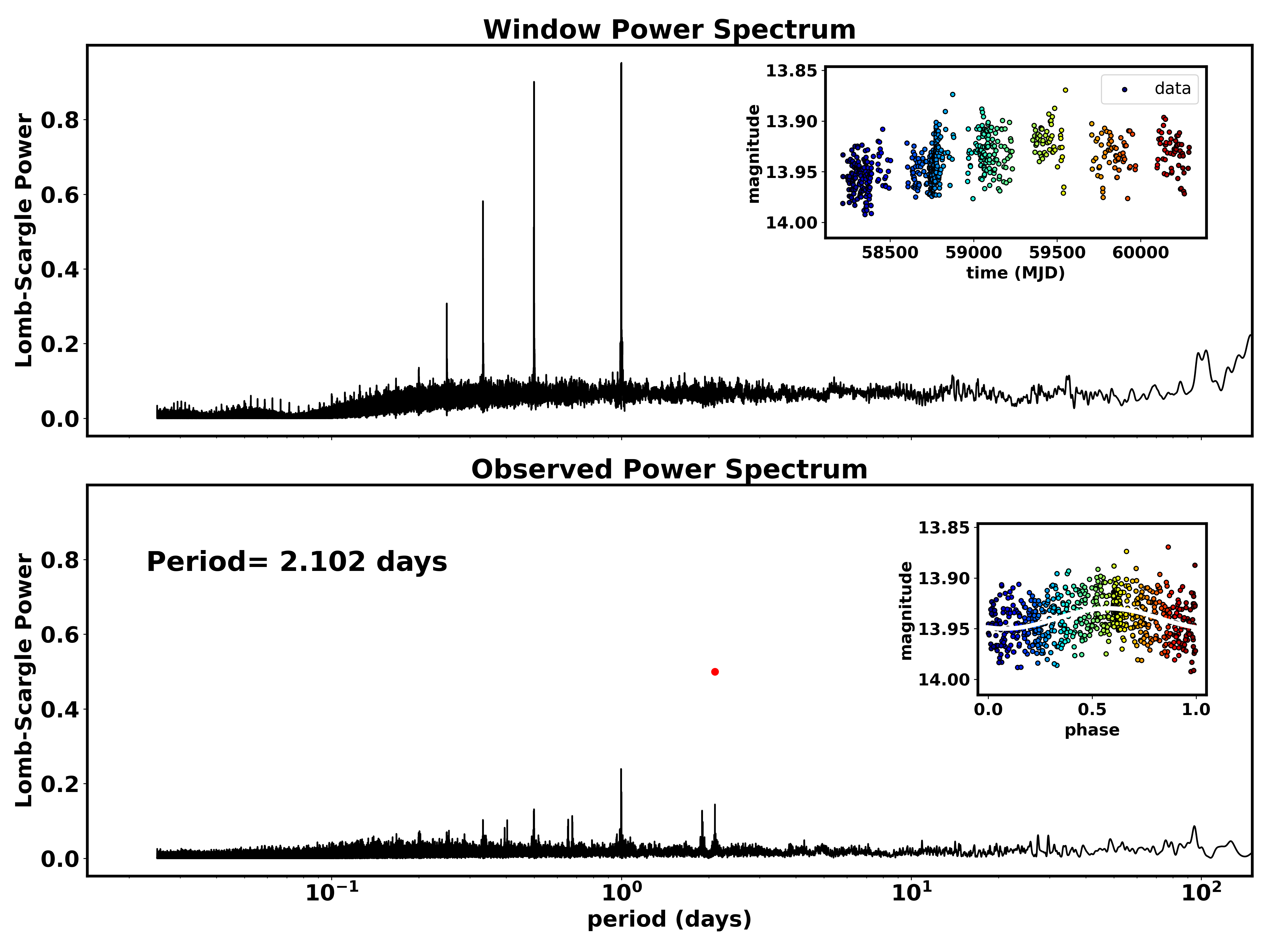}
\caption{An example output for a periodic star UCAC4 755-075651 from ZTF data for NGC 7419. The top panel denotes the window power spectrum, and the bottom panel denotes the observed power spectrum. Inset plots show the data and the phased lightcurve with the best-fit sinusoid, respectively. The red dot indicates the peak corresponding to the best period.}
\label{LSexample}
\end{figure*}

After removing the window aliases and performing a visual inspection to check for good fitting to the lightcurves, out of 288 ZTF sources, 27 show periodic signals. Out of the 33 variables in optical bands we identify within NGC 7419 (see Section \ref{ztfvarsec}), we find 12 to be periodic. More than 50\% of the periodic sources are non-variables in our work. Looking into it further shows that 14 out of 15 sources of this type have a positive Stetson Index value, and 6 have a Stetson Index value greater than the median Stetson Index value of all ZTF sources. A positive Stetson Index indicates some correlation in the variation of magnitudes between the two \textit{pseudo} bands (see Section \ref{ZTFstetsonsection}), which suggests these sources are low amplitude variables that were not identified previously based on our stringent cutoff for high confidence variables. Hence, we define these stars as \textit{sub-threshold} periodic stars. Out of 44 CBe sources in ZTF data, 10 are classified as periodic, and 9 out of these are also included in the list of variables. A deeper look into the multiple periodicity, which is common in CBe stars, is discussed in Section \ref{CBemultperiod}. The relative ratio of periodic sources with respect to their total counterparts is shown in Fig. \ref{19}. 

Table \ref{ZTF variability summary} represents a summary of the variability analysis using ZTF data. A detailed table for all the ZTF sources showing the results of variability and Periodicity analysis can be found in the Appendix (Table \ref{ZTF detailed table}).

\begin{figure}
\includegraphics[width=\columnwidth]{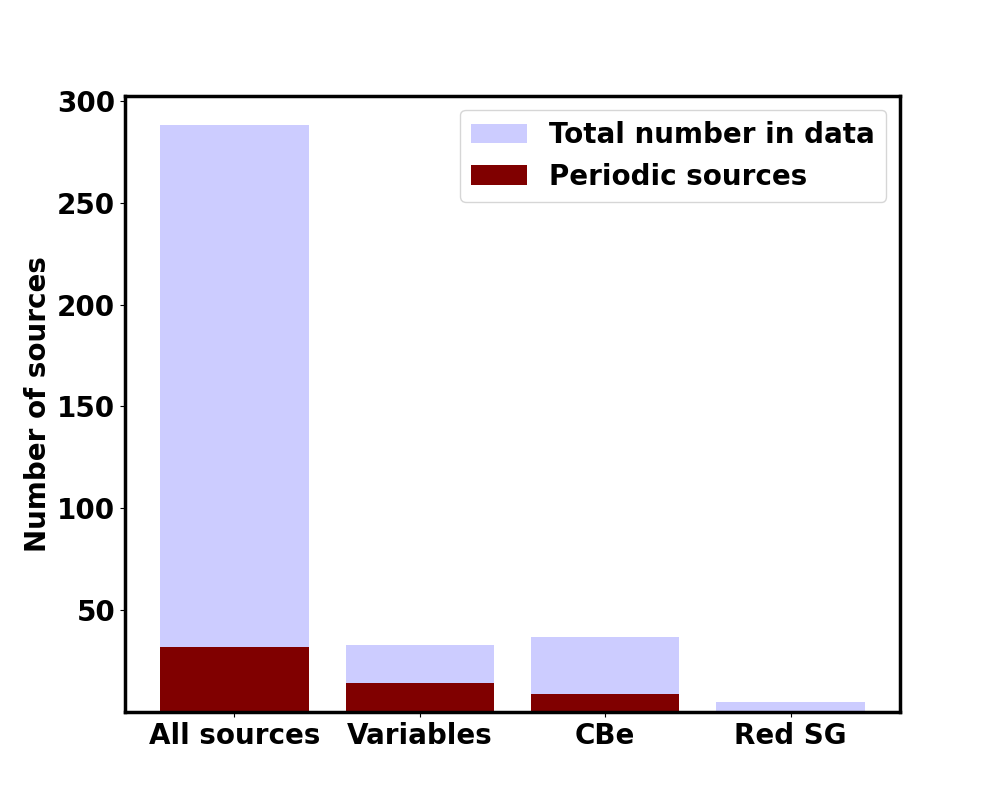}
\caption{Bar plot showing the relative number of periodic sources for all sources, variables, CBe stars, and Red Supergiants in ZTF data.}
\label{19}
\end{figure}

\begin{table}

\centering
\begin{tabular}{cccc}
\hline
\textbf{Type} & \textbf{NGC 7419}  & \textbf{ZTF}  & \textbf{NEOWISE}\\
\hline
\hline
Members & 499 & 288 & 149\\
Classical Be stars &  49 & 44 & 28 \\
 Red Supergiants  &  5 & 1 & 0\\
\hline

\end{tabular}
\caption{\label{member statistics}Statistics of NGC 7419 members and their respective crossmatches in ZTF and NEOWISE data}
\end{table}

\begin{table*}

\centering
\begin{tabular}{ccc}
\hline
\textbf{Criteria} & \textbf{Subtype} &\textbf{Count (Non CBe | CBe | Red SG)}\\
\hline
\hline
& Strong  & 4 | 18 | 1\\
Stetson Index (J) &   Moderate   & 1 | 3 | 0 \\
  & Low  &  3 | 3 | 0\\
\hline
 & Periodic  & 17| 10 | 0\\
Periodic Signal (Lomb Scargle, LS)  \\

& Non Periodic/ Uncertain &  222 | 34 | 5 \\
\hline
\end{tabular}
\caption{\label{ZTF variability summary}ZTF variability summary}
\end{table*}

\section{Identification of variable sources using NEOWISE data}\label{neovarsec}

We obtain a total of 302 sources from the NEOWISE survey having time-series photometry in two bands, i.e., in W1 and W2 (3.4 $\mu$m and 4.6 $\mu$m). After considering sources with  SNR greater than 3 and having at least 30 good quality observations, and after removing the saturated Red Supergiant (Cl* NGC 7419 BNSW e), we obtain 149 NEOWISE counterparts for NGC 7419. 28 CBe counterparts are included in these 149 sources. All 28 of these CBe sources are also present in the ZTF data. Below, we describe the various steps to identify the variable sources in the list. Table \ref{member statistics} summarizes the statistics of NGC 7419 members and their respective crossmatches (satisfying all quality criteria) in ZTF and NEOWISE data. 

\subsection{\textbf{Variability from Standard Deviation and Median Absolute Deviation}}

Similar to the work done for ZTF data, we consider any source lying above the 3-sigma line a variable. For NEOWISE data, specifically, we perform this for both W1 and W2 bands and take the common sources that satisfy the variability criteria across the two bands as variables. For brevity, these plots are shown in Fig. \ref{MAD and SD NEOWISE} in Appendix~\ref{App Figures}.

From the SD method, we obtain a total of 54 probable variables and a total of 53 probable variables from the MAD method, with 51 crossmatches between the two methods.

\subsection{\textbf{Variability from Stetson Index}}

 For the Stetson Index computation, unlike the ZTF data, we did not apply the treatment described in \cite{10.1093/mnras/stw2262} as the NEOWISE data include simultaneous 2-band observations. Sources with Stetson Index (J) greater than 3 times the median J value (i.e., J $>$ 180) are considered variable. This led us to identify 36 candidate variables. Similar to the treatment for the ZTF survey (see Section \ref{ZTFstetsonsection}), based on the Stetson Index value, we subdivide the 36 candidate variables into three categories: Strong, Moderate, and Low. The distribution of the Stetson Index as a function of W1-magnitude is shown in Fig. \ref{24}. Unlike the optical wavelength (see Fig. \ref{18}), most of the variables we identify in the IR wavelength lie towards the brighter end (W1 $<$ 12.5 mag) compared to the non-variables, which are mostly of W1 $>$ 12 mag. Out of these 36 sources, there are 21 and 20 crossmatches, respectively, with the Standard Deviation and MAD methods, and 20 crossmatches across all three. Hence, from here onwards, like before, we consider these 20 sources as the final list of candidate variables from NEOWISE analysis. 

Cross-matching the 20 variables we identify from NEOWISE data with the 49 CBe stars in the region, we find 12 of them to be Be-type stars. Thus, almost 25\% of the CBe type stars in NGC 7419 are showing variability in NIR wavelength. A total of 7 CBe sources are classified as variable in both optical (ZTF) and NIR (NEOWISE) wavelengths.

\begin{figure}
\includegraphics[width=\columnwidth]{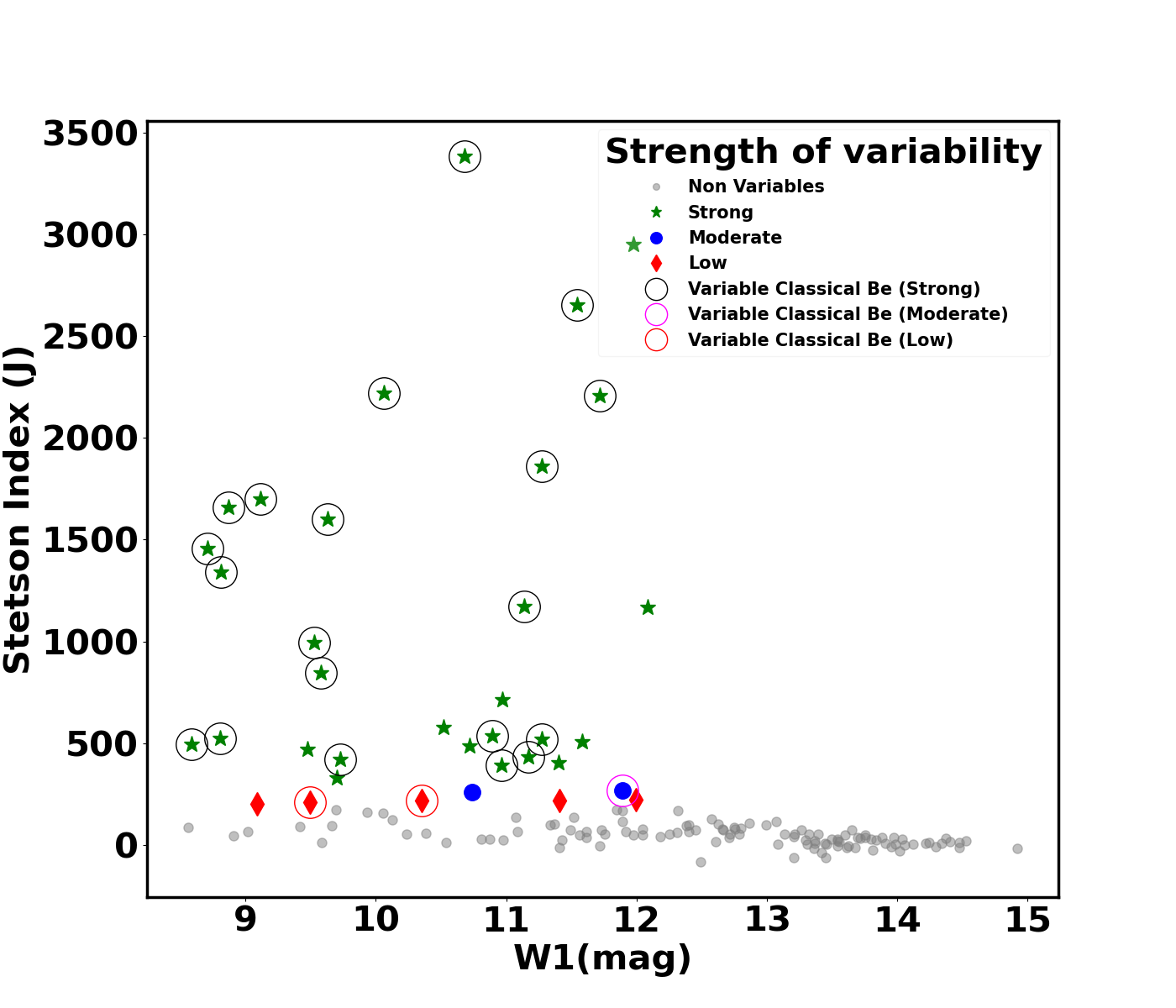}
\caption {Stetson Index (J),  as a function of W1 band magnitude for the NEOWISE data. Stars with J greater than 3 times the median Stetson Index value are classified as variables. Different colors represent the strength of their variability, which are listed in the index.}
\label{24}
\end{figure}

\subsection{ Periodicity Analysis of NEOWISE light curves using Lomb Scargle Periodograms}

Using a prescription similar to that in Section \ref{ztfls}, we perform the periodicity analysis of cluster members using NEOWISE data. NEOWISE, being a space-based telescope, has a near-uniform cadence of about 0.066 days (see Fig. \ref{NEOWISE delt Hist}). The slight non-uniformity present is due to the SNR limit that we enforced when gathering the light curves, which removed some data points. There is also a non-uniformity at 180 days (see inset plot in Fig. \ref{NEOWISE delt Hist}) due to NEOWISE's gap of 6 months between observation cycles, but they are extremely few and far between. Thus the maximum frequency that can be probed is the Nyquist frequency of $1/(2dt)$ where $dt$ is the observation cadence. This is around 7.6 $\text{day}^{-1}$. We kept the lower limit of the frequency range the same as we did for ZTF (0.0067=1/150 $\text{day}^{-1}$). Thus, we probe for frequencies ranging from 0.0067 $\text{day}^{-1}$ to 7.6 $\text{day}^{-1}$, i.e., a period range of 0.132 days to 150 days. Again, we kept $n=10$ for the frequency grid step size. Out of the 149 NEOWISE counterparts, the number of sources that show periodic signals is 15 in the W1 band and 22 in the W2 band, with 7 common between both bands. Of the 20 variables, 2 are identified as periodic in W1, 3 in W2, and 1 in both bands. We classify the remaining variables as non-periodic or uncertain. As in the case of ZTF, many sources we classify as non-variable show some kind of periodic signal from the Lomb Scargle analysis. Upon further investigation, we found 21 out of 26 of these \textit{sub-threshold} periodic sources have a positive Stetson Index value, 4 have a Stetson
Index value greater than the median Stetson Index value of all ZTF sources. Since our classification uses a stringent cutoff of 3 times the median for high-confidence variables, these low-amplitude variables are understandably left out. For the remaining five sources, which have negative Stetson Index values but still classified as periodic, the mean number of observations per lightcurve was $\approx$80 (variable sources have an average of 200 observations in NEOWISE data). We caution the reader that the low statistics of the number of observations for these five sources might cause the Lomb Scargle periodogram to misidentify them. Out of the 28 CBe sources present in NEOWISE data, we classify 4 as periodic in W1, 5 in W2, and 3 in both bands. From our earlier variability analysis, we find that two periodic CBe sources in W1 and two periodic CBe sources in W2 are variables. Section \ref{CBemultperiod}  discusses the variation in different timescales of CBe stars in detail.  The relative ratio of periodic sources with respect to their total counterparts is shown in Fig. \ref{neobar}.

\begin{figure}
\includegraphics[width=\columnwidth]{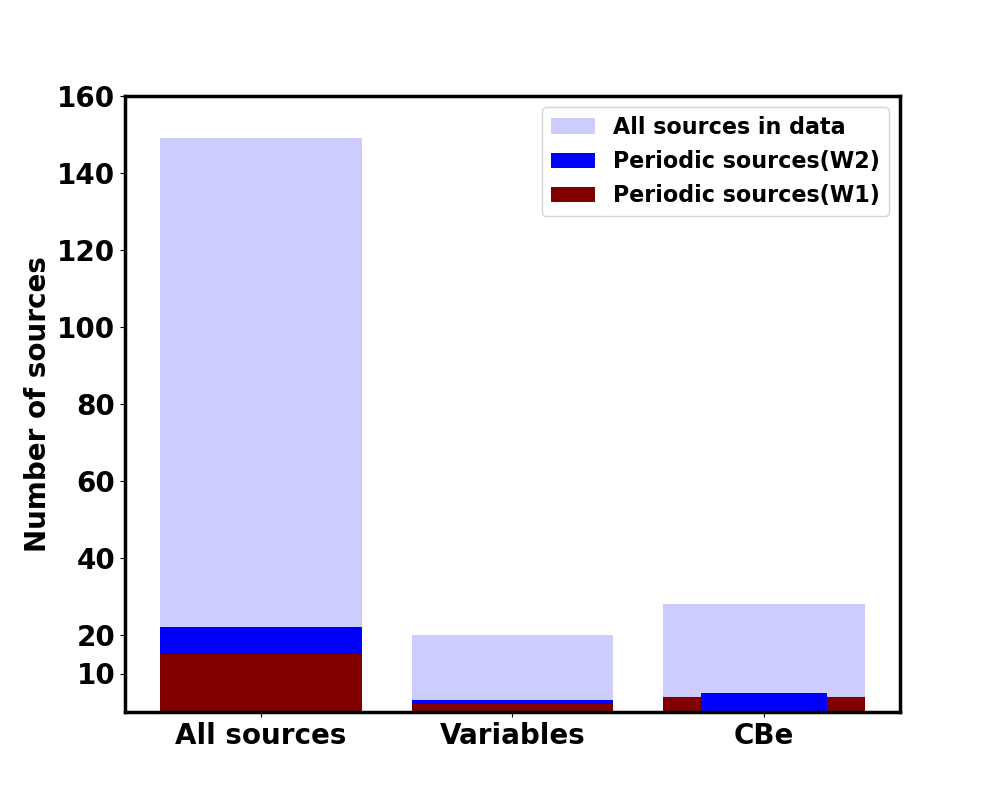}
\caption{Bar plot showing the relative number of periodic sources for all sources, variables, and CBe stars present in NEOWISE data.}
\label{neobar}
\end{figure}

Table \ref{NEO variability summary} represents a summary of the variability analysis using NEOWISE data. A detailed table for all the NEOWISE counterparts showing the variability and Periodicity analysis results can be found in the Appendix (Table \ref{NEOWISE detailed table}).

\begin{table*}

\centering
\begin{tabular}{lcr }
\hline
\textbf{Criteria} & \textbf{Subtype} &\textbf{Count (Non CBe | CBe | Red SG)}\\
\hline
\hline
& Strong  & 6 | 11 | 0\\
Stetson Index (J) & Moderate    & 1 | 1 | 0 \\
  & Low &  1 | 0 | 0\\
\hline
 & Periodic  & 11(W1), 17(W2) | 4(W1), 5(W2) | 0\\
Periodic Signal (Lomb Scargle, LS)  \\

& Non Periodic/ Uncertain &  110(W1), 104(W2)  | 24(W1) , 23(W2) | 0 \\
\hline
\end{tabular}
\caption{\label{NEO variability summary}NEOWISE variability summary}
\end{table*}

\section{Discussion}\label{discussion}

\subsection{Comparison with Gaia multi-band variability}

We compare three-band Gaia-based variability indicators from \cite{2023Maiz} to our work and look for any commonality between the variables in our work and their catalog\footnote{\label{Maiz}\url{https://cdsarc.cds.unistra.fr/viz-bin/cat/J/A+A/677/A137}}. We consider sources with at least \textit{V} variability flag in the G band to be the variable as the G band flags are considered the most reliable among the three bands \citep{2023Maiz}. We find that 93 \% (31 / 33) of ZTF variables and 95 \% (19/20) of NEOWISE variables are classified as variables in this catalog.
Almost all CBe stars in NGC 7419 are also classified as variables in this catalog. In contrast, our method detects high probability variables since it considers three different methods (Standard Deviation, MAD, and Stetson Index). 

\cite{2021Mowlavi} presents an interesting Gaia-based characterization of large amplitude variables (amplitudes greater than 0.2 mag in Gaia G band) based on their 
$A_{\mathrm{proxy, BP}}$ and $A_{\mathrm{proxy, RP}}$ ratio. $A_{\mathrm{proxy}}$ is a measure of the scatter in the light curve of an object, which is heavily dependent on the Standard Deviation of its flux curve \citep{2021Mowlavi}. For variable stars, the Standard Deviation is larger, and hence $A_{\mathrm{proxy}}$ indicates the variability amplitude of astrophysical origin, assuming that the larger Standard Deviation is due to an astrophysical phenomenon. We find 7 crossmatches between stars we classify as variables (ZTF or NEOWISE or both) with the main catalog\footnote{\label{Mowlavi catalog}\url{https://cdsarc.u-strasbg.fr/viz-bin/cat/J/A+A/648/A44}} of this paper. These seven crossmatches also show \textit{strong} type variability based on their Stetson Index values (see tables \ref{ZTF variability summary} and \ref{NEO variability summary}). 5 out of these 7 sources are CBe stars in the cluster. Figure 15 of \cite{2021Mowlavi} shows the correlation of $A_{\mathrm{proxy, BP}}$ / $A_{\mathrm{proxy, RP}}$ with various types of variables based on their origin of variability. 
Even though the statistics are low, we find that the $A_{\mathrm{proxy, BP}}$ / $A_{\mathrm{proxy, RP}}$ range for large amplitude CBe stars matches well with that of \cite{2021Mowlavi}.

\subsection{Multi-timescale variability in Classical Be stars in optical(ZTF) and IR(NEOWISE)}\label{CBemultperiod}

The variability observed in CBe stars can arise due to a variety of reasons. The Balmer emission of such stars is transient \citep{collins_1987}. Thus, they can transition from behaving like a Be star to a normal B-type star during disk dissipation or formation, which manifests, in stars seen pole-on, as a rapid increase (outburst) or decrease (decay) in brightness followed by a slow return to baseline, spanning weeks to decades. The reverse behavior is observed for edge-on (shell) objects \citep{haubois2012dynamical}. Then there are quasi-periodic variations on intermediate timescales of months to years due to wave motions in the \textit{decretion} disk or binarity \citep{okazaki2006, carciofi2009, 2025Rubio}. 
Finally, there is variability due to stellar non-radial pulsation \citep{2022Labadie} and/or rotation \citep{2022Balona}, which can have timescales between 0.2 and 3 days. Sources for which the dominant source of variability is from pulsation/rotation, are defined as "$\lambda$ \textit{Eri variables}". This type of variation due to pulsation/rotation is more easily detectable using space-based observations compared to ground-based observations \citep{baade2016}.
Low-frequency stochastic variation is another feature of the power spectra of CBe stars. These signals typically show up as \textit{red noise} at the lowest frequencies, but they have an astrophysical origin \citep{Naze2020}. 

We performed the Lomb Scargle analysis from Section \ref{ztfls} for just CBe stars using both NEOWISE and ZTF lightcurves to search for the above-mentioned types of variations (periodic or not) in different timescales ranging from 0.2 to 3 days (pulsation), 7 days to 150 days (short term disk formation or dissipation or binarity or wave motions in a \textit{decretion} disk). We also look for CBe sources that show clear variations on timescales of weeks to decades due to disk formation or dissipation by eye. Many CBe stars contain more than one type of these variations, confirming a well-established notion from previous literature (\citealt{Walker2005CBemultperiod,CBemultiperiod}). 

For ZTF $r$ band data, out of 44 CBe sources, 6 CBe stars show a periodic signal in the typical pulsation/rotation range of 0.3 to 2 days, while 9 CBe stars have a periodic signal on timescales characteristic of quasi-cyclic variability due to wave motions in the disk similar to $\omega$ CMa (B2V) \citep{quasicyclicCBe} and/or binarity. In addition, by manually inspecting the lightcurves by eye, we detect 4 additional sources showing variations that can be attributed to disk alternating disk growth and decay /quasi-cyclic variations/binarity. In addition, 6 sources show long-term variations likely associated with disk loss or formation episodes on the timescales of years. \cite{CBemultiperiod} demonstrated that almost all CBe stars exhibit grouped multi-periodicity near the pulsation timescales, which was already believed to be the case \citep{Walker2005CBemultperiod}. Many sources in this work show more than one type of variation on different timescales, which is quite common for CBe stars \citep{2013Rivinius}. Stochastic aperiodic signals are also very common in CBe stars \citep{CBemultiperiod} and are difficult to analyze quantitatively. However, our results, shown in Table \ref{CBevariability}, still provide a useful representation of the timescales of variability observed in CBe stars. Figure \ref{CBelightcurves} shows a few ZTF and NEOWISE lightcurves of CBe stars showing disk dissipation/formation events and quasi-cyclic variations. Panels A, F, B, and G belong to 2 stars, showing clear long-term variations due to disk formation or dissipation. Star 65 is a typical case of a pole-on view, where disk formation is associated with a net brightening, whereas star 213 shows the opposite situation, corresponding to an edge-on orientation. Panels C, H, D, and I are lightcurves of another 2 stars, which show alternating periods of disk growth or decay and/or quasi-cyclic variability. Finally, panels E and J belong to a CBe star, showing clear quasi-cyclic variability on timescales of around 400 days.

\begin{figure*}
\centerline{\includegraphics[width=2.1\columnwidth]{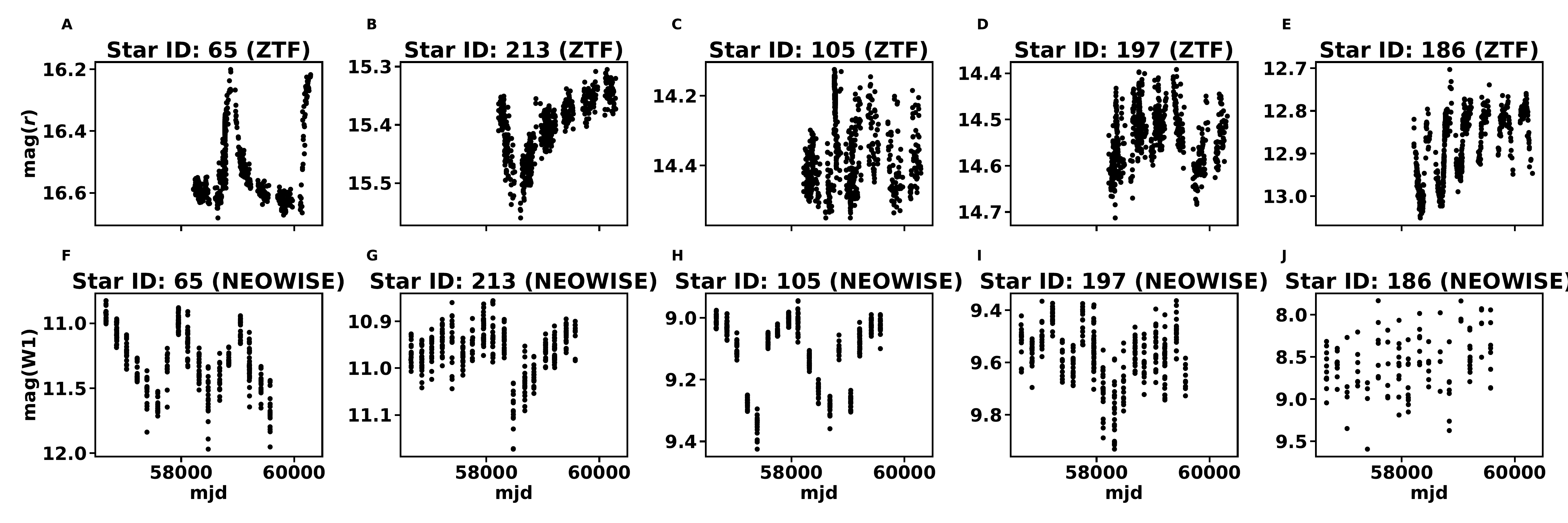}}
\caption{Lightcurves of 5 CBe stars, showing different kinds of magnitude variations. The top and bottom rows indicate ZTF and NEOWISE light curves for each star, respectively. For stars 65 and 213 (panels A, F, and B, G), we see clear long-term variations in timescales seen typically due to disk formation or dissipation. The build-up phase of these 2 stars (brightening for star 65 and dimming for 213) happens on much shorter timescales compared to their return to the baseline, which is typical for these kinds of variations (\citealt{2012Haubois}). Stars 105 and 197 (panels C, H, and D, I) show alternating periods of disk growth or decay and/or quasi-cyclic variability. Star 186 (panels E and J) shows clear quasi-cyclic variability on timescales of around 400 days.  }
\label{CBelightcurves}
\end{figure*}

Based on spectral classification from previous studies (\citealt{2003Caron,subramaniam,2011subramaniam,Marco_2013}) and our own, we find that all of the 6 CBe stars that show periodic non-radial pulsation are early-type stars, which agrees with the fact that only early-type CBe stars pulsate strongly enough to be detected by ground-based telescopes, with late-type stars sometimes showing pulsation with very small amplitudes \citep{2013Rivinius}. Among those stars that show variability on timescales seen due to alternate disk growth/dissipation or wave motion in disk and/or binarity, 11 out of 13 are early type, and 2 are mid type. Of those stars, we visually confirm to show variability due to long-term disk formation or dissipation events, 3 out of the 6 are early type, and the other 3 are mid type. 

Following a similar approach to the one used with the ZTF data, we perform the same analysis using NEOWISE light curves for the 28 CBe crossmatches. Again we use a combination of visual inspection and a Lomb Scargle Periodogram to identify variability at different time scales. Out of the 28 CBe crossmatches in NEOWISE data, 4 stars show a periodic signal on the typical pulsation/rotation timescales of 0.2 and 3 days, all of which are early-types. One star shows a periodic signal on timescales of weeks or years, which can be attributed to variability due to disk changes (formation/dissipation or wave-motion) or binarity. From visual inspection, we find 8 more stars exhibiting distinct variability, which may arise from quasi-cyclic behavior, alternating phases of disk growth and dissipation, or binarity. Thus, 9 CBe stars show infrared variability in these timescales. 6 of these 9 stars are early-type, one is mid-type, and two are late-type. 4 out of these 9 stars also show similar variability in optical wavelengths (ZTF $r$ band). Finally, two stars (mid-type and early type) show long-term variations, typically seen during disk loss or formation from visual inspection.  These two stars also show similar long-term variability in optical wavelength (ZTF $r$ band). All the trends in spectral types seen in infrared wavelength match well with previous findings mentioned in \cite{2013Rivinius} and \cite{2018Granada}. A summary of the results is shown in Table \ref{CBevariability}. A detailed representation of B and CBe stars in the cluster, along with various properties like mass, radius, spectral type, breakup velocities, variability, and periods, is given in Table \ref{B-type star table}. 

To summarise the combined results of the variability of CBe stars identified using the data from ZTF and NEOWISE, 66\% (29/44) of CBe counterparts are found to be variable. Our analysis finds that 23\% (10/44) of CBe crossmatches show a periodic signal due to pulsation/rotation. The phase-folded lightcurves of these stars are shown in Fig. \ref{CBe NRP}. 

There are two classes of variable stars, namely $\beta$ Cephei and Slowly Pulsating B (SPB) stars, which show similar kinds of variability due to pulsation. 
In Fig. \ref{CBe pulsation CMD}, we show the Gaia CMD along with the typical theoretical instability strips for $\beta$ Cephei and SPB stars based on their $T_{\mathrm{eff}}$ and log($L/\mathrm{L_{\odot}}$) values according to Fig. 1.12 of \cite{AertsAstroseismology} and interpolating them with our best-fit isochrone (21.1 Myr) to find the corresponding G and BP-RP value. We find that the CBe stars in NGC 7419 that show periodic pulsation mostly lie around the $\beta$ Cephei instability strip, with our periods matching well with the typical $\beta$ Cephei periods (i.e., 0.1 to 0.6 days) \citep{GAIADR2variablestars}. In Section \ref{distance age section}, we mentioned how our estimates of mass and spectral types of stars are only approximate for stars away from the isochrone. The most extreme outlier in Fig. \ref{CBe pulsation CMD}, at $m_G$ $\sim$ 17 mag and Gaia BP-RP color of 1.21 mag, is classified in our work as a CBe star with an estimated spectral type of B7V–B8V. It also has an IPHAS r'-i' color of 0.93 mag, making it the bluest among all detected CBe stars with $H\alpha$ excess. In comparison, other CBe members have a mean Gaia BP-RP color of 2.26 mag and an IPHAS r'-i' color of 1.15 mag. Its position well to the blue side of the main sequence indicates that the estimate of its spectral type is a lower limit, and the star is likely to have an earlier spectral type. This displacement might also indicate a potential binary product, possibly a blue straggler that has gained mass through past binary interaction. For more examples of such cases, see \cite{Gies} and \cite{Bodensteiner}. 

\begin{figure*}
\includegraphics[width=2\columnwidth]{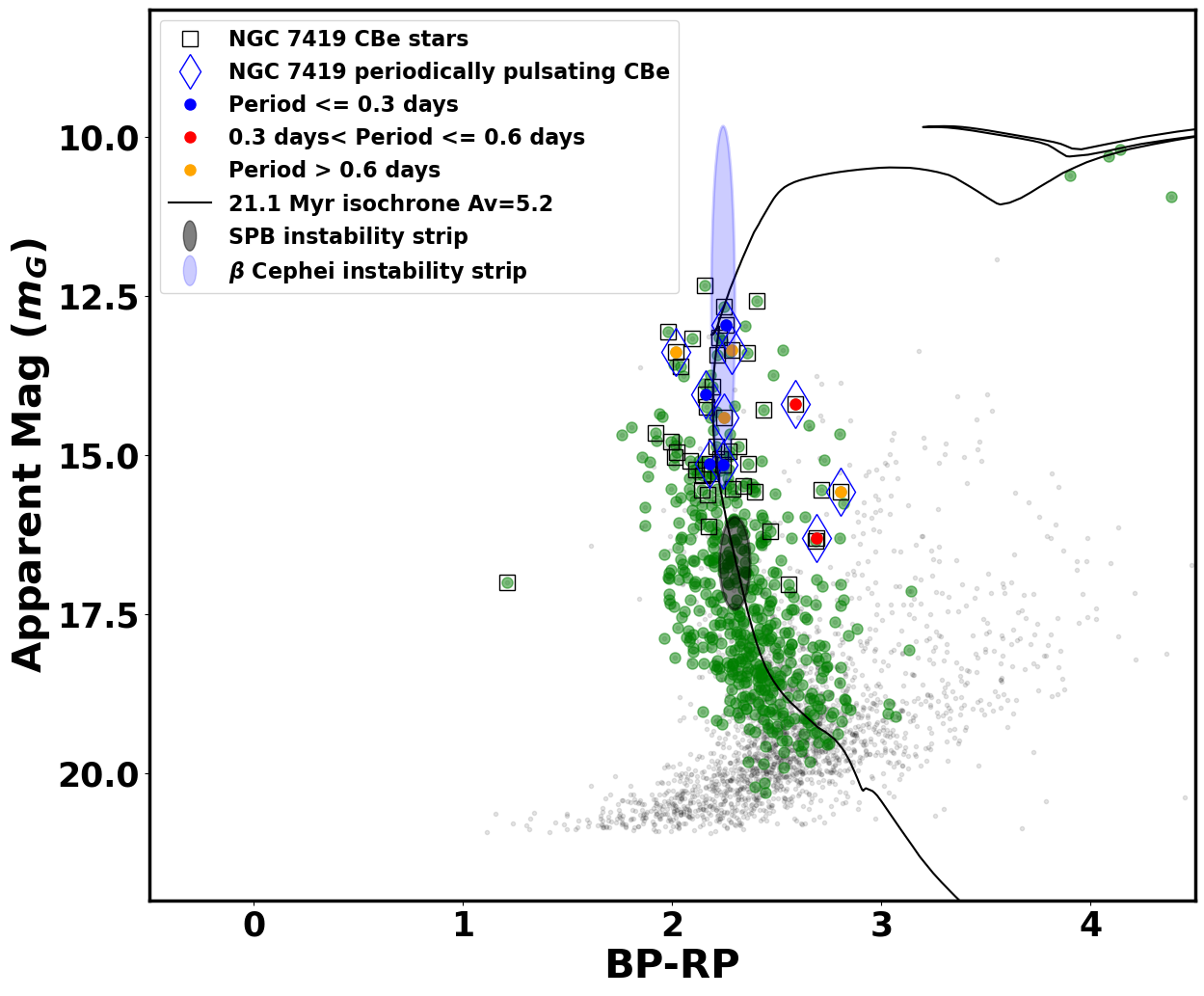}
\caption{Gaia BP-RP versus G band Color-magnitude diagram for main sequence stars in NGC 7419. CBe stars and their periodically pulsating variants are marked by open black squares and blue diamonds, respectively. The blue and black shaded regions indicate the theoretical $\beta$ Cephei and SPB instability strips based on Fig. 1.12 of \protect\cite{AertsAstroseismology}. The different ranges of periods for the pulsating CBe stars are marked by red, orange and green dots.}
\label{CBe pulsation CMD}
\end{figure*}

However, if these pulsations are non-sinusoidal, the individual harmonics may become less significant and thus increase their chances of overlap with the window aliases described in Section \ref{ztfls}, which could hinder their detection by the Lomb-Scargle periodogram used in this study. In addition, ZTF and NEOWISE lightcurves do not have a high enough cadence to successfully detect such high-frequency signals accurately most of the time. We also find that 41\% (18/44) CBe stars show variability due to alternating disk growth and decay, wave motion in disk or binarity, and 14\% (6/44) show long-term variations due to disk dissipation/formation. These 6 stars are shown in Fig. \ref{CBelightcurves} and \ref{CBelongterm}.

We use previous studies (\citealt{2003Caron,subramaniam,2011subramaniam,Marco_2013}) and our own estimation of spectral types of cluster members and find that CBe stars showing periodicity due to non-radial pulsation /rotation are all early type. Although the statistics are poor, the results agree with the fact that late-type CBe stars pulsate with very small amplitudes, and thus, variations are harder to detect. We also find 50\% (3/6) of CBe stars showing long-term disk variations due to dissipation and formation to be early type, and the remaining to be mid type. Even though statistics are poor, \cite{2018Labadie} found that most long-term variations in CBe stars are seen in early-type stars.

\begin{table*}

\centering
\begin{tabular}{ |c|c|c|c| }
\hline
\hline
\textbf{Variability Mechanism} &  
 \textbf{ZTF} &\textbf{NEOWISE} &\textbf{Combined}\\
 & (Total=44) &  (Total=28) &  (Total=44)\\
\hline
(a) Periodic pulsation/ rotation (0.2 to 3 days)) & 6 & 4 & 10 \\

b) Alternating disk growth and decay, Wave motion in disk, binarity (weeks to years) & 13 & 9 & 18  \\

c) Long-term disk dissipation/formation (years to decades) & 6 & 2 & 6 \\

Multi-timescale variation & 5(ab), 1(ac), 3(bc) & 1(ab) & 6(ab), 1(ac), 3(bc)  \\

Dominant & 1(a), 6(b), 2(c) & 3(a), 8(b), 2 (c) &  4(a), 12(b) , 4(c)\\

Irregular, Non-periodic or Uncertain & 28 & 14 & 20\\
\hline
\end{tabular}
\caption{Summary of the variability of CBe stars showing either periodic signals or visually clear variations from ZTF and NEOWISE lightcuves. The letters in parentheses indicate which types of variations are present for sources showing variations on multiple timescales and those showing only one kind of variation (dominant).}
\label{CBevariability}
\end{table*}

\subsection{Correlation between WISE color-magnitude diagram and CBe star properties}\label{WISECMDsection}

Near-IR excess in CBe stars compared to normal B-type stars has been detected for a long time, the first being in \cite{1967Johnson}. This excess was attributed to free–free and bound-free processes in the circumstellar material (gaseous disks) around these Be objects \citep{1994dougherty}. Useful global information and trends for Be star's life stage, spectral type, and variability can be obtained from its color excess in a color-color diagram \citep{2018Granada}. Recent works like  \cite{2024A&A...682A..59J} find results in agreement with \cite{2018Granada}.

We adopt the conditions given in \cite{2018Granada} for the color excess in WISE bands for classifying the active and quiescent CBe stars as well as the conditions from \cite{2014MNRAS.442.3361N}  for the \textit{naked} or normal B-type stars. We also look for any possible trends in the variability of candidate CBe stars from our earlier analysis using both ZTF and NEOWISE data. Figure \ref{WISE cmds} shows the WISE color-magnitude diagram. Crossmatching WISE data with candidate CBe stars in ZTF and NEOWISE data resulted in 29 crossmatches.

\begin{figure}
{\includegraphics[width=\columnwidth]{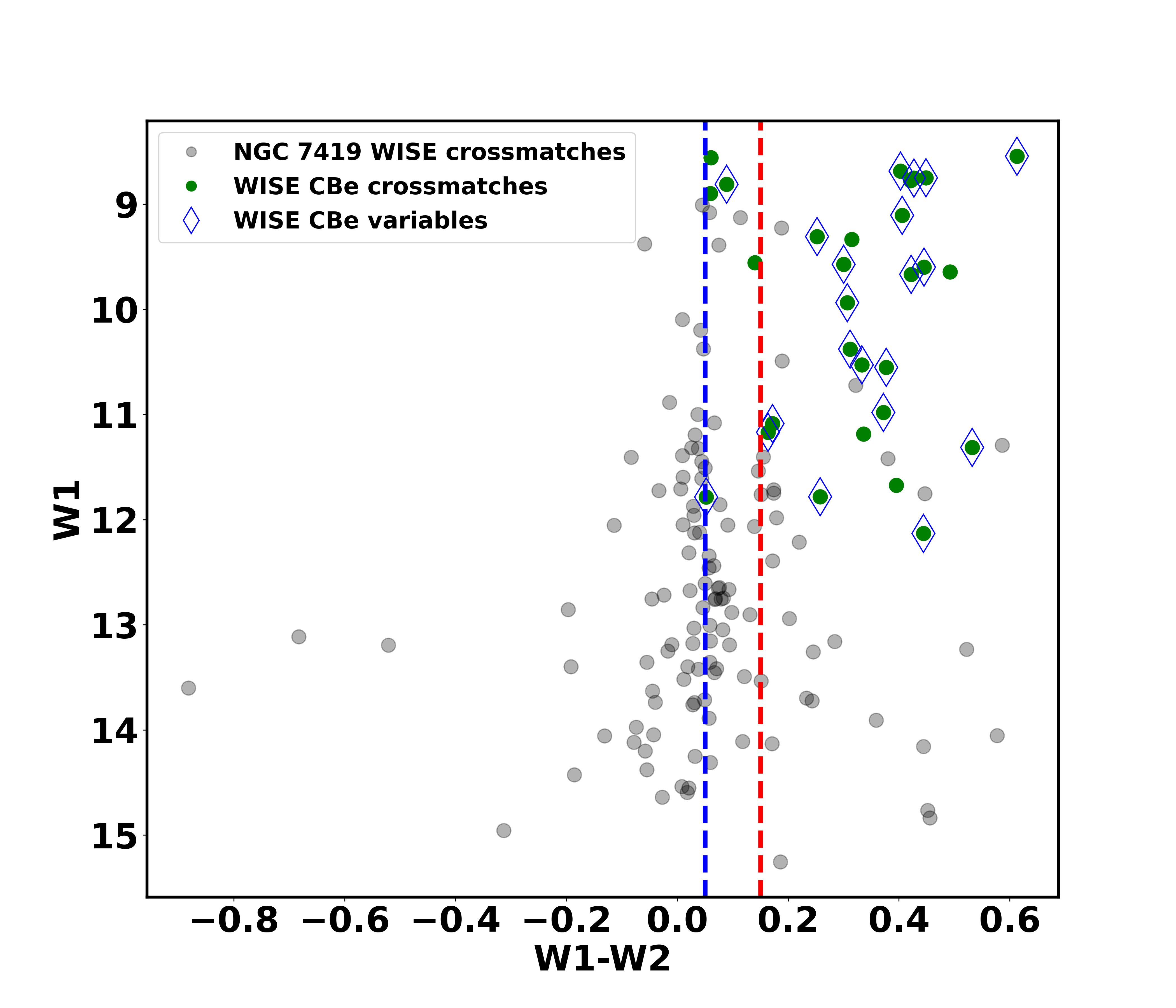}}\par 
\caption{Color magnitude diagram with variability information for known and candidate CBe stars in NGC 7419. All known and candidate Be stars are indicated with green dots. The color limit (W1-W2 \textless{} 0.05) for normal B-type stars is shown by the dashed vertical blue line. The three regions namely, quiescent (no disk), active dissipating disk, and active stable disk for CBe stars are marked with dashed blue and red vertical lines with the W1-W2 limits as follows: Quiescent: W1-W2 <0.05, Active dissipating disk: 0.05 $\leq$ W1-W2 $\leq$ 0.15 and Active stable disk: W1-W2 > 0.15 \citep{2018Granada}.}
\label{WISE cmds}
\end{figure}

We find that in NGC 7419, 83\% of CBe stars have a high likelihood to have well-formed stable disks  (W1-W2 > 0.15) while 17\% lie in the dissipating disk (0.05 $\leq$ W1-W2 $\leq$ 0.15) region (see \citealt{2018Granada}).  None of our candidate CBe stars lie in the quiescent region, which is expected as we used H$\alpha$ excess as the selection criterion for CBe stars. All the sources on the right of the dashed blue line in Fig. \ref{WISE cmds} can be used in the future as a test bed for detecting remaining CBe stars in this cluster, if any.

According to \cite{2018Granada}, variable CBe stars are found in stars undergoing disk changes (0.05 $\leq$ W1-W2 $\leq$ 0.15) but predominantly in the active phase (W1-W2 $\geq$ 0.05). We find that all variable CBe stars belong to the active region.

From our spectral type estimation in Section \ref{distance age section} and previous studies (\citealt{2003Caron,subramaniam,2011subramaniam,Marco_2013}) , we see that, among the CBe stars in the active region (W1-W2>0.15), 83.3\% are early, 8.3\% are mid, and 8.3\% are late-type B stars. 4 out of 5 sources in the disk dissipating region (0.05<W1-W2$\leq$0.15) are early-type and 1 is mid-type. According to \cite{2018Granada}, early-type CBe stars host disks and are more likely to be in the active region. The low number of known late-type CBe stars with disks can be justified because they are harder to detect spectroscopically, since only some show positive equivalent widths in the models described in \cite{2018Granada}. In addition, late-type CBe stars possess less dense disks \citep{Viera2017} and hence exhibit lower IR emission. 

We see similar numbers for variable CBe WISE counterparts. We find that most are early B-type (81\%), with 9.5\% being mid and 9.5\% being late type. Hence, we observe that CBe stars (variable or not) are mostly early B-type stars, although this can be an observational bias due to late-type stars being harder to detect as a result of being less bright and possessing less dense disks. An explanation for why CBe stars are more likely to be early-type stars has been described in \cite{2013Rivinius}. When the rotational velocity ($v_{\mathrm{ rot}}$) at the equator and orbital velocity ($v_{\mathrm{orb}}$) at the equator of a star are equal, it is said to be rotating critically. The ratio of $v_{\mathrm{rot}}$ to  $v_{\mathrm{orb}}$ is defined as $W$, which signifies the velocity boost required for a given star to eject material into orbit. 
$$ W = \frac{v_{\mathrm{rot}}}{v_{\mathrm{orb}}}$$ \,.

According to \citealt{Townsend,Riv2006} and \citealt{Fremat}, the mean value of $W$ is around 0.75 for CBe stars, with the minimum (defined as cutoff $W$) being 0.62 for a B type star to be classified as a CBe star. The mean and the cutoff $W$ value did not show any trend with effective temperature from star to star. Finally, according to \cite{Huang_2010}, low-mass B stars (late types; $ < 4\, \mathrm{M}_{\odot}$) require a high threshold of $W$ > 0.96 to become Be stars. As stellar mass increases, this threshold decreases, dropping to 0.64 for early B stars with $ \mathrm{M} > 8.6\, \mathrm{M}_{\odot}$. This implies that early B-type stars have a much higher probability of being a CBe type since the mean value is much closer to the cut-off value for $W$, which is 0.62. This is in accordance with the distribution of Be stars within NGC 7419, which are generally brighter and early types. 

\section{Summary}\label{Summary}
We studied the open cluster NGC 7419, which stands out due to its high number of CBe stars and 5 Red Supergiants. Even though spectroscopic analysis of cluster members has been done in previous literature (\citealt{Marco_2013,subramaniam}), no study has examined the variability of these members.

Following the footsteps of previous studies on cluster membership using both supervised and unsupervised algorithms, we use a combination of the Gaussian Mixture model and Random Forest algorithms on Gaia DR3 data for the open cluster NGC 7419 to find its high probability members.
Using Gaia parallaxes and later comparing the inverse parallax distance with the Bailer-Jones distance \citep{2021yCat.1352....0B}, we estimate the average distance to the cluster to be around ${3.6^{+1.0}_{-0.6}}$ kpc. Next, using stellar evolutionary tracks from the PARSEC model, we determine the average age of the cluster to be  $\approx$ $21.1 ^{+1.6}_{-0.6}$ Myr with a visual extinction of 5.2 mag. We interpolate over the best-fit stellar isochrone (21.1 Myr) using
the G band magnitudes of the cluster members as input to estimate their stellar masses. We then use the modified table\footref{mamajek} of \cite{2013PecautMamajek} to estimate the spectral type and stellar radii of cluster members. Combining this with previously known B-type stars from \citealt{2003Caron,subramaniam,2011subramaniam,Marco_2013}, we find the majority of our members ($\sim$ 75\%) to be B-type stars.
Following this, we also estimate the critical velocities of B-type members in NGC 7419 using their definition in the framework of the Roche Approximation.

Based on $H\alpha$ photometric excess, we use the synthetic tracks from \cite{Drew}, and adopt a method similar to that described in \cite{Barensten}, to}identify a total of 42 sources showing $H\alpha$ excess in the cluster, which is in agreement with a previous work by Paul Schmidtke and Tim Hunter \citep{Schmidtke_2019}. Given the age of the cluster, these sources cannot be Herbig Ae/Be type stars \citep{2006} and are thus classified as candidate CBe stars as the \textit{decretion} disk is directly related to the emission of $H\alpha$ lines in these stars \citep{2013Rivinius}. Combined with the list of previously known CBe stars, this brought the number of candidate CBe stars to 49 in the cluster, which is high for young open clusters. We find the ratio of  CBe to (B+CBe) members to be 12.7\%. The same fraction is 24.8\% for early-type stars, 10.5\% for mid-type stars, and 1.7\% for late-type stars.
The CBe fraction reported here is likely a lower limit due to two main factors: (1) our identification method depends on H$\alpha$ emission and may overlook stars with weak or absent disks, and (2) CBe stars experience active and inactive phases, so those without disks at the time of observation could be misclassified as normal B-type stars.

Following this, we perform an extensive search for variable stars using data from both ground-based (ZTF, optical) and space-based (NEOWISE, infrared). We use 3 different methods, namely Standard Deviation, Median Absolute Deviation, and Stetson Index \citep{1996stetson}, to identify high-probability variables in the cluster.  We find 66\% (29/44) of the candidate CBe stars in NGC 7419 CBe stars to be variable in either optical, infrared or both. Of the five supergiants, one (V* My Cep) is classified as variable in optical. All supergiants in NEOWISE data have magnitudes beyond the saturation limit and, hence, cannot be analyzed. Our method and thresholds for identifying variables are quite stringent, and hence, we are unable to identify very low amplitude variables, as those sources may have a high Stetson Index due to a high correlation between the variation in the two bands. However, the amplitude of variation being low causes those sources to have quite low values of Standard Deviation and Median Absolute Deviation.

We compare our variability analysis with the three-band Gaia-based variability indicators from \cite{2023Maiz}. We find that over 90\% of the sources we classify as variables are also identified as variables in their catalog\footref{Maiz}. Notably, their method classifies most CBe stars in NGC 7419 as variables, whereas our method identifies only high-probability variables. We also crossmatch our list of variables with the Gaia-based large amplitude variables catalog\footref{Mowlavi catalog} of \cite{2021Mowlavi} and find 7 crossmatches (5 are CBe stars), all of which we classify as strong-type variables in Sections \ref{ztfvarsec} and \ref{neovarsec}. Even though the statistics are poor, we find that the $A_{\mathrm{proxy, BP}}$ / $A_{\mathrm{proxy, RP}}$ ratios of the 5 CBe crossmatches lie within the ranges shown in Fig. 15 of \cite{2021Mowlavi}.

We then use the Generalised Lomb Scargle periodogram (\citealt{1976Ap&SS..39..447L,1982ApJ...263..835S,vanderplas,GLS2009A&A...496..577Z}) to search for stars showing periodic signals. Following the suggestions given in \citealt{vanderplas,Aliasremovalalgo}, and \citealt{Ansdell}, we search for periodic signals in both ZTF and NEOWISE crossmatches in the cluster. The results of this analysis are given in Table \ref{ZTF variability summary} and \ref{NEO variability summary}. Since CBe stars are objects of special interest in this work, we take a deeper look into the various types of variability seen in these stars on different timescales in Section \ref{CBemultperiod}. Due to computational constraints, we only look for periods up to 150 days, and sources with even longer periods will be looked at in future works.

We investigate variability in CBe stars across different timescales. Our analysis reveals that 23\% (10/44) of CBe stars exhibit periodic signals, which can be ascribed to pulsation or rotation. While most known CBe stars pulsate \citep{CBemultiperiod,2013Rivinius}, non-sinusoidal periodic signals may overlap with the window function (Section \ref{ztfls}) and affect the period detection by the Lomb-Scargle periodogram. Additionally, ZTF and NEOWISE lightcurves often lack the cadence to capture high-frequency signals. Using Fig. 1.12 of \cite{AertsAstroseismology}, we find that these periodically pulsating CBe stars mostly lie near the typical theoretical $\beta$ Cephei instability strip, with our periods matching well with typical $\beta$ Cephei periods. We also observe that 41\% (18/44) of CBe stars show variability consistent with disk dynamics or binarity, and 14\% (6/44) exhibit long-term variations likely due to disk formation/dissipation. 

Trends in spectral types align with prior studies \citep{2013Rivinius,2018Granada}, with all stars showing periodicity due to non-radial pulsation being early-type, consistent with the lower pulsation amplitudes seen in late-type CBe stars. Among the stars with long-term disk variations, we find 50\% (3/6) are early-type and the other 3 are mid-type. \cite{2018Labadie} found a similar percentage (57\%) in long-term variations due to outbursts in early types. Future spectroscopic studies on the candidate CBe stars identified in this work, along with better cadence and longer-duration light curves, will improve our understanding of these stars.

Finally, to obtain global information and trends for a Be star's life stage, spectral type, and variability, we use WISE color-magnitude diagrams in a method similar to \cite{2018Granada}. We observe that CBe stars exclusively lie in the region defined by W1-W2>0.05, which comes from the fact that the Be phenomenon is known to redden the star \citep{2013Rivinius}.

A possible direction for future work involves addressing the effect of the disk duty cycle (DDC) -- the fraction of time a Be star is observed with a disk. This aspect has been largely overlooked in the literature. Estimates of the DDC in the SMC and LMC have been obtained by \cite{figueiredo2025}, highlighting its importance in correctly interpreting observed Be star fractions. For the Milky Way galaxy, the DDC remains unknown. The DDC for a large sample of clusters and their evolution under various environmental conditions will also be an interesting aspect to study.

\section*{Acknowledgements}
The authors are grateful to the reviewer for providing very constructive comments. We would also like to thank Zhen Ghuo for his constructive and valuable suggestions and comments, which have greatly improved the quality of the paper. This work has made use of data from the European Space Agency (ESA) mission {\it Gaia} (\url{https://www.cosmos.esa.int/gaia}), processed by the {\it Gaia} Data Processing and Analysis Consortium (DPAC,
\url{https://www.cosmos.esa.int/web/gaia/dpac/consortium}). Funding for the DPAC has been provided by national institutions, in particular, the institutions participating in the {\it Gaia} Multilateral Agreement.
This paper makes use of data obtained as part of the INT Photometric $H\alpha$ Survey of the Northern Galactic Plane (IPHAS, www.iphas.org) carried out at the Isaac Newton Telescope (INT). The INT is operated on the island of La Palma by the Isaac Newton Group in the Spanish Observatorio del Roque de los Muchachos of the Instituto de Astrofisica de Canarias. All IPHAS data are processed by the Cambridge Astronomical Survey Unit, at the Institute of Astronomy in Cambridge. The bandmerged DR2 catalogue was assembled at the Centre for Astrophysics Research, University of Hertfordshire, supported by STFC grant ST/J001333/1.
This paper is based on observations obtained with the Samuel Oschin Telescope 48-inch and the 60-inch Telescope at the Palomar
Observatory as part of the Zwicky Transient Facility project. ZTF is supported by the National Science Foundation under Grant No. AST1440341 and a collaboration including Caltech, IPAC, the Weizmann
Institute for Science, the Oskar Klein Center at Stockholm University, the University of Maryland, the University of Washington, Deutsches Elektronen-Synchrotron and Humboldt University, Los
Alamos National Laboratories, the TANGO Consortium of Taiwan, the University of Wisconsin at Milwaukee, and Lawrence Berkeley National Laboratories. Operations are conducted by COO, IPAC, and UW.
This publication also makes use of data products from NEOWISE, which is a project of the Jet Propulsion Laboratory/California Institute of Technology, funded by the Planetary Science Division of the National Aeronautics and Space Administration.
This publication makes use of data products from the Wide-field Infrared Survey Explorer, which is a joint project of the University of California, Los Angeles, and the Jet Propulsion Laboratory/California Institute of Technology, funded by the National Aeronautics and Space Administration. 
Funding for the Sloan Digital Sky Survey V has been provided by the Alfred P. Sloan Foundation, the Heising-Simons Foundation, the National Science Foundation, and the Participating Institutions. SDSS acknowledges support and resources from the Center for High-Performance Computing at the University of Utah. SDSS telescopes are located at Apache Point Observatory, funded by the Astrophysical Research Consortium and operated by New Mexico State University, and at Las Campanas Observatory, operated by the Carnegie Institution for Science. The SDSS website is \url{www.sdss.org}.

SDSS is managed by the Astrophysical Research Consortium for the Participating Institutions of the SDSS Collaboration, including Caltech, the Carnegie Institution for Science, Chilean National Time Allocation Committee (CNTAC) ratified researchers, The Flatiron Institute, the Gotham Participation Group, Harvard University, Heidelberg University, The Johns Hopkins University, L’Ecole polytechnique fédérale de Lausanne (EPFL), Leibniz-Institut für Astrophysik Potsdam (AIP), Max-Planck-Institut für Astronomie (MPIA Heidelberg), Max-Planck-Institut für Extraterrestrische Physik (MPE), Nanjing University, National Astronomical Observatories of China (NAOC), New Mexico State University, The Ohio State University, Pennsylvania State University, Smithsonian Astrophysical Observatory, Space Telescope Science Institute (STScI), the Stellar Astrophysics Participation Group, Universidad Nacional Autónoma de México, University of Arizona, University of Colorado Boulder, University of Illinois at Urbana-Champaign, University of Toronto, University of Utah, University of Virginia, Yale University, and Yunnan University. 

This research has made use of the VizieR catalogue access tool, CDS,
Strasbourg, France \citep{10.26093/cds/vizier}. The original description 
of the VizieR service was published in \citet{vizier2000}. We acknowledge the partial support from the grant RJF/2020/000071 as a part of the Ramanujan Fellowship (PI: Eswaraiah Chakali) awarded by the Science and Engineering Research Board (SERB). JJ acknowledges the financial support from the DST-SERB grant SPG/2021/003850. 
A. C. C. acknowledges support from CNPq (grant 314545/2023-9) and FAPESP (grants 2018/04055-8 and 2019/13354-1). 

\section*{Data Availability}

The data used in this paper are available as follows:

ZTF (dr18): \url{https://www.ztf.caltech.edu/ztf-public-releases.html}

WISE and NEOWISE: \url{https://irsa.ipac.caltech.edu/cgi-bin/Gator/nph-scan?submit=Select&projshort=WISE}

IPHAS: \url{https://vizier.cds.unistra.fr/viz-bin/VizieR?-source=IPHAS2}

The full version of the various tables in this paper can be found online.



\bibliographystyle{mnras}
\bibliography{b} 

\begin{thebibliography}{}
\makeatletter
\relax
\def\mn@urlcharsother{\let\do\@makeother \do\$\do\&\do\#\do\^\do\_\do\%\do\~}
\def\mn@doi{\begingroup\mn@urlcharsother \@ifnextchar [ {\mn@doi@} {\mn@doi@[]}}
\def\mn@doi@[#1]#2{\def\@tempa{#1}\ifx\@tempa\@empty \href {http://dx.doi.org/#2} {doi:#2}\else \href {http://dx.doi.org/#2} {#1}\fi \endgroup}
\def\mn@eprint#1#2{\mn@eprint@#1:#2::\@nil}
\def\mn@eprint@arXiv#1{\href {http://arxiv.org/abs/#1} {{\tt arXiv:#1}}}
\def\mn@eprint@dblp#1{\href {http://dblp.uni-trier.de/rec/bibtex/#1.xml} {dblp:#1}}
\def\mn@eprint@#1:#2:#3:#4\@nil{\def\@tempa {#1}\def\@tempb {#2}\def\@tempc {#3}\ifx \@tempc \@empty \let \@tempc \@tempb \let \@tempb \@tempa \fi \ifx \@tempb \@empty \def\@tempb {arXiv}\fi \@ifundefined {mn@eprint@\@tempb}{\@tempb:\@tempc}{\expandafter \expandafter \csname mn@eprint@\@tempb\endcsname \expandafter{\@tempc}}}

\bibitem[\protect\citeauthoryear{{Aerts}, {Christensen-Dalsgaard}  \& {Kurtz}}{{Aerts} et~al.}{2010}]{AertsAstroseismology}
{Aerts} C.,  {Christensen-Dalsgaard} J.,   {Kurtz} D.~W.,  2010, {Asteroseismology}, \mn@doi{10.1007/978-1-4020-5803-5.
}

\bibitem[\protect\citeauthoryear{Ansdell et~al.,}{Ansdell et~al.}{2017a}]{aliascleaning}
Ansdell M.,  et~al., 2017a, \mn@doi [Monthly Notices of the Royal Astronomical Society] {10.1093/mnras/stx2293}, 473, 1231

\bibitem[\protect\citeauthoryear{Ansdell et~al.,}{Ansdell et~al.}{2017b}]{Ansdell}
Ansdell M.,  et~al., 2017b, \mn@doi [Monthly Notices of the Royal Astronomical Society] {10.1093/mnras/stx2293}, 473, 1231

\bibitem[\protect\citeauthoryear{{Baade} et~al.,}{{Baade} et~al.}{2016}]{baade2016}
{Baade} D.,  et~al., 2016, \mn@doi [\aap] {10.1051/0004-6361/201528026}, \href {https://ui.adsabs.harvard.edu/abs/2016A&A...588A..56B} {588, A56}

\bibitem[\protect\citeauthoryear{{Bailer-Jones}, {Rybizki}, {Fouesneau}, {Demleitner}  \& {Andrae}}{{Bailer-Jones} et~al.}{2021a}]{2021yCat.1352....0B}
{Bailer-Jones} C.~A.~L.,  {Rybizki} J.,  {Fouesneau} M.,  {Demleitner} M.,   {Andrae} R.,  2021a, VizieR Online Data Catalog, \href {https://ui.adsabs.harvard.edu/abs/2021yCat.1352....0B} {p. I/352}

\bibitem[\protect\citeauthoryear{{Bailer-Jones}, {Rybizki}, {Fouesneau}, {Demleitner}  \& {Andrae}}{{Bailer-Jones} et~al.}{2021b}]{2021bailerjones}
{Bailer-Jones} C.~A.~L.,  {Rybizki} J.,  {Fouesneau} M.,  {Demleitner} M.,   {Andrae} R.,  2021b, \mn@doi [\aj] {10.3847/1538-3881/abd806}, \href {https://ui.adsabs.harvard.edu/abs/2021AJ....161..147B} {161, 147}

\bibitem[\protect\citeauthoryear{{Balona}}{{Balona}}{2022}]{2022Balona}
{Balona} L.~A.,  2022, \mn@doi [\mnras] {10.1093/mnras/stac2515}, \href {https://ui.adsabs.harvard.edu/abs/2022MNRAS.516.3641B} {516, 3641}

\bibitem[\protect\citeauthoryear{{Baluev}}{{Baluev}}{2008}]{baluev2008}
{Baluev} R.~V.,  2008, \mn@doi [\mnras] {10.1111/j.1365-2966.2008.12689.x}, \href {https://ui.adsabs.harvard.edu/abs/2008MNRAS.385.1279B} {385, 1279}

\bibitem[\protect\citeauthoryear{{Barentsen} et~al.,}{{Barentsen} et~al.}{2011}]{Barensten}
{Barentsen} G.,  et~al., 2011, \mn@doi [\mnras] {10.1111/j.1365-2966.2011.18674.x}, \href {https://ui.adsabs.harvard.edu/abs/2011MNRAS.415..103B} {415, 103}

\bibitem[\protect\citeauthoryear{Barentsen et~al.,}{Barentsen et~al.}{2014}]{IPHASDR2}
Barentsen G.,  et~al., 2014, \mn@doi [Monthly Notices of the Royal Astronomical Society] {10.1093/mnras/stu1651}, 444, 3230

\bibitem[\protect\citeauthoryear{{Beauchamp}, {Moffat}  \& {Drissen}}{{Beauchamp} et~al.}{1994}]{Beauchamp}
{Beauchamp} A.,  {Moffat} A. F.~J.,   {Drissen} L.,  1994, \mn@doi [\apjs] {10.1086/192051}, \href {https://ui.adsabs.harvard.edu/abs/1994ApJS...93..187B} {93, 187}

\bibitem[\protect\citeauthoryear{Bellm et~al.,}{Bellm et~al.}{2018}]{Bellm_2018}
Bellm E.~C.,  et~al., 2018, \mn@doi [Publications of the Astronomical Society of the Pacific] {10.1088/1538-3873/aaecbe}, 131, 018002

\bibitem[\protect\citeauthoryear{{Bhatt}, {Pandey}, {Mohan}, {Mahra}  \& {Paliwal}}{{Bhatt} et~al.}{1993}]{bhatt1993}
{Bhatt} B.~C.,  {Pandey} A.~K.,  {Mohan} V.,  {Mahra} H.~S.,   {Paliwal} D.~C.,  1993, Bulletin of the Astronomical Society of India, \href {https://ui.adsabs.harvard.edu/abs/1993BASI...21...33B} {21, 33}

\bibitem[\protect\citeauthoryear{Bishop}{Bishop}{2006}]{Pattern}
Bishop C.~M.,  2006, Pattern Recognition and Machine Learning (Information Science and Statistics).
Springer-Verlag, Berlin, Heidelberg

\bibitem[\protect\citeauthoryear{{Bodensteiner} et~al.,}{{Bodensteiner} et~al.}{2020}]{Bodensteiner}
{Bodensteiner} J.,  et~al., 2020, \mn@doi [\aap] {10.1051/0004-6361/201936743}, \href {https://ui.adsabs.harvard.edu/abs/2020A&A...634A..51B} {634, A51}

\bibitem[\protect\citeauthoryear{{Bossini} et~al.,}{{Bossini} et~al.}{2019}]{2019GaiaageDR2}
{Bossini} D.,  et~al., 2019, \mn@doi [\aap] {10.1051/0004-6361/201834693}, \href {https://ui.adsabs.harvard.edu/abs/2019A&A...623A.108B} {623, A108}

\bibitem[\protect\citeauthoryear{Breiman}{Breiman}{2001}]{Statistics01randomforests}
Breiman L.,  2001, Random Forests

\bibitem[\protect\citeauthoryear{{Bressan}, {Marigo}, {Girardi}, {Salasnich}, {Dal Cero}, {Rubele}  \& {Nanni}}{{Bressan} et~al.}{2012}]{2012MNRAS.427..127B}
{Bressan} A.,  {Marigo} P.,  {Girardi} L.,  {Salasnich} B.,  {Dal Cero} C.,  {Rubele} S.,   {Nanni} A.,  2012, \mn@doi [\mnras] {10.1111/j.1365-2966.2012.21948.x}, \href {https://ui.adsabs.harvard.edu/abs/2012MNRAS.427..127B} {427, 127}

\bibitem[\protect\citeauthoryear{Carciofi}{Carciofi}{2011}]{Carciofi:2011fc}
Carciofi A.~C.,  2011, \mn@doi [IAU Symp.] {10.1017/S1743921311010738}, 272, 325

\bibitem[\protect\citeauthoryear{{Carciofi}, {Okazaki}, {Le Bouquin}, {{\v{S}}tefl}, {Rivinius}, {Baade}, {Bjorkman}  \& {Hummel}}{{Carciofi} et~al.}{2009}]{carciofi2009}
{Carciofi} A.~C.,  {Okazaki} A.~T.,  {Le Bouquin} J.~B.,  {{\v{S}}tefl} S.,  {Rivinius} T.,  {Baade} D.,  {Bjorkman} J.~E.,   {Hummel} C.~A.,  2009, \mn@doi [\aap] {10.1051/0004-6361/200810962}, \href {https://ui.adsabs.harvard.edu/abs/2009A&A...504..915C} {504, 915}

\bibitem[\protect\citeauthoryear{{Caron}, {Moffat}, {St-Louis}, {Wade}  \& {Lester}}{{Caron} et~al.}{2003a}]{NGC7419Radius}
{Caron} G.,  {Moffat} A. F.~J.,  {St-Louis} N.,  {Wade} G.~A.,   {Lester} J.~B.,  2003a, \mn@doi [\aj] {10.1086/377314}, \href {https://ui.adsabs.harvard.edu/abs/2003AJ....126.1415C} {126, 1415}

\bibitem[\protect\citeauthoryear{{Caron}, {Moffat}, {St-Louis}, {Wade}  \& {Lester}}{{Caron} et~al.}{2003b}]{2003Caron}
{Caron} G.,  {Moffat} A. F.~J.,  {St-Louis} N.,  {Wade} G.~A.,   {Lester} J.~B.,  2003b, \mn@doi [\aj] {10.1086/377314}, \href {https://ui.adsabs.harvard.edu/abs/2003AJ....126.1415C} {126, 1415}

\bibitem[\protect\citeauthoryear{{Chen}, {Wang}, {Deng}, {de Grijs}, {Yang}  \& {Tian}}{{Chen} et~al.}{2020}]{2020ZTFPercatalog}
{Chen} X.,  {Wang} S.,  {Deng} L.,  {de Grijs} R.,  {Yang} M.,   {Tian} H.,  2020, {VizieR Online Data Catalog: The ZTF catalog of periodic variable stars (Chen+, 2020)}, VizieR On-line Data Catalog: J/ApJS/249/18. Originally published in: 2020ApJS..249...18C, \mn@doi{10.26093/cds/vizier.22490018}

\bibitem[\protect\citeauthoryear{{Cody} et~al.,}{{Cody} et~al.}{2014}]{cody2014}
{Cody} A.~M.,  et~al., 2014, \mn@doi [\aj] {10.1088/0004-6256/147/4/82}, \href {https://ui.adsabs.harvard.edu/abs/2014AJ....147...82C} {147, 82}

\bibitem[\protect\citeauthoryear{{Collins}}{{Collins}}{1963}]{1963ApJ...138.1134C}
{Collins} II G.~W.,  1963, \mn@doi [\apj] {10.1086/147712}, \href {https://ui.adsabs.harvard.edu/abs/1963ApJ...138.1134C} {138, 1134}

\bibitem[\protect\citeauthoryear{Collins}{Collins}{1987}]{collins_1987}
Collins G.~W.,  1987, \mn@doi [International Astronomical Union Colloquium] {10.1017/S0252921100115969}, 92, 3–21

\bibitem[\protect\citeauthoryear{{Cutri} et~al.,}{{Cutri} et~al.}{2012}]{2012wise.rept....1C}
{Cutri} R.~M.,  et~al., 2012, {Explanatory Supplement to the WISE All-Sky Data Release Products}, Explanatory Supplement to the WISE All-Sky Data Release Products

\bibitem[\protect\citeauthoryear{{Damian} et~al.,}{{Damian} et~al.}{2024}]{2024Damian}
{Damian} B.,  et~al., 2024, \mn@doi [\mnras] {10.1093/mnras/stae2452}, \href {https://ui.adsabs.harvard.edu/abs/2024MNRAS.535.1321D} {535, 1321}

\bibitem[\protect\citeauthoryear{Das, Gupta, Prakash, Samal  \& Jose}{Das et~al.}{2023}]{Das_2023}
Das S.~R.,  Gupta S.,  Prakash P.,  Samal M.,   Jose J.,  2023, \mn@doi [The Astrophysical Journal] {10.3847/1538-4357/acbf54}, 948, 7

\bibitem[\protect\citeauthoryear{{Das}, {Gupta}, {Jose}, {Samal}, {Herczeg}, {Guo}, {More}  \& {Prakash}}{{Das} et~al.}{2025}]{2025Das}
{Das} S.~R.,  {Gupta} S.,  {Jose} J.,  {Samal} M.,  {Herczeg} G.~J.,  {Guo} Z.,  {More} S.,   {Prakash} P.,  2025, \mn@doi [arXiv e-prints] {10.48550/arXiv.2503.16205}, \href {https://ui.adsabs.harvard.edu/abs/2025arXiv250316205D} {p. arXiv:2503.16205}

\bibitem[\protect\citeauthoryear{{Dougherty}, {Waters}, {Burki}, {Cote}, {Cramer}, {van Kerkwijk}  \& {Taylor}}{{Dougherty} et~al.}{1994}]{1994dougherty}
{Dougherty} S.~M.,  {Waters} L.~B.~F.~M.,  {Burki} G.,  {Cote} J.,  {Cramer} N.,  {van Kerkwijk} M.~H.,   {Taylor} A.~R.,  1994, \aap, \href {https://ui.adsabs.harvard.edu/abs/1994A&A...290..609D} {290, 609}

\bibitem[\protect\citeauthoryear{{Drew} et~al.,}{{Drew} et~al.}{2005}]{Drew}
{Drew} J.~E.,  et~al., 2005, \mn@doi [\mnras] {10.1111/j.1365-2966.2005.09330.x}, \href {https://ui.adsabs.harvard.edu/abs/2005MNRAS.362..753D} {362, 753}

\bibitem[\protect\citeauthoryear{{Dutta}, {Mondal}, {Jose}, {Das}, {Samal}  \& {Ghosh}}{{Dutta} et~al.}{2015}]{2015Dutta}
{Dutta} S.,  {Mondal} S.,  {Jose} J.,  {Das} R.~K.,  {Samal} M.~R.,   {Ghosh} S.,  2015, \mn@doi [\mnras] {10.1093/mnras/stv2190}, \href {https://ui.adsabs.harvard.edu/abs/2015MNRAS.454.3597D} {454, 3597}

\bibitem[\protect\citeauthoryear{Figueiredo, Carciofi, Labadie-Bartz, Pinho, de Amorin, dos Santos, Soszyński  \& Udalski}{Figueiredo et~al.}{2025}]{figueiredo2025}
Figueiredo A.~L.,  Carciofi A.~C.,  Labadie-Bartz J.,  Pinho M.~L.,  de Amorin T.~H.,  dos Santos P.~T.,  Soszyński I.,   Udalski A.,  2025, Be star demographics: a comprehensive study of thousands of lightcurves in the Magellanic Clouds (\mn@eprint {arXiv} {2505.08714}), \url {https://arxiv.org/abs/2505.08714}

\bibitem[\protect\citeauthoryear{{Fr{\'e}mat}, {Zorec}, {Hubert}  \& {Floquet}}{{Fr{\'e}mat} et~al.}{2005}]{Fremat}
{Fr{\'e}mat} Y.,  {Zorec} J.,  {Hubert} A.~M.,   {Floquet} M.,  2005, \mn@doi [\aap] {10.1051/0004-6361:20042229}, \href {https://ui.adsabs.harvard.edu/abs/2005A&A...440..305F} {440, 305}

\bibitem[\protect\citeauthoryear{{Gaia Collaboration} et~al.,}{{Gaia Collaboration} et~al.}{2016}]{gaia1}
{Gaia Collaboration} et~al., 2016, \mn@doi [\aap] {10.1051/0004-6361/201629272}, \href {https://ui.adsabs.harvard.edu/abs/2016A&A...595A...1G} {595, A1}

\bibitem[\protect\citeauthoryear{{Gaia Collaboration} et~al.,}{{Gaia Collaboration} et~al.}{2018a}]{gaiadr2paper}
{Gaia Collaboration} et~al., 2018a, \mn@doi [\aap] {10.1051/0004-6361/201833051}, \href {https://ui.adsabs.harvard.edu/abs/2018A&A...616A...1G} {616, A1}

\bibitem[\protect\citeauthoryear{{Gaia Collaboration} et~al.,}{{Gaia Collaboration} et~al.}{2018b}]{gaiacoefficients}
{Gaia Collaboration} et~al., 2018b, \mn@doi [\aap] {10.1051/0004-6361/201832843}, \href {https://ui.adsabs.harvard.edu/abs/2018A&A...616A..10G} {616, A10}

\bibitem[\protect\citeauthoryear{{Gaia Collaboration} et~al.,}{{Gaia Collaboration} et~al.}{2019}]{GAIADR2variablestars}
{Gaia Collaboration} et~al., 2019, \mn@doi [\aap] {10.1051/0004-6361/201833304}, \href {https://ui.adsabs.harvard.edu/abs/2019A&A...623A.110G} {623, A110}

\bibitem[\protect\citeauthoryear{{Gaia Collaboration} et~al.,}{{Gaia Collaboration} et~al.}{2023}]{gaiadr3paper}
{Gaia Collaboration} et~al., 2023, \mn@doi [\aap] {10.1051/0004-6361/202243940}, \href {https://ui.adsabs.harvard.edu/abs/2023A&A...674A...1G} {674, A1}

\bibitem[\protect\citeauthoryear{{Galli} et~al.,}{{Galli} et~al.}{2020}]{galli2020}
{Galli} P.~A.~B.,  et~al., 2020, \mn@doi [\aap] {10.1051/0004-6361/202038717}, \href {https://ui.adsabs.harvard.edu/abs/2020A&A...643A.148G} {643, A148}

\bibitem[\protect\citeauthoryear{{Gies}, {Bagnuolo}, {Ferrara}, {Kaye}, {Thaller}, {Penny}  \& {Peters}}{{Gies} et~al.}{1998}]{Gies}
{Gies} D.~R.,  {Bagnuolo} Jr. W.~G.,  {Ferrara} E.~C.,  {Kaye} A.~B.,  {Thaller} M.~L.,  {Penny} L.~R.,   {Peters} G.~J.,  1998, \mn@doi [\apj] {10.1086/305113}, \href {https://ui.adsabs.harvard.edu/abs/1998ApJ...493..440G} {493, 440}

\bibitem[\protect\citeauthoryear{Granada, Jones, Sigut, Semaan, Georgy, Meynet  \& Ekström}{Granada et~al.}{2018}]{2018Granada}
Granada A.,  Jones C.~E.,  Sigut T. A.~A.,  Semaan T.,  Georgy C.,  Meynet G.,   Ekström S.,  2018, \mn@doi [The Astronomical Journal] {10.3847/1538-3881/aa9f1d}, 155, 50

\bibitem[\protect\citeauthoryear{{Grundstrom}, {McSwain}, {Aragona}, {Boyajian}, {Marsh}  \& {Roettenbacher}}{{Grundstrom} et~al.}{2011}]{Grundstrom2011}
{Grundstrom} E.~D.,  {McSwain} M.~V.,  {Aragona} C.,  {Boyajian} T.~S.,  {Marsh} A.~N.,   {Roettenbacher} R.~M.,  2011, Bulletin de la Societe Royale des Sciences de Liege, \href {https://ui.adsabs.harvard.edu/abs/2011BSRSL..80..371G} {80, 371}

\bibitem[\protect\citeauthoryear{Gupta, Jose, Das, Guo, Damian, Prakash  \& Samal}{Gupta et~al.}{2024}]{gupta2024}
Gupta S.,  Jose J.,  Das S.~R.,  Guo Z.,  Damian B.,  Prakash P.,   Samal M.~R.,  2024, \mn@doi [Monthly Notices of the Royal Astronomical Society] {10.1093/mnras/stae369}, 528, 5633

\bibitem[\protect\citeauthoryear{Haubois, Carciofi, Rivinius, Okazaki  \& Bjorkman}{Haubois et~al.}{2012a}]{haubois2012dynamical}
Haubois X.,  Carciofi A.~C.,  Rivinius T.,  Okazaki A.,   Bjorkman J.,  2012a, The Astrophysical Journal, 756, 156

\bibitem[\protect\citeauthoryear{{Haubois}, {Carciofi}, {Rivinius}, {Okazaki}  \& {Bjorkman}}{{Haubois} et~al.}{2012b}]{2012Haubois}
{Haubois} X.,  {Carciofi} A.~C.,  {Rivinius} T.,  {Okazaki} A.~T.,   {Bjorkman} J.~E.,  2012b, \mn@doi [\apj] {10.1088/0004-637X/756/2/156}, \href {https://ui.adsabs.harvard.edu/abs/2012ApJ...756..156H} {756, 156}

\bibitem[\protect\citeauthoryear{{Hayasaki} \& {Okazaki}}{{Hayasaki} \& {Okazaki}}{2006}]{okazaki2006}
{Hayasaki} K.,  {Okazaki} A.~T.,  2006, \mn@doi [\mnras] {10.1111/j.1365-2966.2006.10917.x}, \href {https://ui.adsabs.harvard.edu/abs/2006MNRAS.372.1140H} {372, 1140}

\bibitem[\protect\citeauthoryear{Huang, Gies  \& McSwain}{Huang et~al.}{2010}]{Huang_2010}
Huang W.,  Gies D.~R.,   McSwain M.~V.,  2010, \mn@doi [The Astrophysical Journal] {10.1088/0004-637x/722/1/605}, 722, 605

\bibitem[\protect\citeauthoryear{{Jian}, {Matsunaga}, {Jiang}, {Yuan}  \& {Zhang}}{{Jian} et~al.}{2024}]{2024A&A...682A..59J}
{Jian} M.,  {Matsunaga} N.,  {Jiang} B.,  {Yuan} H.,   {Zhang} R.,  2024, \mn@doi [\aap] {10.1051/0004-6361/202347911}, \href {https://ui.adsabs.harvard.edu/abs/2024A&A...682A..59J} {682, A59}

\bibitem[\protect\citeauthoryear{{Johnson}}{{Johnson}}{1967}]{1967Johnson}
{Johnson} H.~L.,  1967, \mn@doi [\apjl] {10.1086/180088}, \href {https://ui.adsabs.harvard.edu/abs/1967ApJ...150L..39J} {150, L39}

\bibitem[\protect\citeauthoryear{Joshi, Kumar, Singh, Sagar, Sharma  \& Pandey}{Joshi et~al.}{2008}]{Joshi}
Joshi H.,  Kumar B.,  Singh K.~P.,  Sagar R.,  Sharma S.,   Pandey J.~C.,  2008, \mn@doi [Monthly Notices of the Royal Astronomical Society] {https://doi.org/10.1111/j.1365-2966.2008.13936.x}, 391, 1279

\bibitem[\protect\citeauthoryear{Kee, Owocki, Townsend  \& Müller}{Kee et~al.}{2015}]{kee2015pulsationalmassejectionstar}
Kee N.,  Owocki S.,  Townsend R.,   Müller H.-R.,  2015, Pulsational Mass Ejection in Be Star Disks (\mn@eprint {arXiv} {1412.8511}), \url {https://arxiv.org/abs/1412.8511}

\bibitem[\protect\citeauthoryear{Kiss, Szabo  \& Bedding}{Kiss et~al.}{2006}]{2006}
Kiss L.~L.,  Szabo G.~M.,   Bedding T.~R.,  2006, \mn@doi [Monthly Notices of the Royal Astronomical Society] {10.1111/j.1365-2966.2006.10973.x}, 372, 1721–1734

\bibitem[\protect\citeauthoryear{{Kozhurina-Platais}, {Girard}, {Platais}, {van Altena}, {Ianna}  \& {Cannon}}{{Kozhurina-Platais} et~al.}{1995}]{kozhurina}
{Kozhurina-Platais} V.,  {Girard} T.~M.,  {Platais} I.,  {van Altena} W.~F.,  {Ianna} P.~A.,   {Cannon} R.~D.,  1995, \mn@doi [\aj] {10.1086/117310}, \href {https://ui.adsabs.harvard.edu/abs/1995AJ....109..672K} {109, 672}

\bibitem[\protect\citeauthoryear{{Kramer}, {Gowanlock}, {Trilling}, {McNeill}  \& {Erasmus}}{{Kramer} et~al.}{2023}]{Aliasremovalalgo}
{Kramer} D.,  {Gowanlock} M.,  {Trilling} D.,  {McNeill} A.,   {Erasmus} N.,  2023, \mn@doi [Astronomy and Computing] {10.1016/j.ascom.2023.100711}, \href {https://ui.adsabs.harvard.edu/abs/2023A&C....4400711K} {44, 100711}

\bibitem[\protect\citeauthoryear{{Kroll} \& {Hanuschik}}{{Kroll} \& {Hanuschik}}{1997}]{1997kroll}
{Kroll} P.,  {Hanuschik} R.~W.,  1997, in {Wickramasinghe} D.~T.,  {Bicknell} G.~V.,   {Ferrario} L.,  eds,  Astronomical Society of the Pacific Conference Series Vol. 121, IAU Colloq. 163: Accretion Phenomena and Related Outflows. p.~494

\bibitem[\protect\citeauthoryear{{Labadie-Bartz} et~al.,}{{Labadie-Bartz} et~al.}{2018}]{2018Labadie}
{Labadie-Bartz} J.,  et~al., 2018, \mn@doi [\aj] {10.3847/1538-3881/aa9c7e}, \href {https://ui.adsabs.harvard.edu/abs/2018AJ....155...53L} {155, 53}

\bibitem[\protect\citeauthoryear{{Labadie-Bartz}, {Carciofi}, {Henrique de Amorim}, {Rubio}, {Luiz Figueiredo}, {Ticiani dos Santos}  \& {Thomson-Paressant}}{{Labadie-Bartz} et~al.}{2022a}]{CBemultiperiod}
{Labadie-Bartz} J.,  {Carciofi} A.~C.,  {Henrique de Amorim} T.,  {Rubio} A.,  {Luiz Figueiredo} A.,  {Ticiani dos Santos} P.,   {Thomson-Paressant} K.,  2022a, \mn@doi [\aj] {10.3847/1538-3881/ac5abd}, \href {https://ui.adsabs.harvard.edu/abs/2022AJ....163..226L} {163, 226}

\bibitem[\protect\citeauthoryear{{Labadie-Bartz}, {Carciofi}, {Henrique de Amorim}, {Rubio}, {Luiz Figueiredo}, {Ticiani dos Santos}  \& {Thomson-Paressant}}{{Labadie-Bartz} et~al.}{2022b}]{2022Labadie}
{Labadie-Bartz} J.,  {Carciofi} A.~C.,  {Henrique de Amorim} T.,  {Rubio} A.,  {Luiz Figueiredo} A.,  {Ticiani dos Santos} P.,   {Thomson-Paressant} K.,  2022b, \mn@doi [\aj] {10.3847/1538-3881/ac5abd}, \href {https://ui.adsabs.harvard.edu/abs/2022AJ....163..226L} {163, 226}

\bibitem[\protect\citeauthoryear{{Labadie-Bartz} et~al.,}{{Labadie-Bartz} et~al.}{2025}]{2025Labadie}
{Labadie-Bartz} J.,  et~al., 2025, \mn@doi [arXiv e-prints] {10.48550/arXiv.2504.07571}, \href {https://ui.adsabs.harvard.edu/abs/2025arXiv250407571L} {p. arXiv:2504.07571}

\bibitem[\protect\citeauthoryear{{Lee}, {Osaki}  \& {Saio}}{{Lee} et~al.}{1991}]{1991Lee}
{Lee} U.,  {Osaki} Y.,   {Saio} H.,  1991, \mn@doi [\mnras] {10.1093/mnras/250.2.432}, \href {https://ui.adsabs.harvard.edu/abs/1991MNRAS.250..432L} {250, 432}

\bibitem[\protect\citeauthoryear{{Lindegren} et~al.,}{{Lindegren} et~al.}{2021}]{lindegren2021}
{Lindegren} L.,  et~al., 2021, \mn@doi [\aap] {10.1051/0004-6361/202039709}, \href {https://ui.adsabs.harvard.edu/abs/2021A&A...649A...2L} {649, A2}

\bibitem[\protect\citeauthoryear{{Lomb}}{{Lomb}}{1976}]{1976Ap&SS..39..447L}
{Lomb} N.~R.,  1976, \mn@doi [\apss] {10.1007/BF00648343}, \href {https://ui.adsabs.harvard.edu/abs/1976Ap&SS..39..447L} {39, 447}

\bibitem[\protect\citeauthoryear{{Mainzer} et~al.,}{{Mainzer} et~al.}{2011}]{Mainzer2011}
{Mainzer} A.,  et~al., 2011, \mn@doi [\apj] {10.1088/0004-637X/743/2/156}, \href {https://ui.adsabs.harvard.edu/abs/2011ApJ...743..156M} {743, 156}

\bibitem[\protect\citeauthoryear{{Mainzer} et~al.,}{{Mainzer} et~al.}{2014}]{2014ApJ...792...30M}
{Mainzer} A.,  et~al., 2014, \mn@doi [\apj] {10.1088/0004-637X/792/1/30}, \href {https://ui.adsabs.harvard.edu/abs/2014ApJ...792...30M} {792, 30}

\bibitem[\protect\citeauthoryear{{Ma{\'\i}z Apell{\'a}niz}, {Holgado}, {Pantaleoni Gonz{\'a}lez}  \& {Caballero}}{{Ma{\'\i}z Apell{\'a}niz} et~al.}{2023}]{2023Maiz}
{Ma{\'\i}z Apell{\'a}niz} J.,  {Holgado} G.,  {Pantaleoni Gonz{\'a}lez} M.,   {Caballero} J.~A.,  2023, \mn@doi [\aap] {10.1051/0004-6361/202346759}, \href {https://ui.adsabs.harvard.edu/abs/2023A&A...677A.137M} {677, A137}

\bibitem[\protect\citeauthoryear{Manoj, Bhatt, Gopinathan  \& Salim}{Manoj et~al.}{2006}]{Herbig}
Manoj P.,  Bhatt H.~C.,  Gopinathan M.,   Salim M.,  2006, \mn@doi [The Astrophysical Journal] {10.1086/508764}, 653

\bibitem[\protect\citeauthoryear{Marco \& Negueruela}{Marco \& Negueruela}{2013}]{Marco_2013}
Marco A.,  Negueruela I.,  2013, \mn@doi [Astronomy & Astrophysics] {10.1051/0004-6361/201220750}, 552, A92

\bibitem[\protect\citeauthoryear{{Marton} et~al.,}{{Marton} et~al.}{2023}]{2023Marton}
{Marton} G.,  et~al., 2023, \mn@doi [\aap] {10.1051/0004-6361/202244101}, \href {https://ui.adsabs.harvard.edu/abs/2023A&A...674A..21M} {674, A21}

\bibitem[\protect\citeauthoryear{Masci et~al.,}{Masci et~al.}{2018}]{2018}
Masci F.~J.,  et~al., 2018, \mn@doi [Publications of the Astronomical Society of the Pacific] {10.1088/1538-3873/aae8ac}, 131, 018003

\bibitem[\protect\citeauthoryear{{Mathew} \& {Subramaniam}}{{Mathew} \& {Subramaniam}}{2011}]{2011subramaniam}
{Mathew} B.,  {Subramaniam} A.,  2011, \mn@doi [Bulletin of the Astronomical Society of India] {10.48550/arXiv.1108.5850}, \href {https://ui.adsabs.harvard.edu/abs/2011BASI...39..517M} {39, 517}

\bibitem[\protect\citeauthoryear{{Mowlavi} et~al.,}{{Mowlavi} et~al.}{2021}]{2021Mowlavi}
{Mowlavi} N.,  et~al., 2021, \mn@doi [\aap] {10.1051/0004-6361/202039450}, \href {https://ui.adsabs.harvard.edu/abs/2021A&A...648A..44M} {648, A44}

\bibitem[\protect\citeauthoryear{Murphy}{Murphy}{2012}]{Probabilistic}
Murphy K.~P.,  2012, Machine Learning: A Probabilistic Perspective.
The MIT Press

\bibitem[\protect\citeauthoryear{{Navarete}, {Ticiani dos Santos}, {Carciofi}  \& {Figueiredo}}{{Navarete} et~al.}{2024}]{2024Navarete}
{Navarete} F.,  {Ticiani dos Santos} P.,  {Carciofi} A.~C.,   {Figueiredo} A.~L.,  2024, \mn@doi [\apj] {10.3847/1538-4357/ad500f}, \href {https://ui.adsabs.harvard.edu/abs/2024ApJ...970..113N} {970, 113}

\bibitem[\protect\citeauthoryear{{Naz{\'e}}, {Rauw}  \& {Pigulski}}{{Naz{\'e}} et~al.}{2020}]{Naze2020}
{Naz{\'e}} Y.,  {Rauw} G.,   {Pigulski} A.,  2020, \mn@doi [\mnras] {10.1093/mnras/staa2553}, \href {https://ui.adsabs.harvard.edu/abs/2020MNRAS.498.3171N} {498, 3171}

\bibitem[\protect\citeauthoryear{{Nikutta}, {Hunt-Walker}, {Nenkova}, {Ivezi{\'c}}  \& {Elitzur}}{{Nikutta} et~al.}{2014}]{2014MNRAS.442.3361N}
{Nikutta} R.,  {Hunt-Walker} N.,  {Nenkova} M.,  {Ivezi{\'c}} {\v{Z}}.,   {Elitzur} M.,  2014, \mn@doi [\mnras] {10.1093/mnras/stu1087}, \href {https://ui.adsabs.harvard.edu/abs/2014MNRAS.442.3361N} {442, 3361}

\bibitem[\protect\citeauthoryear{Ochsenbein}{Ochsenbein}{1996}]{10.26093/cds/vizier}
Ochsenbein F.,  1996, The VizieR database of astronomical catalogues, \mn@doi{10.26093/CDS/VIZIER}, \url {https://vizier.cds.unistra.fr}

\bibitem[\protect\citeauthoryear{{Ochsenbein}, {Bauer}  \& {Marcout}}{{Ochsenbein} et~al.}{2000}]{vizier2000}
{Ochsenbein} F.,  {Bauer} P.,   {Marcout} J.,  2000, \mn@doi [\aaps] {10.1051/aas:2000169}, \href {https://ui.adsabs.harvard.edu/abs/2000A&AS..143...23O} {143, 23}

\bibitem[\protect\citeauthoryear{{Panoglou}, {Carciofi}, {Vieira}, {Cyr}, {Jones}, {Okazaki}  \& {Rivinius}}{{Panoglou} et~al.}{2016}]{2016Panoglou}
{Panoglou} D.,  {Carciofi} A.~C.,  {Vieira} R.~G.,  {Cyr} I.~H.,  {Jones} C.~E.,  {Okazaki} A.~T.,   {Rivinius} T.,  2016, \mn@doi [\mnras] {10.1093/mnras/stw1508}, \href {https://ui.adsabs.harvard.edu/abs/2016MNRAS.461.2616P} {461, 2616}

\bibitem[\protect\citeauthoryear{{Panoglou}, {Faes}, {Carciofi}, {Okazaki}, {Baade}, {Rivinius}  \& {Borges Fernandes}}{{Panoglou} et~al.}{2018}]{2018Panoglou}
{Panoglou} D.,  {Faes} D.~M.,  {Carciofi} A.~C.,  {Okazaki} A.~T.,  {Baade} D.,  {Rivinius} T.,   {Borges Fernandes} M.,  2018, \mn@doi [\mnras] {10.1093/mnras/stx2497}, \href {https://ui.adsabs.harvard.edu/abs/2018MNRAS.473.3039P} {473, 3039}

\bibitem[\protect\citeauthoryear{{Pecaut} \& {Mamajek}}{{Pecaut} \& {Mamajek}}{2013}]{2013PecautMamajek}
{Pecaut} M.~J.,  {Mamajek} E.~E.,  2013, \mn@doi [\apjs] {10.1088/0067-0049/208/1/9}, \href {https://ui.adsabs.harvard.edu/abs/2013ApJS..208....9P} {208, 9}

\bibitem[\protect\citeauthoryear{Pedregosa et~al.,}{Pedregosa et~al.}{2018}]{pedregosa2018scikitlearnmachinelearningpython}
Pedregosa F.,  et~al., 2018, Scikit-learn: Machine Learning in Python (\mn@eprint {arXiv} {1201.0490}), \url {https://arxiv.org/abs/1201.0490}

\bibitem[\protect\citeauthoryear{{Porter} \& {Rivinius}}{{Porter} \& {Rivinius}}{2003a}]{2003PorterCBe}
{Porter} J.~M.,  {Rivinius} T.,  2003a, \mn@doi [\pasp] {10.1086/378307}, \href {https://ui.adsabs.harvard.edu/abs/2003PASP..115.1153P} {115, 1153}

\bibitem[\protect\citeauthoryear{{Porter} \& {Rivinius}}{{Porter} \& {Rivinius}}{2003b}]{2003Rivinius}
{Porter} J.~M.,  {Rivinius} T.,  2003b, \mn@doi [\pasp] {10.1086/378307}, \href {https://ui.adsabs.harvard.edu/abs/2003PASP..115.1153P} {115, 1153}

\bibitem[\protect\citeauthoryear{{R{\'\i}mulo} et~al.,}{{R{\'\i}mulo} et~al.}{2018}]{Rimulo2018}
{R{\'\i}mulo} L.~R.,  et~al., 2018, \mn@doi [\mnras] {10.1093/mnras/sty431}, \href {https://ui.adsabs.harvard.edu/abs/2018MNRAS.476.3555R} {476, 3555}

\bibitem[\protect\citeauthoryear{{Rivinius, Th.}, {Stefl, S.}  \& {Baade, D.}}{{Rivinius, Th.} et~al.}{2006}]{Riv2006}
{Rivinius, Th.} {Stefl, S.}  {Baade, D.} 2006, \mn@doi [A\&A] {10.1051/0004-6361:20053008}, 459, 137

\bibitem[\protect\citeauthoryear{{Rivinius}, {Baade}, {Stefl}, {Stahl}, {Wolf}  \& {Kaufer}}{{Rivinius} et~al.}{1998}]{1998rivinius}
{Rivinius} T.,  {Baade} D.,  {Stefl} S.,  {Stahl} O.,  {Wolf} B.,   {Kaufer} A.,  1998, \aap, \href {https://ui.adsabs.harvard.edu/abs/1998A&A...336..177R} {336, 177}

\bibitem[\protect\citeauthoryear{Rivinius, Carciofi  \& Martayan}{Rivinius et~al.}{2013}]{2013Rivinius}
Rivinius T.,  Carciofi A.~C.,   Martayan C.,  2013, \mn@doi [The Astronomy and Astrophysics Review] {10.1007/s00159-013-0069-0}, 21

\bibitem[\protect\citeauthoryear{{Rubio}, {Carciofi}, {Bjorkman}, {de Amorim}, {Okazaki}, {Suffak}, {Jones}  \& {Candido}}{{Rubio} et~al.}{2025}]{2025Rubio}
{Rubio} A.~C.,  {Carciofi} A.~C.,  {Bjorkman} J.~E.,  {de Amorim} T.~H.,  {Okazaki} A.~T.,  {Suffak} M.~W.,  {Jones} C.~E.,   {Candido} P.~P.,  2025, \mn@doi [arXiv e-prints] {10.48550/arXiv.2502.11626}, \href {https://ui.adsabs.harvard.edu/abs/2025arXiv250211626R} {p. arXiv:2502.11626}

\bibitem[\protect\citeauthoryear{{Sanders}}{{Sanders}}{1971}]{sanders1971}
{Sanders} W.~L.,  1971, \aap, \href {https://ui.adsabs.harvard.edu/abs/1971A&A....14..226S} {14, 226}

\bibitem[\protect\citeauthoryear{{Sarro} et~al.,}{{Sarro} et~al.}{2014}]{Sarro2014}
{Sarro} L.~M.,  et~al., 2014, \mn@doi [\aap] {10.1051/0004-6361/201322413}, \href {https://ui.adsabs.harvard.edu/abs/2014A&A...563A..45S} {563, A45}

\bibitem[\protect\citeauthoryear{{Scargle}}{{Scargle}}{1982}]{1982ApJ...263..835S}
{Scargle} J.~D.,  1982, \mn@doi [\apj] {10.1086/160554}, \href {https://ui.adsabs.harvard.edu/abs/1982ApJ...263..835S} {263, 835}

\bibitem[\protect\citeauthoryear{Schmidtke \& Hunter}{Schmidtke \& Hunter}{2019}]{Schmidtke_2019}
Schmidtke P.,  Hunter T.,  2019, \mn@doi [Research Notes of the {AAS}] {10.3847/2515-5172/ab1d61}, 3, 67

\bibitem[\protect\citeauthoryear{Sokolovsky et~al.,}{Sokolovsky et~al.}{2016a}]{Sokolovsky_2016}
Sokolovsky K.~V.,  et~al., 2016a, \mn@doi [Monthly Notices of the Royal Astronomical Society] {10.1093/mnras/stw2262}, 464, 274

\bibitem[\protect\citeauthoryear{Sokolovsky et~al.,}{Sokolovsky et~al.}{2016b}]{10.1093/mnras/stw2262}
Sokolovsky K.~V.,  et~al., 2016b, \mn@doi [Monthly Notices of the Royal Astronomical Society] {10.1093/mnras/stw2262}, 464, 274

\bibitem[\protect\citeauthoryear{{Stefl}, {Rivinius}, {Le Bouquin}, {Carciofi}, {Baade}, {Otero}  \& {Rantakyr{\"o}}}{{Stefl} et~al.}{2010}]{quasicyclicCBe}
{Stefl} S.,  {Rivinius} T.,  {Le Bouquin} J.~B.,  {Carciofi} A.,  {Baade} D.,  {Otero} S.,   {Rantakyr{\"o}} F.,  2010, in Revista Mexicana de Astronomia y Astrofisica Conference Series. pp 89--91

\bibitem[\protect\citeauthoryear{{Stetson}}{{Stetson}}{1996}]{1996stetson}
{Stetson} P.~B.,  1996, \mn@doi [\pasp] {10.1086/133808}, \href {https://ui.adsabs.harvard.edu/abs/1996PASP..108..851S} {108, 851}

\bibitem[\protect\citeauthoryear{Subramaniam, Mathew, Bhatt  \& Ramya}{Subramaniam et~al.}{2006}]{subramaniam}
Subramaniam A.,  Mathew B.,  Bhatt B.~C.,   Ramya S.,  2006, \mn@doi [Monthly Notices of the Royal Astronomical Society] {10.1111/j.1365-2966.2006.10481.x}, 370, 743

\bibitem[\protect\citeauthoryear{Townsend, Owocki  \& Howarth}{Townsend et~al.}{2004}]{Townsend}
Townsend R. H.~D.,  Owocki S.~P.,   Howarth I.~D.,  2004, \mn@doi [Monthly Notices of the Royal Astronomical Society] {10.1111/j.1365-2966.2004.07627.x}, 350, 189

\bibitem[\protect\citeauthoryear{{VanderPlas}}{{VanderPlas}}{2018}]{vanderplas}
{VanderPlas} J.~T.,  2018, \mn@doi [\apjs] {10.3847/1538-4365/aab766}, \href {https://ui.adsabs.harvard.edu/abs/2018ApJS..236...16V} {236, 16}

\bibitem[\protect\citeauthoryear{{Vasilevskis}, {Klemola}  \& {Preston}}{{Vasilevskis} et~al.}{1958}]{vasilevskis}
{Vasilevskis} S.,  {Klemola} A.,   {Preston} G.,  1958, \mn@doi [\aj] {10.1086/107787}, \href {https://ui.adsabs.harvard.edu/abs/1958AJ.....63..387V} {63, 387}

\bibitem[\protect\citeauthoryear{{Vieira}, {Carciofi}  \& {Bjorkman}}{{Vieira} et~al.}{2015}]{2015Viera}
{Vieira} R.~G.,  {Carciofi} A.~C.,   {Bjorkman} J.~E.,  2015, \mn@doi [\mnras] {10.1093/mnras/stv2074}, \href {https://ui.adsabs.harvard.edu/abs/2015MNRAS.454.2107V} {454, 2107}

\bibitem[\protect\citeauthoryear{{Vieira}, {Carciofi}, {Bjorkman}, {Rivinius}, {Baade}  \& {R{\'\i}mulo}}{{Vieira} et~al.}{2017}]{Viera2017}
{Vieira} R.~G.,  {Carciofi} A.~C.,  {Bjorkman} J.~E.,  {Rivinius} T.,  {Baade} D.,   {R{\'\i}mulo} L.~R.,  2017, \mn@doi [\mnras] {10.1093/mnras/stw2542}, \href {https://ui.adsabs.harvard.edu/abs/2017MNRAS.464.3071V} {464, 3071}

\bibitem[\protect\citeauthoryear{Virtanen et~al.,}{Virtanen et~al.}{2020}]{2020SciPy-NMeth}
Virtanen P.,  et~al., 2020, \mn@doi [Nature Methods] {10.1038/s41592-019-0686-2}, \href {https://rdcu.be/b08Wh} {17, 261}

\bibitem[\protect\citeauthoryear{{Walker} et~al.,}{{Walker} et~al.}{2005}]{Walker2005CBemultperiod}
{Walker} G.~A.~H.,  et~al., 2005, \mn@doi [\apjl] {10.1086/430254}, \href {https://ui.adsabs.harvard.edu/abs/2005ApJ...623L.145W} {623, L145}

\bibitem[\protect\citeauthoryear{{Welch} \& {Stetson}}{{Welch} \& {Stetson}}{1993}]{1993AJ....105.1813W}
{Welch} D.~L.,  {Stetson} P.~B.,  1993, \mn@doi [\aj] {10.1086/116556}, \href {https://ui.adsabs.harvard.edu/abs/1993AJ....105.1813W} {105, 1813}

\bibitem[\protect\citeauthoryear{{Wright} et~al.,}{{Wright} et~al.}{2010}]{WISE}
{Wright} E.~L.,  et~al., 2010, \mn@doi [\aj] {10.1088/0004-6256/140/6/1868}, \href {https://ui.adsabs.harvard.edu/abs/2010AJ....140.1868W} {140, 1868}

\bibitem[\protect\citeauthoryear{{Zechmeister} \& {K{\"u}rster}}{{Zechmeister} \& {K{\"u}rster}}{2009}]{GLS2009A&A...496..577Z}
{Zechmeister} M.,  {K{\"u}rster} M.,  2009, \mn@doi [\aap] {10.1051/0004-6361:200811296}, \href {https://ui.adsabs.harvard.edu/abs/2009A&A...496..577Z} {496, 577}

\makeatother
\end{thebibliography}

\clearpage



\appendix

\section{Figures}\label{App Figures}

\begin{figure}
\includegraphics[width=\columnwidth]{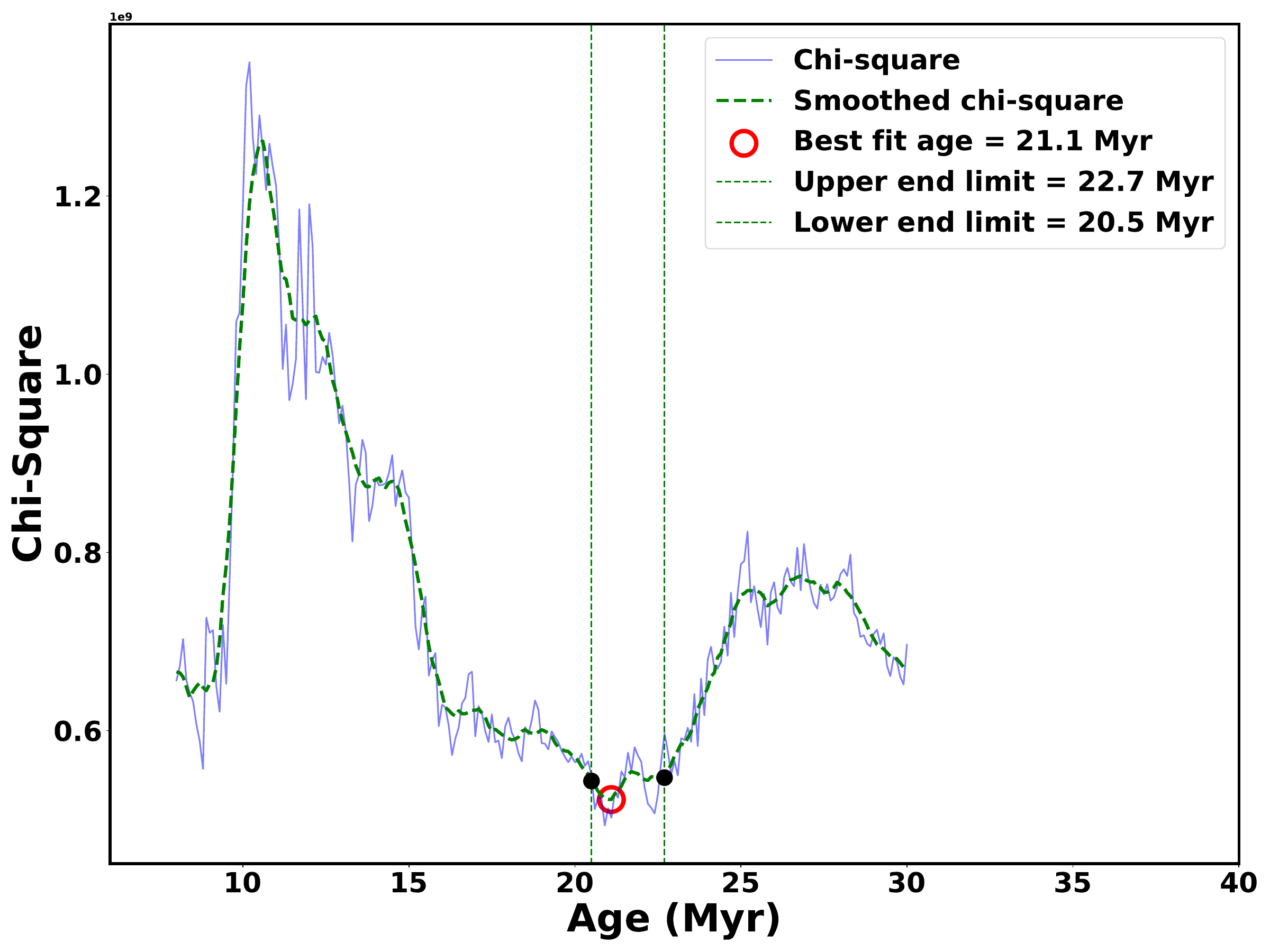}
\caption{Chi-square values for isochrones of ages in the range of 8-30 Myr for NGC 7419. The red open circle indicates the lowest chi-square value (best fit) corresponding to the 21.1 Myr isochrone. The green dashed vertical lines indicate the upper and lower limits within 5\% of this age.}
\label{chi-square isochrone}
\end{figure}

\begin{figure}
\centerline{\includegraphics[width=\columnwidth]{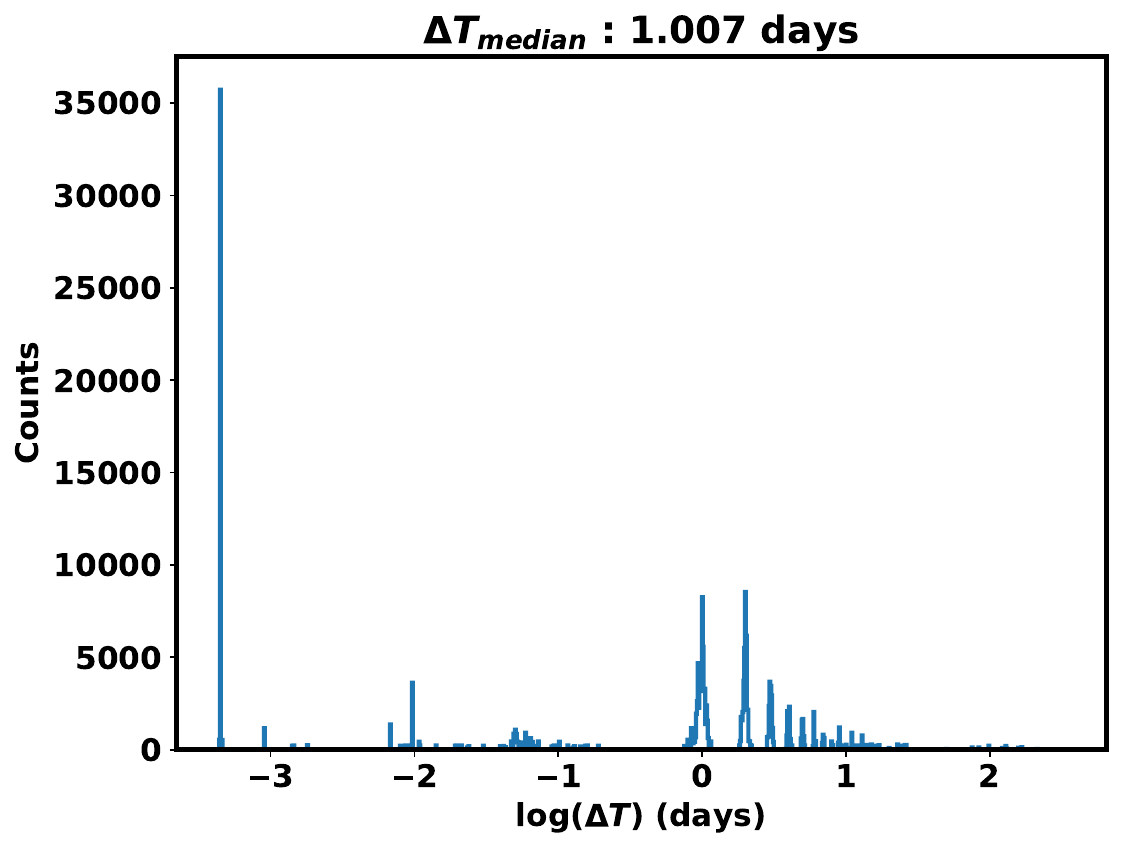}}
\caption{Histogram of $\mathrm{log_{10}}$($\Delta$T) values for ZTF lightcurves. $\Delta$T is the time difference between two consecutive observations in a lightcurve.}
\label{ZTF delt Hist}
\end{figure}

\begin{figure}
\centerline{\includegraphics[width=\columnwidth]{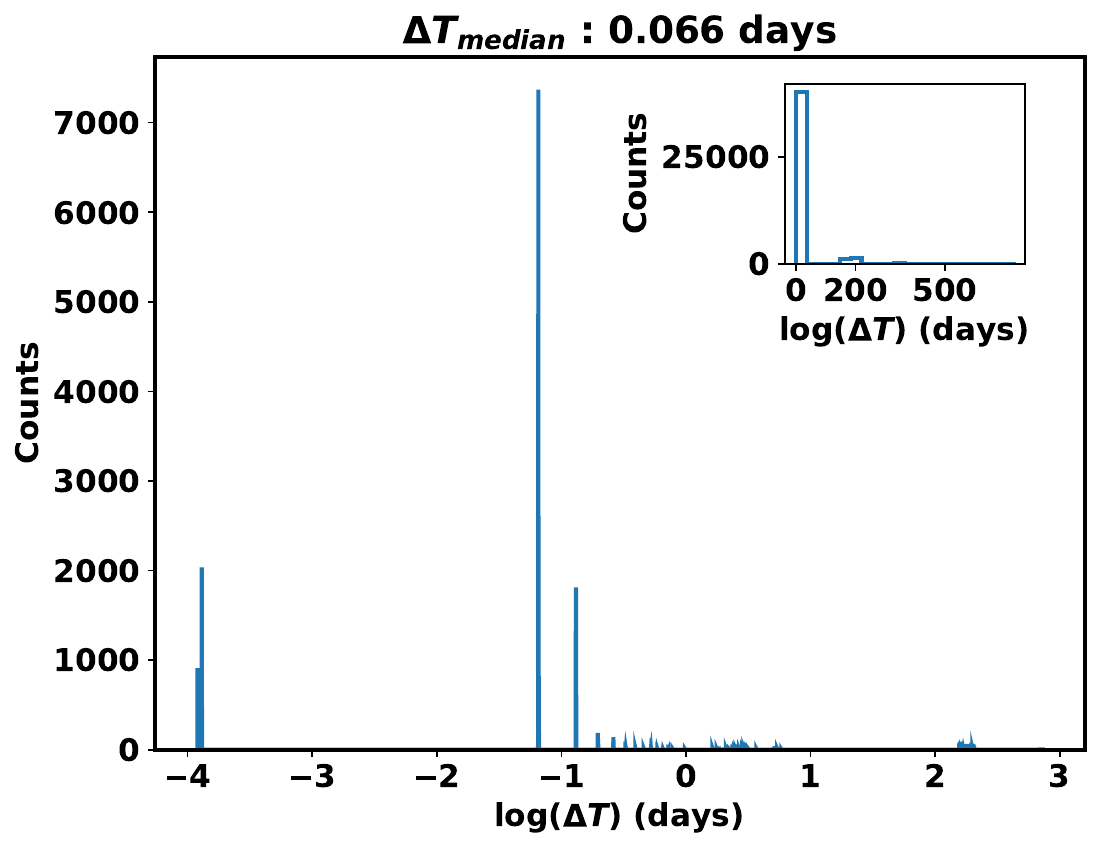}}
\caption{Histogram of $\mathrm{log_{10}}$($\Delta$T) values for NEOWISE lightcurves. $\Delta$T is the time difference between two consecutive observations in a lightcurve.}
\label{NEOWISE delt Hist}
\end{figure}

\begin{figure*}
\begin{multicols}{2}
     \subcaptionbox{Standard Deviation (W1)}{\includegraphics[width=\columnwidth]{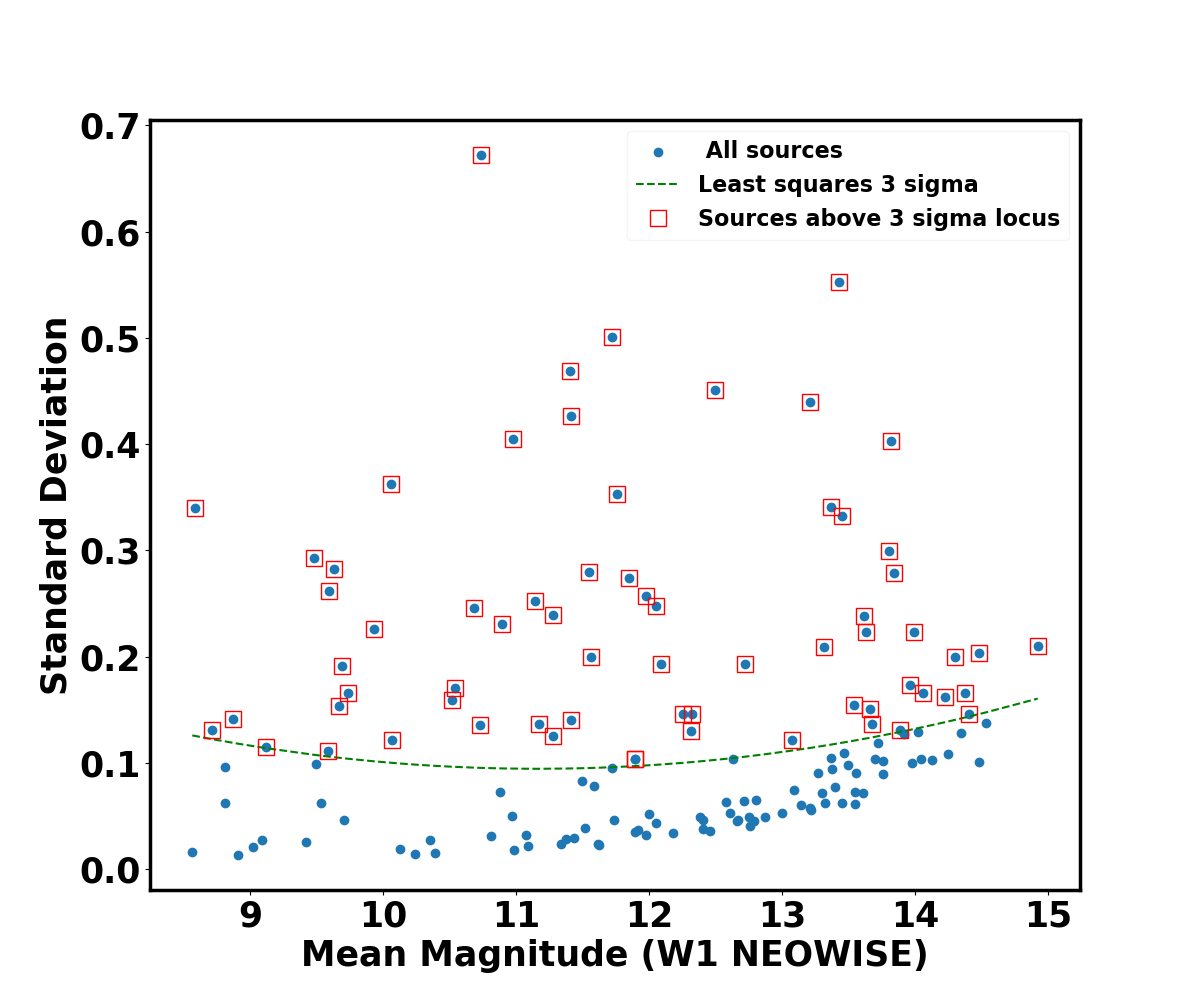}}\par 
     \subcaptionbox{Standard Deviation (W2)}{\includegraphics[width=\columnwidth]{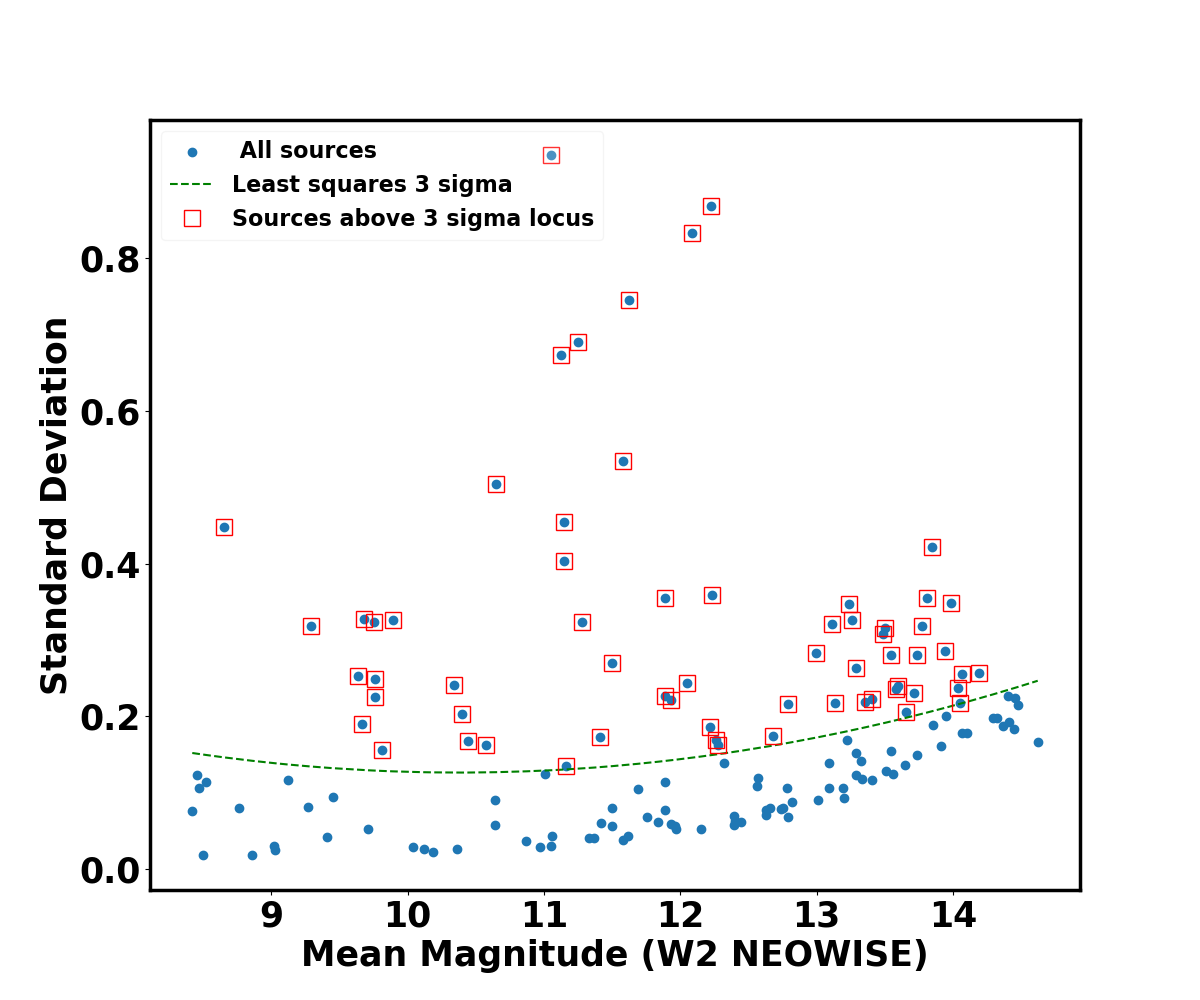}}\par
   \end{multicols}     
   \begin{multicols}{2}
      \subcaptionbox{MAD (W1)}{\includegraphics[width=\columnwidth]{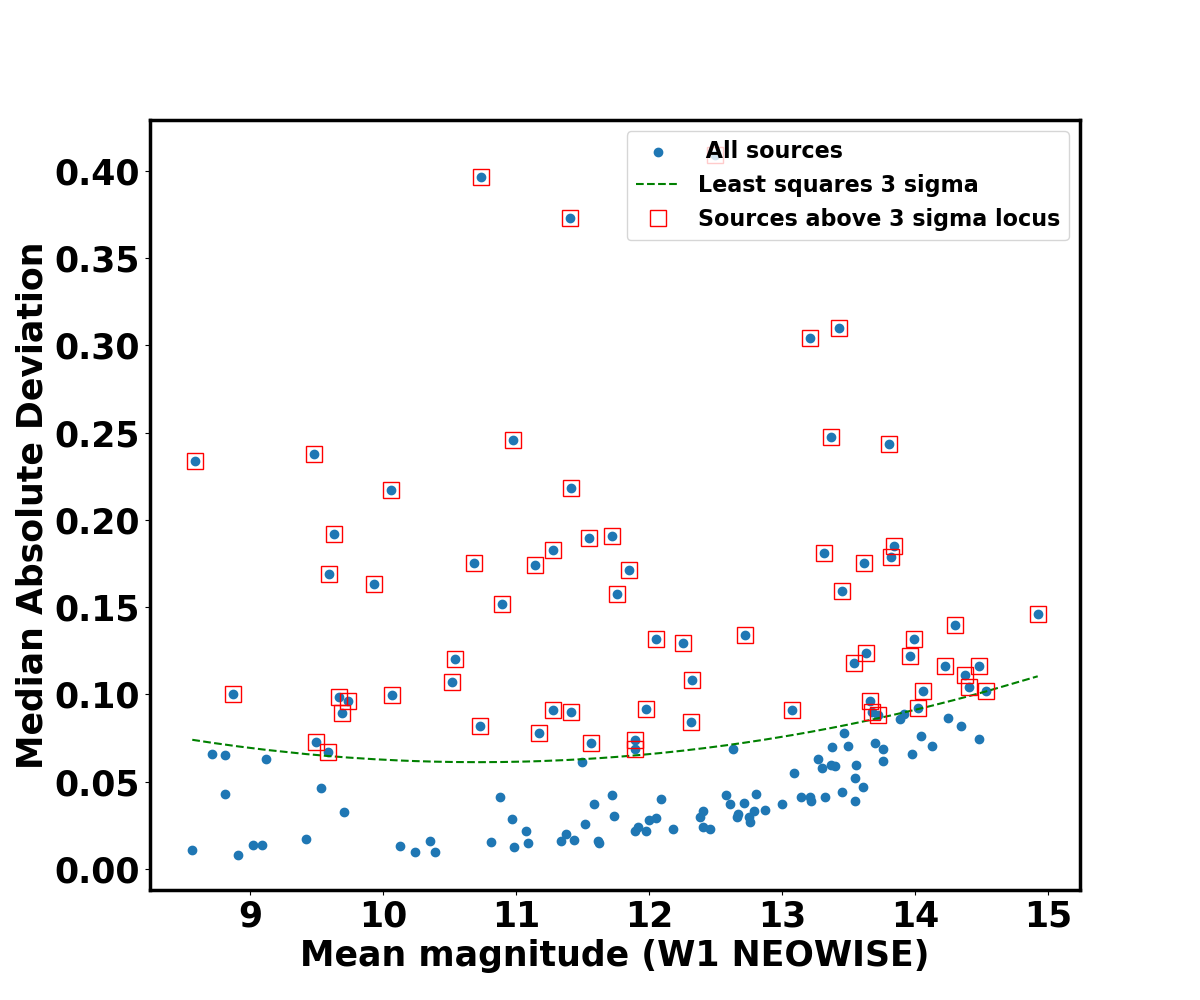}}\par
      \subcaptionbox{MAD (W2)}{\includegraphics[width=\columnwidth]{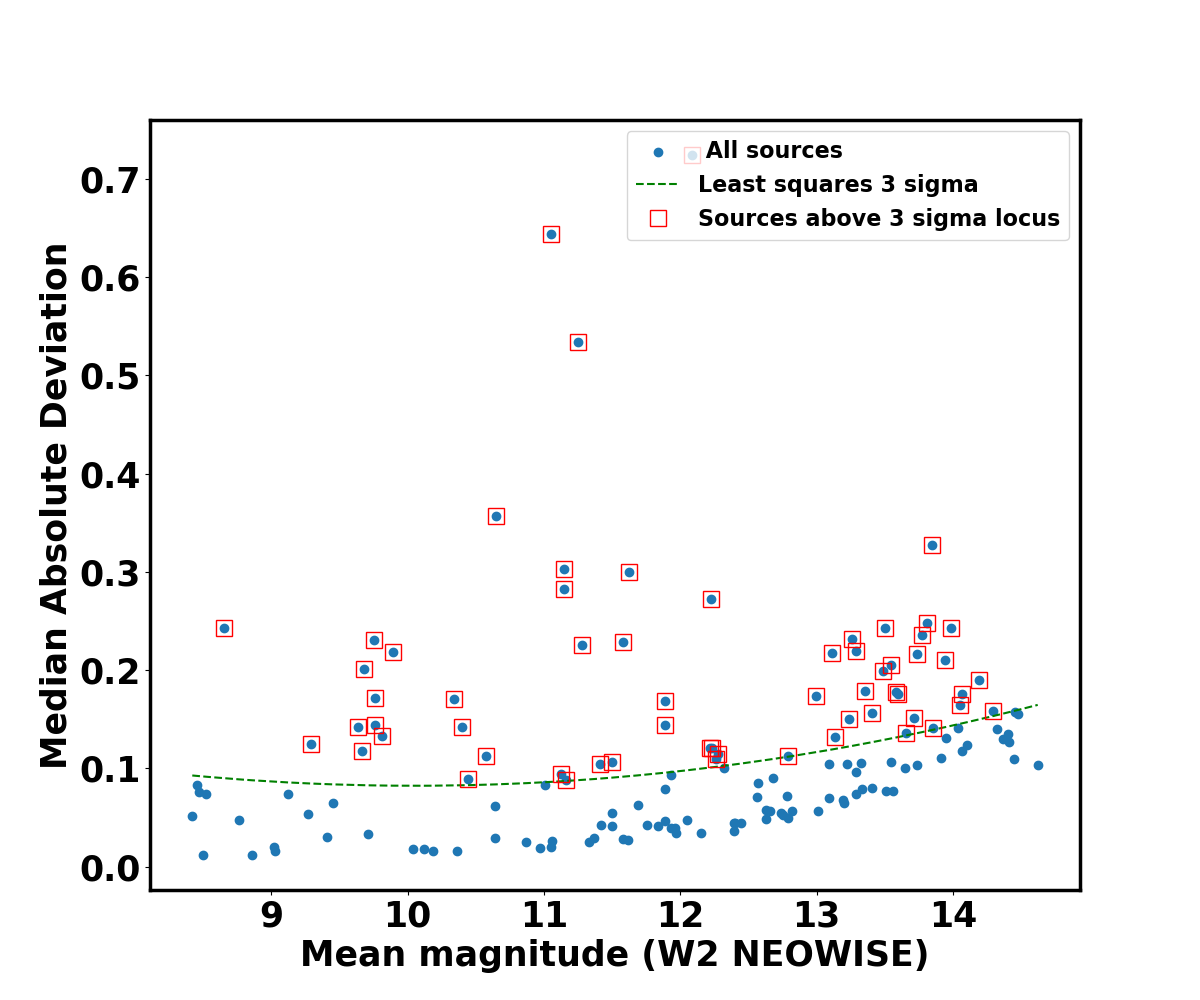}}\par
    \end{multicols}

\caption{Standard Deviation and MAD of sources in NEOWISE data as a function of W1 and W2 band magnitude, along with 3$\sigma$ locus, where $\sigma$ is the median SD/MAD value for the NEOWISE counterparts.}
\clearpage
\label{MAD and SD NEOWISE}
\end{figure*}

\begin{figure*}
    \centering
    \begin{minipage}{\textwidth}
        \centering
        \includegraphics[width=\textwidth]{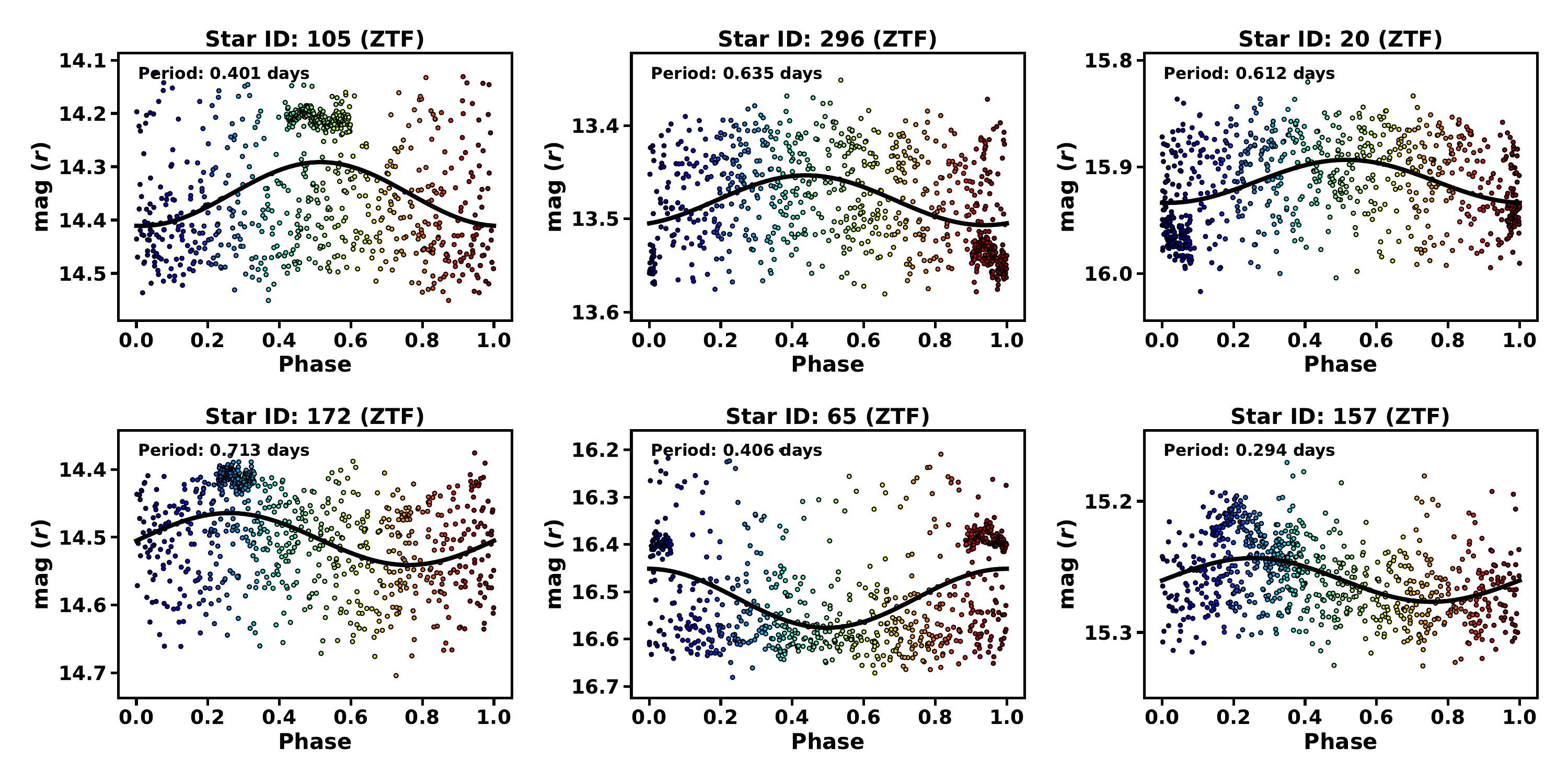}
    \end{minipage}

    \vspace{0.3cm} 

    \begin{minipage}{\textwidth}
        \centering
        \includegraphics[width=0.95\textwidth]{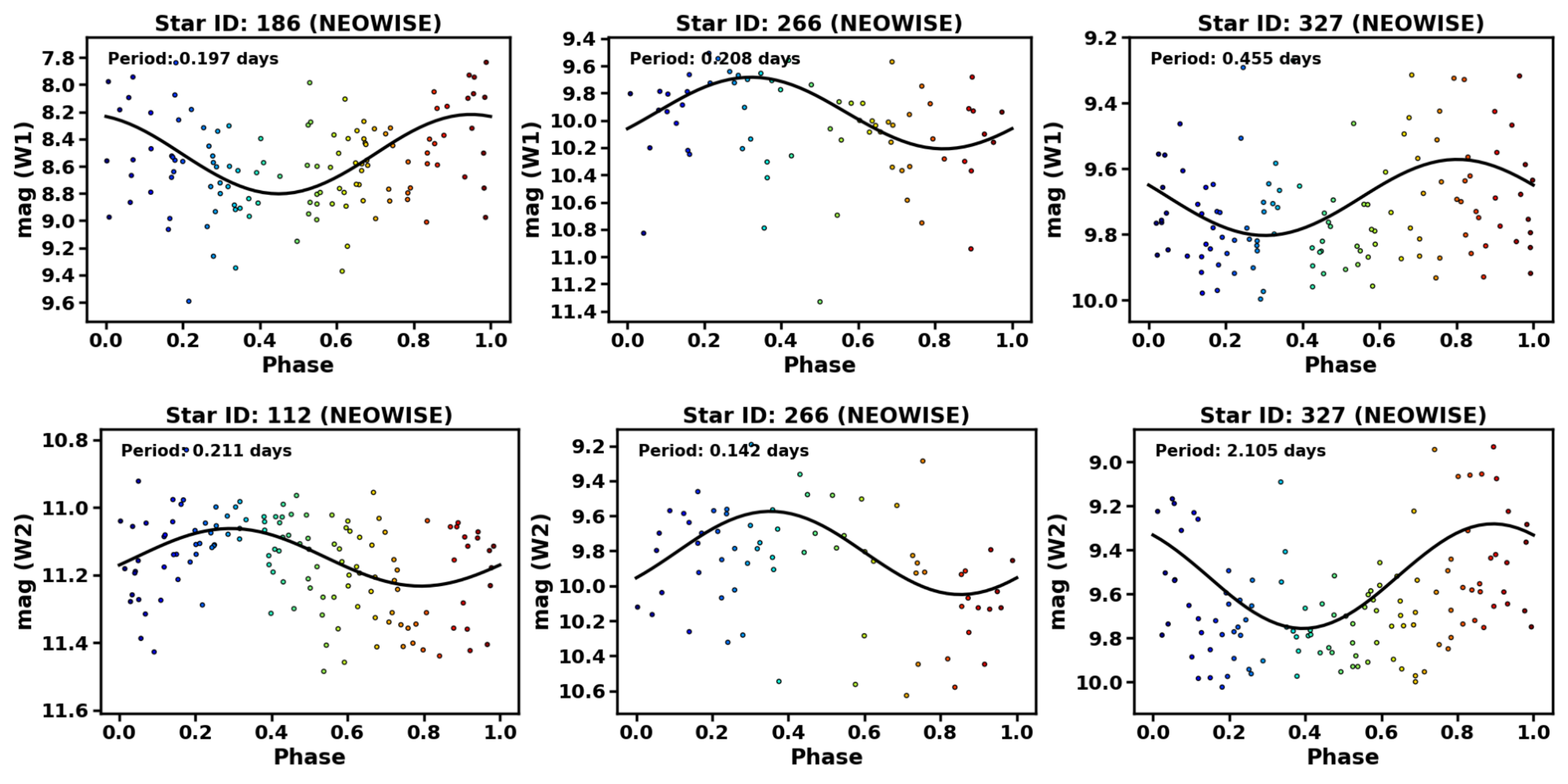}
    \end{minipage}

    \caption{Phase folded lightcurves of CBe stars showing periodic signal due to pulsation/rotation. The respective periods are given inside each plot.}
    \label{CBe NRP}
\end{figure*}

\begin{figure*}
\centerline{\includegraphics[width=2.1\columnwidth]{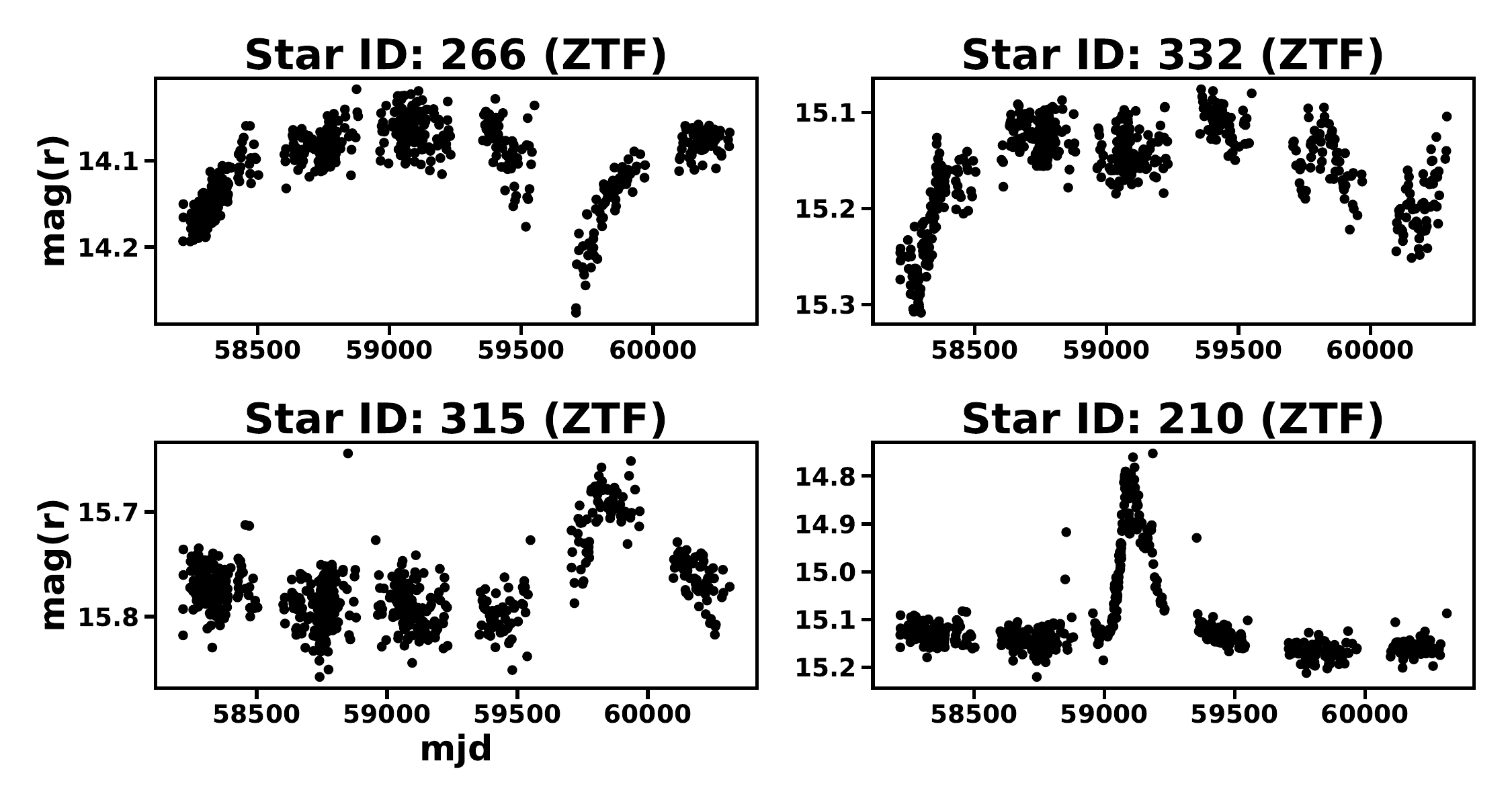}}
\caption{Lightcurves of 4 CBe stars, showing long-term variability due to disk formation/dissipation. The other 2 CBe stars of this type are shown in Fig. \ref{CBelightcurves}.}
\label{CBelongterm}
\end{figure*}

\section{Tables}

\begin{table*}
\centering
\scriptsize
\begin{tabular}{|c|c|c|c|c|c|c|c|c|c|c|}
\hline
\hline
\textbf{StarID} & \textbf{Gaia DR3 ID} & \textbf{ra}$_{\textbf{J2016}}$ & \textbf{dec}$_{\textbf{J2016}}$ & \textbf{$\pi$ (mas)} & \textbf{$\mu_\text{ra}$ (mas/yr)} & \textbf{$\mu_\text{dec}$ (mas/yr)} & \textbf{mag$_G$} & \textbf{mag$_{BP}$} & \textbf{mag$_{RP}$} &  \textbf{Membership probability}\\

\hline

1 & Gaia DR3 2014644948102398592 & 343.17621 & 60.83888 & 0.3405 & -2.744 & -1.685 & 19.09 & 20.58 & 17.91 & 0.915 \\
2 & Gaia DR3 2014646498587993856 & 343.17645 & 60.84831 & 0.2748 & -2.626 & -1.807 & 16.33 & 17.64 & 15.2 & 0.976 \\
3 & Gaia DR3 2014646258069829504 & 343.23409 & 60.85325 & 0.4212 & -2.711 & -1.61 & 17.37 & 18.97 & 16.16 & 0.908 \\
4 & Gaia DR3 2014643234412862080 & 343.24672 & 60.82561 & 0.2676 & -2.749 & -1.729 & 15.75 & 17.35 & 14.53 & 0.93 \\
5 & Gaia DR3 2014647495020396800 & 343.2514 & 60.91139 & 0.2295 & -2.889 & -1.636 & 16.61 & 17.73 & 15.53 & 0.933 \\
6 & Gaia DR3 2014642061884240768 & 343.26184 & 60.76568 & 0.1711 & -2.896 & -1.951 & 17.99 & 19.51 & 16.8 & 0.931 \\
7 & Gaia DR3 2014648074835632640 & 343.29404 & 60.94917 & 0.2488 & -2.929 & -1.756 & 17.76 & 18.82 & 16.74 & 0.928 \\
8 & Gaia DR3 2014647117063666176 & 343.35317 & 60.91576 & 0.2409 & -2.664 & -1.632 & 16.71 & 17.89 & 15.62 & 0.941 \\
9 & Gaia DR3 2014642581577864192 & 343.36459 & 60.80258 & 0.2685 & -2.708 & -1.546 & 17.35 & 18.87 & 16.19 & 0.972 \\
10 & Gaia DR3 2014642581577864064 & 343.3708 & 60.80499 & 0.278 & -2.693 & -1.454 & 13.36 & 14.73 & 12.21 & 0.962 \\
11 & Gaia DR3 2014642680359627904 & 343.37969 & 60.81559 & 0.2412 & -2.612 & -1.66 & 18.84 & 20.34 & 17.64 & 0.949 \\
12 & Gaia DR3 2014642615937604736 & 343.39324 & 60.8086 & 0.245 & -2.783 & -1.619 & 16.41 & 17.71 & 15.28 & 0.974 \\
13 & Gaia DR3 2014646940964249472 & 343.39678 & 60.91958 & 0.2818 & -2.535 & -1.666 & 18.36 & 19.66 & 17.24 & 0.91 \\
14 & Gaia DR3 2014647971756419584 & 343.39711 & 60.95327 & 0.2668 & -2.818 & -1.486 & 18.19 & 19.15 & 17.14 & 0.918 \\
15 & Gaia DR3 2014642611640147456 & 343.39741 & 60.81313 & 0.3344 & -2.52 & -1.323 & 19.02 & 20.51 & 17.85 & 0.92 \\
16 & Gaia DR3 2014642611640160256 & 343.40341 & 60.82153 & 0.3988 & -2.685 & -1.782 & 18.99 & 20.53 & 17.82 & 0.912 \\
17 & Gaia DR3 2014642336762184960 & 343.40753 & 60.78614 & 0.3147 & -2.625 & -1.746 & 17.58 & 19.05 & 16.41 & 0.967 \\
18 & Gaia DR3 2014642714719368832 & 343.41131 & 60.8178 & 0.3632 & -2.607 & -1.498 & 18.23 & 19.73 & 17.03 & 0.966 \\
19 & Gaia DR3 2014630246431830144 & 343.42137 & 60.73835 & 0.2815 & -2.74 & -1.814 & 13.58 & 14.58 & 12.57 & 0.912 \\
20 & Gaia DR3 2014642753376553472 & 343.43839 & 60.83333 & 0.3262 & -2.69 & -1.722 & 15.59 & 17.17 & 14.36 & 0.968 \\
21 & Gaia DR3 2014642817798656000 & 343.44087 & 60.85238 & 0.1648 & -3.118 & -1.549 & 20.22 & 21.4 & 19 & 0.932 \\
22 & Gaia DR3 2014630413932929536 & 343.44198 & 60.76308 & 0.1611 & -2.873 & -1.578 & 18.55 & 19.88 & 17.42 & 0.952 \\
23 & Gaia DR3 2014643646730108672 & 343.44434 & 60.86986 & 0.2899 & -2.725 & -1.529 & 14.22 & 15.43 & 13.13 & 0.964 \\
24 & Gaia DR3 2014643578006806400 & 343.44444 & 60.86876 & 0.3255 & -2.946 & -1.52 & 18.28 & 19.44 & 17.04 & 0.916 \\
25 & Gaia DR3 2014642817798641152 & 343.44788 & 60.84399 & 0.3365 & -2.838 & -1.481 & 18.75 & 20.22 & 17.55 & 0.966 \\
26 & Gaia DR3 2014642542920679680 & 343.44989 & 60.81892 & 0.2863 & -2.875 & -1.803 & 18.17 & 19.68 & 16.99 & 0.979 \\
27 & Gaia DR3 2014630620091398400 & 343.45053 & 60.7838 & 0.3134 & -2.683 & -1.514 & 17.74 & 19.06 & 16.61 & 0.994 \\
28 & Gaia DR3 2014642478498660224 & 343.45099 & 60.80874 & 0.2456 & -2.655 & -2.026 & 18.47 & 19.67 & 17.32 & 0.977 \\
29 & Gaia DR3 2014642474208701312 & 343.45291 & 60.80851 & 0.3043 & -2.799 & -1.779 & 18.47 & 19.83 & 17.32 & 0.984 \\
30 & Gaia DR3 2014630620091384704 & 343.45386 & 60.77529 & 0.339 & -2.509 & -1.449 & 18.36 & 19.68 & 17.24 & 0.989 \\
31 & Gaia DR3 2014642474201180160 & 343.45527 & 60.80658 & 0.3948 & -2.535 & -1.799 & 18.63 & 19.98 & 17.49 & 0.955 \\
32 & Gaia DR3 2014630723170639616 & 343.45531 & 60.79803 & 0.3519 & -2.782 & -1.598 & 19.54 & 21.09 & 18.35 & 0.952 \\
33 & Gaia DR3 2014653056997696512 & 343.45975 & 60.94957 & 0.3522 & -2.691 & -1.43 & 17.66 & 18.84 & 16.58 & 0.952 \\
34 & Gaia DR3 2014649522245365120 & 343.46037 & 60.89032 & 0.2974 & -2.681 & -1.52 & 15.74 & 16.91 & 14.66 & 0.96 \\
35 & Gaia DR3 2014630658748675072 & 343.4615 & 60.78477 & 0.2158 & -2.856 & -1.923 & 18.53 & 19.82 & 17.32 & 0.963 \\
36 & Gaia DR3 2014630723172351232 & 343.46301 & 60.79751 & 0.3829 & -2.814 & -1.658 & 19 & 20.64 & 17.79 & 0.963 \\
37 & Gaia DR3 2014630418230515072 & 343.4653 & 60.76785 & 0.1925 & -2.495 & -1.28 & 17.95 & 19.29 & 16.78 & 0.975 \\
38 & Gaia DR3 2014642547218131712 & 343.46664 & 60.82503 & 0.2782 & -2.663 & -1.626 & 15.08 & 16.61 & 13.88 & 0.995 \\
39 & Gaia DR3 2014642547218131200 & 343.46687 & 60.8266 & 0.2994 & -2.731 & -1.724 & 14.68 & 16.26 & 13.46 & 0.982 \\
40 & Gaia DR3 2014630418230518144 & 343.46798 & 60.76442 & 0.2829 & -2.86 & -1.744 & 15.61 & 16.96 & 14.47 & 0.997 \\
41 & Gaia DR3 2014630723170650240 & 343.47187 & 60.80232 & 0.2806 & -2.742 & -1.582 & 16.35 & 17.84 & 15.15 & 0.998 \\
42 & Gaia DR3 2014642474201188992 & 343.47491 & 60.8107 & 0.3749 & -2.48 & -1.494 & 19.19 & 20.78 & 17.97 & 0.928 \\
43 & Gaia DR3 2014648418435967616 & 343.4754 & 60.83323 & 0.3316 & -2.943 & -1.566 & 18.34 & 19.92 & 17.11 & 0.98 \\
44 & Gaia DR3 2014648349716463744 & 343.47752 & 60.81868 & 0.1793 & -2.895 & -1.394 & 19.37 & 20.82 & 18.14 & 0.934 \\
45 & Gaia DR3 2014648422733752192 & 343.47848 & 60.83771 & 0.1685 & -2.44 & -1.478 & 17.28 & 18.88 & 16.06 & 0.958 \\
46 & Gaia DR3 2014630727468146304 & 343.47965 & 60.80357 & 0.2093 & -2.75 & -1.768 & 16.66 & 18.03 & 15.51 & 0.988 \\
47 & Gaia DR3 2014648422733753984 & 343.48064 & 60.83338 & 0.3352 & -2.708 & -1.36 & 15.97 & 17.43 & 14.79 & 0.996 \\
48 & Gaia DR3 2014648354013920768 & 343.48269 & 60.82119 & 0.3899 & -2.628 & -1.565 & 17.62 & 19.2 & 16.42 & 0.989 \\
49 & Gaia DR3 2014630654451148544 & 343.48517 & 60.78986 & 0.2654 & -2.708 & -1.364 & 19.11 & 20.92 & 17.85 & 0.939 \\
50 & Gaia DR3 2014630658748674176 & 343.48768 & 60.79573 & 0.2846 & -2.639 & -1.493 & 17.04 & 18.63 & 15.82 & 0.964 \\
\hline
\end{tabular}
\caption{Sample table of NGC 7419 members from our membership analysis using Gaia DR3 data.\\
(The full table will be available online.)}
\label{members}
\end{table*}

\begin{table*}
\centering

\begin{tabular}{|c|c|c|c|c|c|c|c|}
\hline
\hline
\textbf{StarID} & \textbf{oid} & \textbf{ra}$_{\textbf{J2000}}$ & \textbf{dec}$_{\textbf{J2000}}$ & \textbf{mag$_r$} & \textbf{Stetson\_Index} & \textbf{Variable} & \textbf{period (days)} \\

\hline
2 & 831206400008687 & 343.17646 & 60.84832 & 16.51 & 69.07 & 0 & - \\
3 & 831206400008544 & 343.23408 & 60.85326 & 17.68 & 45.11 & 0 & - \\
4 & 831206400009263 & 343.24673 & 60.82562 & 16.07 & 63.4 & 0 & - \\
5 & 831206400007191 & 343.2514 & 60.9114 & 16.72 & 97.35 & 0 & 2.45 \\
7 & 831206400006239 & 343.29403 & 60.94918 & 17.85 & 44.92 & 0 & - \\
8 & 831206400007031 & 343.35317 & 60.91577 & 16.84 & 41.26 & 0 & - \\
9 & 831206400009812 & 343.3646 & 60.80259 & 17.59 & 55.66 & 0 & - \\
10 & 831206400009734 & 343.37081 & 60.805 & 13.57 & 69.18 & 0 & - \\
12 & 831206400009594 & 343.39325 & 60.80861 & 16.56 & 149.72 & 1 & - \\
17 & 831206400031822 & 343.40753 & 60.78615 & 17.85 & 57.93 & 0 & - \\
19 & 1837214300027110 & 343.42137 & 60.73836 & 13.64 & 14.74 & 0 & - \\
20 & 831206400008914 & 343.4384 & 60.83334 & 15.92 & 328.13 & 1 & 2.7105, 0.6121 \\
23 & 831206400008010 & 343.44435 & 60.86987 & 14.36 & 73.26 & 0 & - \\
27 & 831206400010299 & 343.45053 & 60.78381 & 17.93 & 31.9 & 0 & - \\
33 & 831206400006113 & 343.45974 & 60.94958 & 17.78 & 34.88 & 0 & - \\
34 & 831206400007521 & 343.46038 & 60.89033 & 15.86 & 24.53 & 0 & - \\
38 & 831206400009115 & 343.46665 & 60.82504 & 15.36 & 100.07 & 0 & - \\
39 & 831206400009066 & 343.46688 & 60.82661 & 14.99 & 97.13 & 0 & - \\
40 & 831206400010824 & 343.46799 & 60.76443 & 15.82 & 48.93 & 0 & - \\
41 & 831206400009729 & 343.47188 & 60.80233 & 16.59 & 196.18 & 1 & - \\
45 & 831206400008752 & 343.47848 & 60.83772 & 17.58 & 28.27 & 0 & - \\
46 & 831206400031554 & 343.47965 & 60.80358 & 16.87 & 50.28 & 0 & - \\
47 & 831206400008884 & 343.48065 & 60.83339 & 16.21 & 10.56 & 0 & - \\
48 & 831206400009207 & 343.48268 & 60.8212 & 17.96 & 33.36 & 0 & - \\
50 & 831206400009915 & 343.48768 & 60.79574 & 17.34 & 53.73 & 0 & - \\
53 & 831206400009455 & 343.49104 & 60.81177 & 17.66 & 51.59 & 0 & - \\
59 & 831206400009672 & 343.50006 & 60.80298 & 16.48 & 54.76 & 0 & - \\
60 & 831206400012445 & 343.50022 & 60.70457 & 16.61 & 32.49 & 0 & - \\
61 & 831206400008183 & 343.50028 & 60.8617 & 12.77 & 32.22 & 0 & - \\
62 & 831206400008560 & 343.50052 & 60.84548 & 16 & 8.84 & 0 & - \\
65 & 831206400010604 & 343.50607 & 60.7716 & 16.51 & 830.09 & 1 & 0.4062, 9.5742 \\
66 & 831206400009567 & 343.50797 & 60.80672 & 17.51 & 60.01 & 0 & - \\
67 & 831206400054424 & 343.50806 & 60.80443 & 16.2 & 61.04 & 0 & - \\
71 & 831206400012136 & 343.51038 & 60.71466 & 17.2 & 42.96 & 0 & - \\
72 & 831206400031912 & 343.51042 & 60.77723 & 16.63 & 66.73 & 0 & - \\
74 & 831206400009618 & 343.51199 & 60.80451 & 13.94 & 105.32 & 0 & 2.1015 \\
76 & 831206400011065 & 343.5127 & 60.75437 & 17.23 & 36.4 & 0 & - \\
77 & 831206400008437 & 343.51536 & 60.85033 & 17.6 & 67.1 & 0 & - \\
80 & 831206400008546 & 343.51765 & 60.84568 & 17.98 & 13.07 & 0 & - \\
81 & 831206400009970 & 343.51814 & 60.79327 & 17.37 & 79.32 & 0 & - \\
82 & 831206400009216 & 343.51834 & 60.81988 & 13.12 & 90.95 & 0 & - \\
83 & 831206400032114 & 343.51924 & 60.76122 & 16.15 & 109.36 & 0 & - \\
84 & 831206400011437 & 343.51936 & 60.74037 & 17.96 & 31.1 & 0 & - \\
86 & 831206400008902 & 343.52016 & 60.8318 & 16.54 & 48.58 & 0 & - \\
87 & 831206400031461 & 343.52021 & 60.80849 & 17.36 & 26.74 & 0 & - \\
88 & 831206400043581 & 343.52077 & 60.71451 & 16.26 & 45.1 & 0 & - \\
89 & 831206400030865 & 343.52148 & 60.85359 & 17.38 & 24.63 & 0 & - \\
92 & 831206400008979 & 343.52242 & 60.82831 & 16.47 & 23.02 & 0 & - \\
94 & 831206400007101 & 343.523 & 60.90815 & 16.3 & 56.35 & 0 & - \\
97 & 831206400008593 & 343.52716 & 60.8431 & 16.99 & 63.65 & 0 & -\\
\hline
\end{tabular}
\caption{Sample variability and periodicity table for NGC 7419 members using ZTF $r$ band data. StarID corresponds to the ID assigned during membership in Section \ref{membershipsection}.  Variable and non-variable stars are indicated by 1 and 0, respectively, in the variable column. \\(The full table will be available online.)}
\label{ZTF detailed table}
\end{table*}

\begin{table*}
\centering

\begin{tabular}{|c|c|c|c|c|c|c|c|c|}
\hline
\hline
\textbf{StarID}  & \textbf{ra}$_{\textbf{J2000}}$ & \textbf{dec}$_{\textbf{J2000}}$ & \textbf{mag$_\text{W1}$} &  \textbf{mag$_\text{W2}$} & \textbf{Stetson\_Index} & \textbf{Variable} & \textbf{period$_\text{W1}$ (days)} &\textbf{period$_\text{W2}$ (days)} \\

\hline

1 & 343.17617 & 60.83897 & 14.24 & 14.1 & 10.46 & 0 & - & - \\
2 & 343.17649 & 60.84832 & 12.18 & 12.15 & 41.34 & 0 & - & - \\
3 & 343.23413 & 60.85326 & 12.79 & 12.75 & 53.33 & 0 & - & - \\
4 & 343.24672 & 60.82562 & 11.09 & 11.05 & 63.99 & 0 & - & - \\
5 & 343.25144 & 60.91136 & 12.75 & 12.74 & 87.64 & 0 & - & - \\
6 & 343.26189 & 60.76569 & 13.6 & 13.65 & 48.33 & 0 & - & - \\
7 & 343.29399 & 60.94918 & 14.34 & 14.36 & 9.63 & 0 & - & 0.1651 \\
8 & 343.35304 & 60.91573 & 12.66 & 12.63 & 77.25 & 0 & - & - \\
10 & 343.37082 & 60.80501 & 9.09 & 9.03 & 200.8 & 0 & - & - \\
12 & 343.39327 & 60.80861 & 12.38 & 12.39 & 93.9 & 0 & - & - \\
13 & 343.39705 & 60.9196 & 14.04 & 14.07 & 29.74 & 0 & - & 103.9236 \\
15 & 343.39731 & 60.81297 & 14.41 & 14.44 & 16.48 & 0 & - & 0.6096 \\
16 & 343.4035 & 60.82162 & 13.88 & 13.29 & 37.36 & 0 & - & 0.142 \\
17 & 343.40754 & 60.78625 & 13.14 & 13.09 & 52.57 & 0 & - & - \\
18 & 343.41168 & 60.81773 & 13.5 & 13.51 & 28.01 & 0 & - & - \\
19 & 343.42137 & 60.73837 & 10.13 & 10.12 & 122.39 & 0 & - & - \\
20 & 343.43842 & 60.83334 & 10.35 & 10.04 & 216.62 & 0 & - & - \\
22 & 343.44211 & 60.76299 & 14.02 & 13.81 & -28.48 & 0 & - & - \\
23 & 343.44437 & 60.86984 & 10.39 & 10.36 & 57.08 & 0 & - & - \\
25 & 343.44789 & 60.84399 & 14.22 & 14.32 & 8.14 & 0 & - & - \\
26 & 343.44994 & 60.81895 & 13.67 & 13.72 & -13.44 & 0 & 0.2327 & 1.0129 \\
27 & 343.45058 & 60.78369 & 13.61 & 13.66 & -10.91 & 0 & - & - \\
28 & 343.4512 & 60.80854 & 13.31 & 13.36 & 6.01 & 0 & 1.9536 & 0.2177 \\
30 & 343.45394 & 60.77512 & 13.96 & 13.85 & -9.52 & 0 & - & - \\
33 & 343.45969 & 60.94959 & 13.76 & 13.73 & 51.02 & 0 & - & - \\
34 & 343.4604 & 60.89033 & 11.97 & 11.97 & 47.83 & 0 & - & - \\
37 & 343.46499 & 60.76789 & 13.37 & 13.29 & 9.25 & 0 & - & - \\
39 & 343.46689 & 60.82643 & 9.69 & 9.66 & 175.23 & 0 & - & - \\
40 & 343.46798 & 60.76444 & 11.43 & 11.42 & 25.19 & 0 & - & - \\
41 & 343.47187 & 60.80236 & 11.54 & 11.28 & 2653.07 & 1 & - & - \\
43 & 343.47519 & 60.83321 & 13.3 & 13.29 & 25.18 & 0 & - & - \\
45 & 343.4785 & 60.83771 & 12.58 & 12.56 & 129.24 & 0 & - & - \\
46 & 343.47971 & 60.80353 & 12.49 & 12.23 & -82.48 & 0 & - & - \\
47 & 343.48063 & 60.83338 & 11.52 & 11.5 & 135.18 & 0 & - & - \\
48 & 343.48259 & 60.82108 & 12.72 & 12.79 & 52.81 & 0 & 0.1519 & - \\
51 & 343.48807 & 60.75563 & 13.7 & 13.73 & 36.22 & 0 & - & - \\
52 & 343.48983 & 60.81768 & 13.45 & 13.5 & -61.77 & 0 & - & - \\
53 & 343.49112 & 60.81175 & 13.21 & 12.99 & -59.82 & 0 & - & - \\
54 & 343.49315 & 60.81591 & 13.63 & 13.77 & -2.92 & 0 & - & - \\
60 & 343.50031 & 60.70458 & 12.4 & 12.39 & 98.15 & 0 & - & - \\
61 & 343.50029 & 60.86169 & 8.56 & 8.5 & 87.4 & 0 & - & - \\
62 & 343.50038 & 60.84546 & 12.05 & 11.96 & 78.65 & 0 & - & - \\
65 & 343.50609 & 60.77159 & 11.28 & 11.15 & 1860.56 & 1 & - & - \\
71 & 343.51041 & 60.71462 & 12.87 & 12.82 & 108.09 & 0 & - & - \\
72 & 343.51041 & 60.77717 & 11.84 & 12.09 & 173.09 & 0 & - & - \\
74 & 343.5119 & 60.80454 & 9.67 & 9.76 & 95.06 & 0 & - & - \\
76 & 343.51279 & 60.75438 & 13.07 & 13.11 & 116.85 & 0 & - & - \\
77 & 343.51537 & 60.85025 & 13.08 & 13.09 & 5.11 & 0 & - & - \\
80 & 343.51776 & 60.84556 & 13.72 & 13.59 & 32.34 & 0 & - & - \\
82 & 343.51836 & 60.81988 & 9.02 & 9.02 & 67.17 & 0 & - & -\\
\hline
\end{tabular}
\caption{Sample variability and periodicity table for NGC 7419 members using NEOWISE 2 band data. StarID corresponds to the ID assigned during membership in Section \ref{membershipsection}.  Variable and non-variable stars are indicated by 1 and 0, respectively, in the variable column.\\
(The full table will be available online.)}
\label{NEOWISE detailed table}
\end{table*}

\begin{table*}
\scriptsize
\centering
\begin{tabular}{|c|c|c|c|c|c|c|c|c|c|c|c|}
\hline
\hline
\textbf{StarID} & \textbf{ra}$_{\textbf{J2000}}$ & \textbf{dec}$_{\textbf{J2000}}$ & \textbf{sp\_type} & \textbf{Radius (R$_\odot$)} & \textbf{Mass (M$_\odot$)} & \textbf{v\_breakup (km/s)} & \textbf{Variable} & \textbf{period\_ZTF} & \textbf{period\_W1} & \textbf{period\_W2 }  & \textbf{CBe}\\
&&&&&& \textbf{(km/s)} & \textbf{(ZTF, NEOWISE)} & \textbf{(days)} & \textbf{(days)} & \textbf{(days)} &\\

\hline
2 & 343.17645 & 60.84831 & B4V-B5V$^{(b)}$ & 3.44 & 5.02 & 430.92 & 0,0 & - & - & - & 0 \\
3 & 343.23409 & 60.85325 & B8V-B9V$^{(b)}$ & 2.81 & 3.3 & 386.25 & 0,0 & - & - & - & 0 \\
4 & 343.24672 & 60.82561 & B2V-B2.5V$^{(b)}$ & 3.91 & 6.26 & 451.18 & 0,0 & - & - & - & 0 \\
5 & 343.2514 & 60.91139 & B5V-B6V$^{(b)}$ & 3.31 & 4.49 & 415.16 & 0,0 & 2.45 & - & - & 0 \\
6 & 343.26184 & 60.76568 & B9.5V-A0V$^{(b)}$ & 2.39 & 2.57 & 369.6 & -,0 & - & - & - & 0 \\
7 & 343.29404 & 60.94917 & B8V-B9V$^{(b)}$ & 2.53 & 2.83 & 376.59 & 0,0 & - & - & 0.1651 & 0 \\
8 & 343.35317 & 60.91576 & B5V-B6V$^{(b)}$ & 3.27 & 4.31 & 409.46 & 0,0 & - & - & - & 0 \\
9 & 343.36459 & 60.80258 & B8V-B9V$^{(b)}$ & 2.83 & 3.32 & 386.75 & 0,- & - & - & - & 0 \\
10 & 343.3708 & 60.80499 & B1V-B1.5V$^{(b)}$ & 5.64 & 11.6 & 511.63 & 0,0 & - & - & - & 0 \\
12 & 343.39324 & 60.8086 & B4V-B5V$^{(b)}$ & 3.4 & 4.87 & 426.76 & 1,0 & - & - & - & 0 \\
13 & 343.39678 & 60.91958 & B9.5V-A0V$^{(b)}$ & 2.21 & 2.21 & 356.86 & -,0 & - & - & 103.9236 & 0 \\
14 & 343.39711 & 60.95327 & B9.5V-A0V$^{(b)}$ & 2.29 & 2.37 & 362.96 & -,- & - & - & - & 0 \\
17 & 343.40753 & 60.78614 & B8V-B9V$^{(b)}$ & 2.66 & 3.03 & 381.12 & 0,0 & - & - & - & 0 \\
18 & 343.41131 & 60.8178 & B9.5V-A0V$^{(b)}$ & 2.27 & 2.34 & 361.62 & -,0 & - & - & - & 0 \\
19 & 343.42137 & 60.73835 & B1V-B1.5V$^{(b)}$ & 5.51 & 11.24 & 509.56 & 0,0 & - & - & - & 0 \\
20 & 343.43839 & 60.83333 & B2V-B2.5V$^{(b)}$ & 3.97 & 6.66 & 461.94 & 1,0 & 2.7105, 0.6121 & - & - & 1 \\
23 & 343.44434 & 60.86986 & B1V-B1.5V$^{(b)}$ & 5.03 & 9.93 & 501.12 & 0,0 & - & - & - & 0 \\
24 & 343.44444 & 60.86876 & B9.5V-A0V$^{(b)}$ & 2.25 & 2.29 & 359.85 & -,- & - & - & - & 0 \\
26 & 343.44989 & 60.81892 & B9.5V-A0V$^{(b)}$ & 2.3 & 2.39 & 363.55 & -,0 & - & 0.2327 & 1.0129 & 0 \\
27 & 343.45053 & 60.7838 & B8V-B9V$^{(b)}$ & 2.55 & 2.85 & 377.05 & 0,0 & - & - & - & 0 \\
30 & 343.45386 & 60.77529 & B9.5V-A0V$^{(b)}$ & 2.21 & 2.22 & 357.13 & -,0 & - & - & - & 0 \\
33 & 343.45975 & 60.94957 & B8V-B9V$^{(b)}$ & 2.6 & 2.94 & 379.21 & 0,0 & - & - & - & 0 \\
34 & 343.46037 & 60.89032 & B2V-B2.5V$^{(b)}$ & 3.92 & 6.29 & 451.92 & 0,0 & - & - & - & 0 \\
37 & 343.4653 & 60.76785 & B9.5V-A0V$^{(b)}$ & 2.42 & 2.62 & 371.12 & -,0 & - & - & - & 0 \\
38 & 343.46664 & 60.82503 & B1.5V-B2V$^{(b)}$ & 4.27 & 7.87 & 484.22 & 0,- & - & - & - & 0 \\
39 & 343.46687 & 60.8266 & B1.5V-B2V$^{(b)}$ & 4.64 & 8.87 & 493.17 & 0,0 & - & - & - & 0 \\
40 & 343.46798 & 60.76442 & B2V-B2.5V$^{(b)}$ & 3.96 & 6.59 & 460.26 & 0,0 & - & - & - & 0 \\
41 & 343.47187 & 60.80232 & B1-3Ve$^{(a)}$ & 3.43 & 4.98 & 429.88 & 1,1 & - & - & - & 1 \\
43 & 343.4754 & 60.83323 & B9.5V-A0V$^{(b)}$ & 2.22 & 2.23 & 357.75 & -,0 & - & - & - & 0 \\
45 & 343.47848 & 60.83771 & B7V-B8V$^{(b)}$ & 2.87 & 3.42 & 389.78 & 0,0 & - & - & - & 0 \\
46 & 343.47965 & 60.80357 & B5V-B6V$^{(b)}$ & 3.29 & 4.4 & 412.33 & 0,0 & - & - & - & 0 \\
47 & 343.48064 & 60.83338 & B2.5V-B3V$^{(b)}$ & 3.76 & 5.78 & 442.06 & 0,0 & - & - & - & 0 \\
48 & 343.48269 & 60.82119 & B8V-B9V$^{(b)}$ & 2.63 & 2.99 & 380.14 & 0,0 & - & 0.1519 & - & 0 \\
50 & 343.48768 & 60.79573 & B7V-B8V$^{(b)}$ & 2.92 & 3.78 & 405.76 & 0,- & - & - & - & 0 \\
51 & 343.48802 & 60.75561 & B9.5V-A0V$^{(b)}$ & 2.36 & 2.51 & 367.59 & -,0 & - & - & - & 0 \\
52 & 343.48989 & 60.81777 & B9.5V-A0V$^{(b)}$ & 2.29 & 2.37 & 362.85 & -,0 & - & - & - & 0 \\
53 & 343.49104 & 60.81176 & B8V-B9V$^{(b)}$ & 2.77 & 3.23 & 385.06 & 0,0 & - & - & - & 0 \\
55 & 343.49392 & 60.84363 & B9.5V-A0V$^{(b)}$ & 2.37 & 2.52 & 368.08 & -,- & - & - & - & 0 \\
56 & 343.49546 & 60.82108 & B9V-B9.5V$^{(b)}$ & 2.49 & 2.74 & 374.66 & -,- & - & - & - & 0 \\
59 & 343.50005 & 60.80297 & B1.5V$^{(a)}$ & 3.46 & 5.08 & 432.61 & 0,- & - & - & - & 0 \\
60 & 343.50022 & 60.70457 & B4V-B5V$^{(b)}$ & 3.38 & 4.79 & 424.4 & 0,0 & - & - & - & 0 \\
61 & 343.50027 & 60.86168 & (B0.5-B1)III$^{(a)}$ & 5.93 & 12.67 & 521.18 & 0,0 & - & - & - & 1 \\
62 & 343.50052 & 60.84547 & B1.5V$^{(a)}$ & 3.83 & 5.95 & 444.59 & 0,0 & - & - & - & 0 \\
65 & 343.50606 & 60.77159 & B1.5V$^{(a)}$ & 3.45 & 5.07 & 432.22 & 1,1 & 0.4062, 9.5742 & - & - & 1 \\
66 & 343.50798 & 60.80671 & B7V-B8V$^{(b)}$ & 2.87 & 3.42 & 389.44 & 0,- & - & - & - & 0 \\
67 & 343.50806 & 60.80442 & (B1-B1.5)V$^{(a)}$ & 3.76 & 5.77 & 441.99 & 0,- & - & - & - & 0 \\
68 & 343.50818 & 60.85458 & B7V-B8V$^{(b)}$ & 2.93 & 3.82 & 407.59 & -,- & - & - & - & 1 \\
69 & 343.50832 & 60.86127 & B9.5V-A0V$^{(b)}$ & 2.26 & 2.31 & 360.51 & -,- & - & - & - & 0 \\
71 & 343.51038 & 60.71465 & B7V-B8V$^{(b)}$ & 2.92 & 3.82 & 407.5 & 0,0 & - & - & - & 0 \\
72 & 343.51042 & 60.77722 & B4V-B5V$^{(b)}$ & 3.45 & 5.07 & 432.29 & 0,0 & - & - & - & 0\\
\hline
\end{tabular}
\caption{Sample table of B-type stars in NGC 7419 with their masses, radii, and break-up velocities. StarID corresponds to the ID assigned during membership in Section \ref{membershipsection}. The superscripts (a) and (b) against each spectral type indicate spectral types from previous literature and our estimates, respectively. Stars classified as variable and non-variable stars in our work are indicated by 1 and 0, respectively, in the variable column. A blank in the variability column means no data is available for that star. The CBe column indicates whether the B-type star is also classified as a CBe star or not. The last 2 stars in the full table are CBe stars, which have no masses/radii/breakup velocities reported, are the CBe members from previous literature \protect\citep{2011subramaniam} that don't have Gaia Data available. \\
(The full table will be available online.)}
\label{B-type star table}
\end{table*}

\bsp	
\label{lastpage}
\end{document}